\documentclass[preprint,journal]{vgtc}            
\graphicspath{{figs/}{figures/}{pictures/}{images/}{./}} 

\usepackage{microtype}                 
\PassOptionsToPackage{warn}{textcomp}  
\usepackage{textcomp}                  
\usepackage{mathptmx}                  
\usepackage{times}                     
\usepackage{cite}                      
\usepackage{tabu}                      
\usepackage{booktabs}                  

\usepackage{float}
\usepackage{cool}

\newcommand{\mySetNotation}[1]{{\mathbb{#1}}}
\newcommand{\RRSet}{{\mySetNotation{R}}}

\newcommand{\myBoldNotation}[1]{{\mathbf #1}}

\newcommand{\vp}   {{\myBoldNotation{p}}}

\newcommand{\vr}   {{\myBoldNotation{r}}}

\newcommand{\vv}   {{\myBoldNotation{v}}}

\newcommand{\vx}   {{\myBoldNotation{x}}}
\newcommand{\xx}   {{\myBoldNotation{x}}}

\newcommand{\vNull}{{\myBoldNotation{0}}}

\newcommand{\mC}   {{\myBoldNotation{C}}}
\newcommand{\mD}   {{\myBoldNotation{D}}}

\newcommand{\mI}   {{\myBoldNotation{I}}}

\newcommand{\mNull}{{\myBoldNotation{0}}}

\newcommand{\invn}[2][0cm]{\mathopen{}\left|\left|{#2}\parbox[h][#1]{0cm}{}\right|\right|}

\Style{IdentityMatrixSymb=\mI,%
       DSymb={\mathrm d},%
       IntegrateDifferentialDSymb={\mathrm d},%
       TrParen=p,%
       DetParen=p}
\Style{DDisplayFunc=outset}

\newcommand*{\diff}{\mathop{}\!\mathrm{d}} 

\let\oldmb\mathbold
\protected\def\mathbold{\oldmb}

\newcommand{\etal}{et al.\ }

\usepackage{bm}
\usepackage[svgnames]{xcolor}
\usepackage{pgf,tikz}
\usepackage{pgfplots}
\usepackage{subcaption}
\pgfplotsset{compat=newest}
\usetikzlibrary{positioning,arrows,shapes.multipart,shapes,shadows,patterns,calc}
\usetikzlibrary{spy}

\usepackage{comment}

\usepackage{todonotes}

\DeclareMathOperator*{\argmin}{arg\,min}

\title{FTLE for Flow Ensembles by Optimal Domain Displacement}

\author{J. Zimmermann \& M. Motejat \& C. Rössl \& H. Theisel}

\abstract{
  FTLE (Finite Time Lyapunov Exponent) computation is one of the
  standard approaches to Lagrangian flow analysis.
  The main features of interest in FTLE fields are ridges that
  represent hyperbolic Lagrangian Coherent Structures.
  FTLE ridges tend to become sharp and crisp with increasing
  integration time, where the sharpness of the ridges is an indicator
  of the strength of separation.
  The additional consideration of uncertainty in flows leads to more
  blurred ridges in the FTLE fields.
  There are multiple causes for such blurred ridges:
  either the locations of the ridges are uncertain, or the strength of
  the ridges is uncertain, or there is low uncertainty but weak
  separation.
  Existing approaches for uncertain FTLE computation are unable to
  distinguish these different sources of uncertainty in the ridges.

  We introduce a new approach to define and visualize FTLE fields
  for flow ensembles.
  Before computing and comparing FTLE fields for the ensemble members,
  we compute optimal displacements of the domains to mutually align
  the ridges of the ensemble members as much as possible.
  We do so in a way that an explicit geometry extraction and alignment of the ridges is not necessary.
  The additional consideration of these displacements allows for a
  visual distinction between uncertainty in ridge location, ridge
  sharpness, and separation strength.
  We apply the approach to several synthetic and real ensemble data
  sets.
}

\teaser{
  \centering
  \vspace{-0.1cm}
  \includegraphics[width=\linewidth, interpolate,trim={8cm 21cm 20cm 19cm},clip]{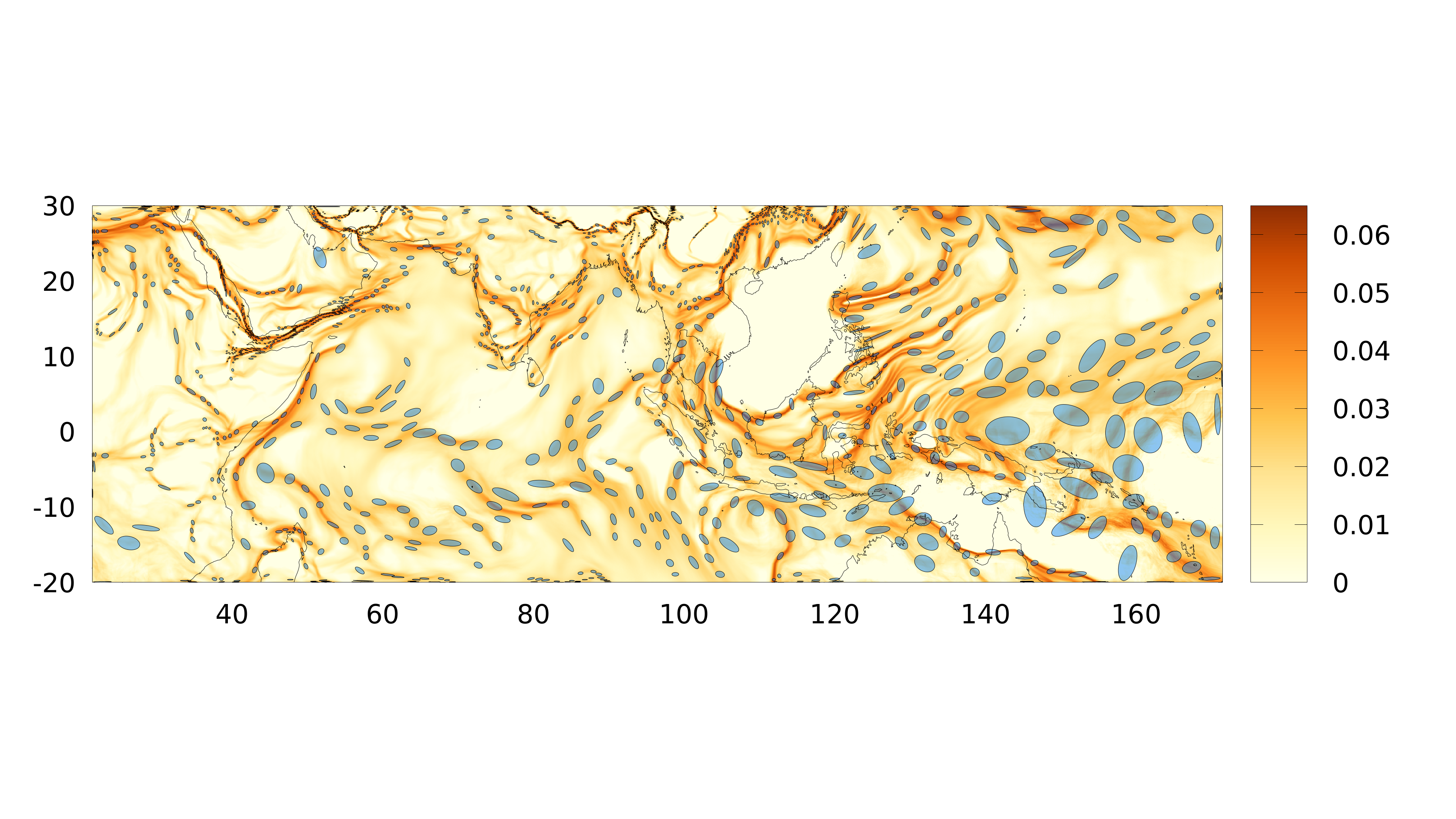}
  \caption{%
  	D-FTLE optimal domain displaced flow maps of wind flow above the Indian Sea. Uncertainty is depicted with ellipses. %
  }
  \label{fig:teaser}
}

\begin{document}

\maketitle


\section{Introduction}
FTLE (Finite Time Lyapunov Exponent) is probably the most popular
approach to computing Lagrangian Coherent Structures (LCS) in
time-dependent (unsteady) flows.
FTLE is particularly tailored towards detecting hyperbolic LCS:
regions of similar hyperbolic Lagrangian flow behavior are separated
by ridge structures in the FTLE field.
For increasing integration times, the increase in hyperbolic
separation tends to lead to thin, crisp and sharp FTLE ridges.
While the ridges are the main objective in analyzing FTLE fields,
their numerical extraction is challenging in various aspects like
accuracy, robustness and computational cost.
For this reason, FTLE computation is still an active area of research.

In recent years, FTLE computation in the presence of uncertainty has
moved into the focus of research.
Several approaches exist for incorporating uncertainty into the
analysis of FTLE fields.
Although their basic concepts are diverse, existing approaches share
one common behavior:
under uncertainty, FTLE ridges tend to become more blurry, unsharp,
and weak.
We show that such blurry ridges in uncertain FTLE fields can have
several reasons:
either they are due to the presence of only a weak separation, or they
are the result of some uncertainty about the ridges.
In the latter case, two different kinds of uncertainty are possible:
strength or location of the FTLE ridges.
There may be a high uncertainty of the ridge location -- we do not
know where exactly the ridge is located -- while the strength of the
ridge is certain.
Conversely, there may be no uncertainty about the exact location of
the ridge, but the strength is uncertain.

We show that existing approaches for uncertain FTLE computation are
unable to distinguish between the different kinds of the uncertainty
about ridges as they produce a similar visual output: blurred ridges.
We argue that FTLE ridges are the main objective of FTLE analysis and
that a better understanding of the kind of uncertainty of FTLE ridges
is necessary for an in-depth uncertain FTLE analysis.

In this paper, we present an approach to computing uncertain
FTLE.
Our approach works under the assumption that the input is an ensemble
of velocity fields and enables distinguishing different kinds of
uncertainty of the ridges.
The main idea is to first find an optimal domain alignment of the flow
maps, \emph{before} FTLE is computed and compared for different ensemble
members.

The input of our approach is an ensemble of $n$ velocity fields
$\vv_1(\vx,t),\dots,\vv_n(\vx,t)$.
We assume that $\vv_1,\dots,\vv_n$ describe the same flow phenomenon
with slightly different parameters.
In particular, we assume that the LCS of $\vv_1,\dots,\vv_n$ are similar
and related to each other:
for an FTLE ridge in $\vv_i$, we can expect a similar FTLE ridge in
$\vv_j$ that is moved, continuously deformed, sharpened or weakened
from the one in $\vv_i$.
This assumption is reasonable due to the nature of flow ensembles:
they are typically produced by observing/simulating the same
phenomenon under slightly different observation/simulation
parameters.
Our approach consists in computing small displacements of the domains
of $\vv_1,\dots,\vv_n$.
We find a map from domain points to domain points, which leads to
displacements (or ``deformations'') of corresponding flow maps such that
their FTLE ridges become aligned as much as possible.
Based on this, we apply a visual analysis of the uncertain FTLE and
the local displacement.
This provides a tool for distinguishing the
sources of FTLE uncertainty mentioned above.

\section{Related Work}

Before reviewing related work, we summarize different ways of representing uncertainty in flows. This is necessary to
give a classification of existing work.
There are two common ways to represent uncertainty: either by having an ensemble $\vv_1,...,\vv_n$ of flows, or by having a velocity field $\vv(\vx,t)$ and a positive semidefinite matrix $\mD(\vx,t)$ describing the local diffusion of the flow. With $\vv$ and  $\mD$ given, Monte Carlo integration techniques can be defined that assume an underlying Wiener process: the result of a stochastic integration step is independent of the steps in the past. While some existing approaches use estimations of  $\mD$ from the ensemble members $\vv_1,...,\vv_n$, we argue that in general it is unclear if a valid estimation  is possible. This is because a Monte Carlo integration may create trajectories that are not physically relevant, i.e., there is no underlying velocity field that fulfills the Navier Stokes equations. In fact, in general a barycentric combination of   $\vv_1,...,\vv_n$ does not fulfill the  Navier Stokes equations, even if each ensemble member $\vv_i$ does. We are not aware of reliable conditions under which $\mD$ can be reliably estimated from $\vv_1,...,\vv_n$. Because of this, our approach restricts to the trajectories of the ensemble members without having a Monte Carlo integration involved.

\subsection{FTLE and FTLE ridges}
\label{sec_ftlerelatedwork}
One of the most prominent approaches to find LCS is the computation of
ridge structures in scalar (FTLE) fields, as introduced by
Haller~\cite{haller01,Haller:2000:Physica}, see also \cite{Haller2015}
for an introduction to LCS, their meaning for describing flow
dynamics and their extraction via FTLE.
FTLE ridges have been used for a variety of applications
\cite{haller02,lekien:2005:Physica,weldon08,Shadden:2009}.
Shadden~\etal~\cite{Shadden:PD:2005} have shown that ridges of FTLE
are approximate material structures, i.e., they converge to material
structures for increasing integration times.
This fact was used in \cite{sadlo2009_timeTopo} to extract
topology structures.
\cite{lipinski:2010} and \cite{Schindler:2012:ridge} introduce
methods for tracking FTLE ridges by locally sampling the FTLE field
and estimating the ridge direction and location.
Due to the discrete sampling used, the accuracy is limited, especially
for very sharp ridges.
In \cite{farazmand2012} the minor eigenvector of the Cauchy-Green
tensor is integrated to track ridge structures.
This approach is, however, prone to accumulating integration errors.
Also in the visualization community, different approaches have been
proposed to increase performance, accuracy and usefulness of FTLE as a
visualization tool
\cite{SCI:Gar2007a,garth07,sadlo07,sadlo07b,sadlo:2009:advection,Germer:2011:EUROVIS,Pobitzer:2011:TOPO}.
Haller and Sapsis \cite{haller:2011:Chaos} additionally explore
the smallest FTLE values.

While most approaches mentioned above restrict themselves to ridge
curves in 2D flows, there are a few approaches that extract ridge
surfaces in 3D flows for moderate integration times.
Schindler~\etal~\cite{Schindler:2012:ridge} show both, standard
height ridge extraction and C-ridge tracking to get 3D surfaces.
\cite{schindler2012doors} show C-ridge surfaces for an analysis of
revolving doors.
Sadlo and Peikert~\cite{sadlo07} present FTLE ridge surfaces where
with focus on an adaptive grid generation.
{\"U}ffinger~\etal~\cite{Uffinger:2013:TVCG} present streak surfaces
as approximations to FTLE ridges.
Depending on the accuracy of the seed structures (obtained by
extremely high sampling), streak surfaces and FTLE ridges show strong
agreement.
\cite{DBLP:journals/tvcg/BarakatT13} propose an adaptive smooth
reconstruction of the flow map field from the sample points based on
Sibson's interpolant on which the ridge extraction is more stable than
on the original sampling.

\subsection{Local uncertainty in vector fields}
Local approaches describe uncertainty as a feature that can be
evaluated at a point inside the vector field domain without
considering the field's ``long-term'' integral behavior.
Sanderson~\etal~\cite{Sanderson04} describe patterns of uncertainty
using a reaction-diffusion model, while
Botchen~\etal~\cite{Botchen-05-VIS} introduce a texture-based
visualization technique, that represents local reliabilities by cross
advection and error diffusion.
The same authors used additional color schemes to emphasize
uncertainty \cite{isfv06-botchen}.
Another approach by Zuk~\etal~\cite{zuk08} uses bidirectional vector
fields to illustrate the impact of uncertainty.

\subsection{Uncertainty for stream lines and path lines}
There exist several approaches for capturing global behavior of stream
lines and path lines for ensembles of vector fields in the
literature.
The perhaps simplest and most direct approach are spaghetti plots
which provide straightforward overviews but tend to produce visual
clutter~\cite{Ferstl2016StreamlineVP}.
Mirzargar~\etal~\cite{Mirzargar2014CurveBG} introduce curve box-plots
for the visualization of curve-like features based on the concept of
statistical data depth.
Otto~\etal~\cite{Otto:2010:EG,Otto:2011:PVIS} present a topological
approach that is based on the integration of vector PDF.
A related method by He~\etal~\cite{wenbinhe2016} for integrating
uncertain stream lines is based on a Bayesian model.
Hollister and Pang~\cite{hollister2020} present a method to measure
uncertainty and to visualize member stream lines from an ensemble of
vector fields by incorporating velocity probability density as a
feature along each member stream line.
Ferstl~\etal~\cite{Ferstl2016StreamlineVP} derive stream line
variability plots by computing a probabilistic mixture model for the
stream line distribution, from which confidence regions can be derived
in which the stream lines are most likely to reside.
For a detailed overview of visualization on ensemble data we refer to
the recent survey by Wang~\etal~\cite{Wang2018VisualizationAV}.

\subsection{FTLE and uncertainty}

Schneider~\etal~\cite{Schneider:2012} introduce Finite Time Variance Analysis (FTVA), a variance based FTLE-like method. The main idea is to integrate several runs of particles along with their spatial neighbors, and apply PCA to the resulting point sets.
Guo~\etal~\cite{Guo:2016} introduce several approaches for uncertain FTLE computation: D-FTLE computes the FTLE field for every run and then considers the mean and the variance of the resulting scalar fields. FTLE-D computes the flow map gradient for each run and then considers its average for FTLE computation.
Guo~\etal~\cite{Guo2019} introduce an approach for particle tracing in uncertain flows.
Haller~\etal~\cite{doi:10.1137/19M1238666} develop a mathematical theory for weakly diffusive tracers to predict transport barriers and enhancers solely from the flow velocity, without reliance on diffusive or stochastic simulations.
A fast implementation of it is presented by  Rapp~\etal~\cite{ Rapp2020}.

\section{Notation and Problem Setting}
Given is an ensemble of $n$ time-dependent $q$-dimensional ($q=2,3$)
flow data sets over the same spatial domain $D \subset \RRSet^q$ and
temporal domain $T=[t_s,t_e]$
\begin{eqnarray*}
  \vv_i: D \times T &\to& \RRSet^q\\
  (\xx,t) &\to& \vv_i(\xx,t)\qquad\text{for}~i=1,\dots,n\,.
\end{eqnarray*}

We consider the flow maps
\begin{eqnarray*}
  \phi_i:  D \times T \times T &\to& D\\
  (\xx,t,\tau) &\to& \phi_i(\xx,t,\tau)~,
\end{eqnarray*}
which describe the location of a massless particle that is seeded at
$(\xx,t)$ and integrated in $\vv_i$ over a time interval $\tau$ such
that $t+\tau \in T$.
As a consequence, $\phi_i$ must satisfy
\begin{equation*}
  \frac{\partial \phi_i(\xx,t,\tau)}{\partial \tau} = \vv_i\left(\phi_i(\xx,t,\tau),t+\tau\right)
  \quad\text{with}\quad
  \phi_i(\xx,t,0) = \xx~
\end{equation*}
for all $\xx \in D$, $t \in T$, $t+\tau \in T$.
Note that we assume for simplicity that the integration of $\vv_i$
does not leave the domain $D$.

Based on this, we consider the standard definition of FTLE as
\begin{equation*}
  \mbox{FTLE}_i(\xx,t,\tau) = \frac{1}{|\tau|} \ln \sqrt{\lambda_{\max}\left( (\nabla \phi_i)^T \, \nabla \phi_i\right)}
\end{equation*}
for $i=1,\dots,n$, where $\nabla \phi_i = \nabla \phi_i(\xx,t,\tau)$ is
the spatial gradient of $\phi_i$ and $\lambda_{\max}$ denotes the largest
eigenvalue of a matrix.

A common approach in the literature is creating more than $n$ flow
maps by a probabilistic process like repeated Monte Carlo
integration.
Such approaches consider a local distribution of the velocity field
and use the assumption that the Monte Carlo integration is a Wiener
process, i.e., the probability distribution for each integration step
is independent of the integration results from the past.
For flow ensembles we do not make this assumption because only the
presence of ensemble members does not allow to estimate a local
distribution of a vector field under consideration of a Wiener
process.

Within this setup, existing approaches for uncertain FTLE computation
can be summarized as follows:
provide a visual representation and analysis of the distribution of
\begin{equation}
  \label{eq_existingapproaches}
  \phi_1,\dots,\phi_n
  \quad\text{or}\quad
  \nabla \phi_1,\dots, \nabla \phi_n
  \quad\text{or}\quad
  \mbox{FTLE}_1,\dots,\mbox{FTLE}_n~,
\end{equation}
where different assumptions on the distribution and different options
for a visual representation exist.

\section{An Introductory Example}
\label{sec:example1}
To illustrate the issues with existing FTLE approaches,
we construct three ensembles containing 50 members each.
We define a family of steady template vector fields
\begin{equation*}
  \vv^{k,r}(\xx) = k \; \vv_{DG}(\xx,r)
\end{equation*}
where $\vv_{DG}$ is the double gyre field from \cite{Shadden2005} (see \eqref{eq_doublegyre} in the 2D domain $\xx \in [0,2] \times [0,1]$ with two
parameters $k>0$ and $r$, where time $t=r$.
For computing the flow map and FTLE of $\vv^{k,r}(\xx)$, we choose $\tau=1$.
This way, $\vv^{k,r}$ consists of an LCS that is represented by the
FTLE ridge whose location depends on $r$.
Further, the strength of separation (corresponding to the sharpness of
FTLE ridges) is controlled by $k$:
the larger $k$, the sharper the FTLE ridge.

We define three ensembles with 50 members each:
\begin{eqnarray}
  \nonumber
  E_1&:& \vv_1 = \vv^{5,\,-0.4} ,\,\dots ,\,\vv_{25} = \vv^{5,\,0} ,\,\dots ,\,\vv_{50} = \vv^{5, \,0.4}  \\
  \nonumber
  E_2&:& \vv_1 = \vv^{2,\,0} ,\, \dots  ,\,\vv_{25} = \vv^{3,\,0} ,\,\dots ,\,\vv_{50} = \vv^{4, \,0}  \\
  \nonumber
  E_3&:& \vv_1 = \vv^{2,\,0} ,\, \dots  ,\,\vv_{25} = \vv^{2,\,0} ,\,\dots  ,\,\vv_{50} = \vv^{2, \,0}.
\end{eqnarray}
This definition implies the following properties of uncertainty:
\begin{itemize}[leftmargin=*]
\item In $E_1$ there is \emph{low}\/ uncertainty about the
  \emph{strength} of the FTLE ridge:
  each ensemble member has an FTLE ridge of approximately the same
  strength.
  But there is \emph{high}\/ uncertainty about the \emph{location}\/
  of the ridges:
  the ridges of the ensemble members are all in different locations.
\item In $E_2$ there is \emph{high}\/ uncertainty about the
  \emph{strength}\/ of the FTLE ridge:
  the ensemble members have an FTLE ridges of different strength.
  But there is a \emph{low}\/ uncertainty about the \emph{location}\/
  of the ridges:
  all ensemble members have the ridge at the same location.
\item In $E_3$ there is low uncertainty about the strength of the FTLE
  ridge and low uncertainty about the location of the ridges.
  In fact, this ensemble does \emph{not}\/ have \emph{any}\/
  uncertainty at all because all ensemble members are identical.
  However, the strength of \emph{separation is weaker}\/ than in $E_1$ and
  $E_2$.
\end{itemize}

\begin{figure}[h]
  \centering
  \begin{subfigure}{0.25\textwidth}
    \includegraphics[width= \textwidth]{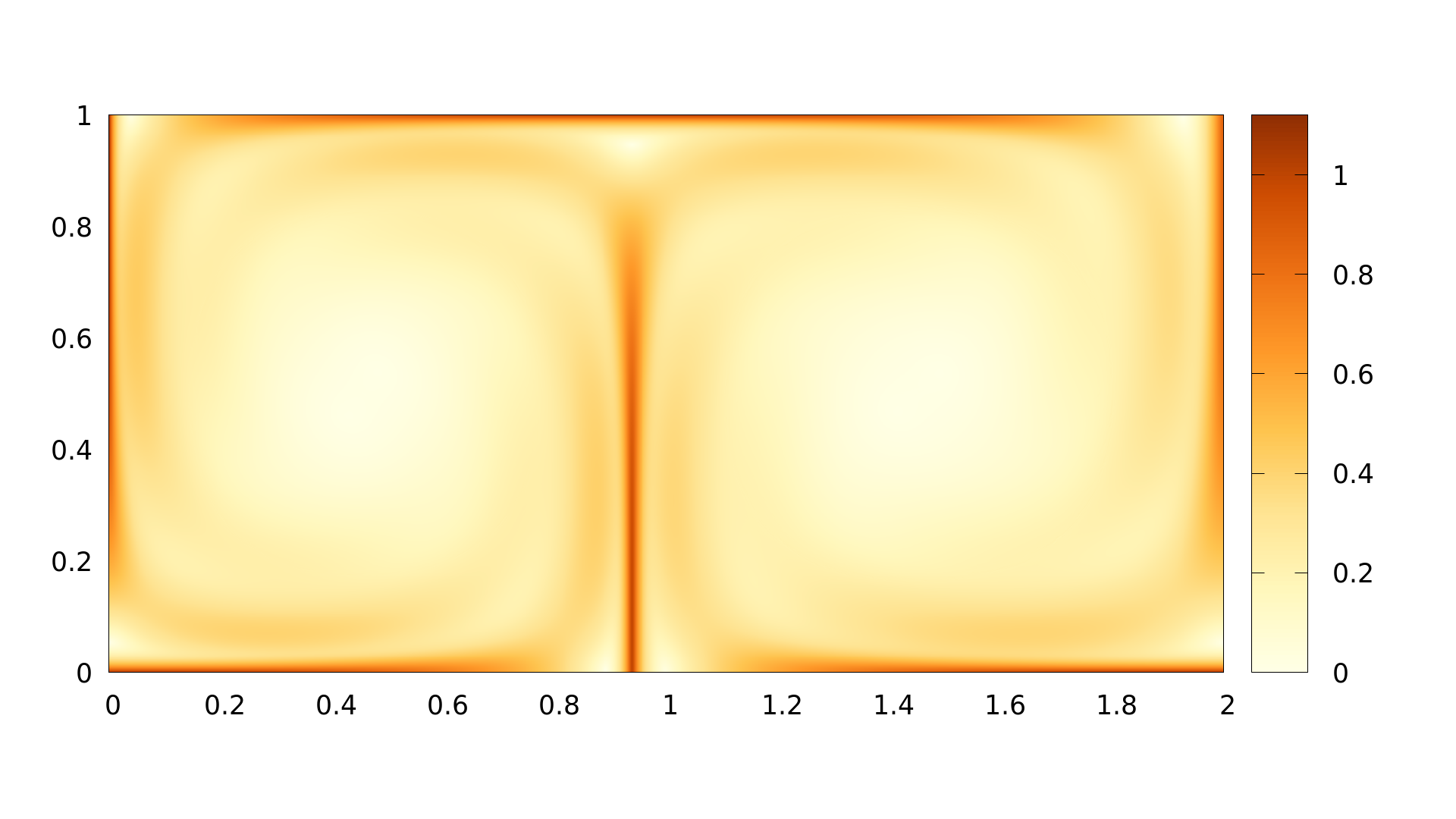}
  \label{fig:sdg_3_ftle_0}
  \end{subfigure}
 \hspace{-0.38cm}
  \vspace{-0.9cm}
      \begin{subfigure}{0.25\textwidth}
    \includegraphics[width= \textwidth]{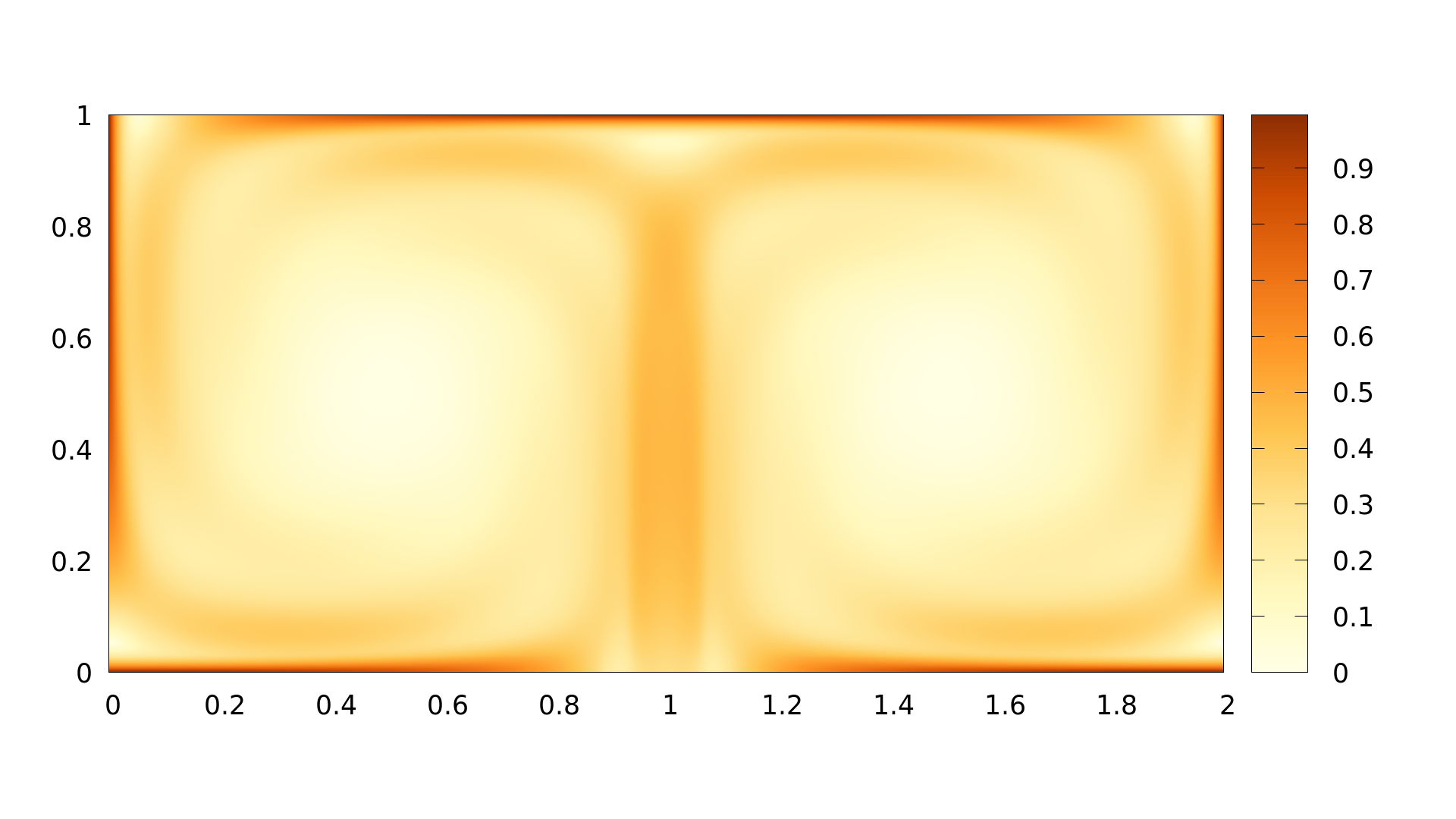}
  \label{fig:sdg_3_d-ftle}
  \end{subfigure}
  \vspace{-0.9cm}
    \begin{subfigure}{0.25\textwidth}
    \includegraphics[width= \textwidth]{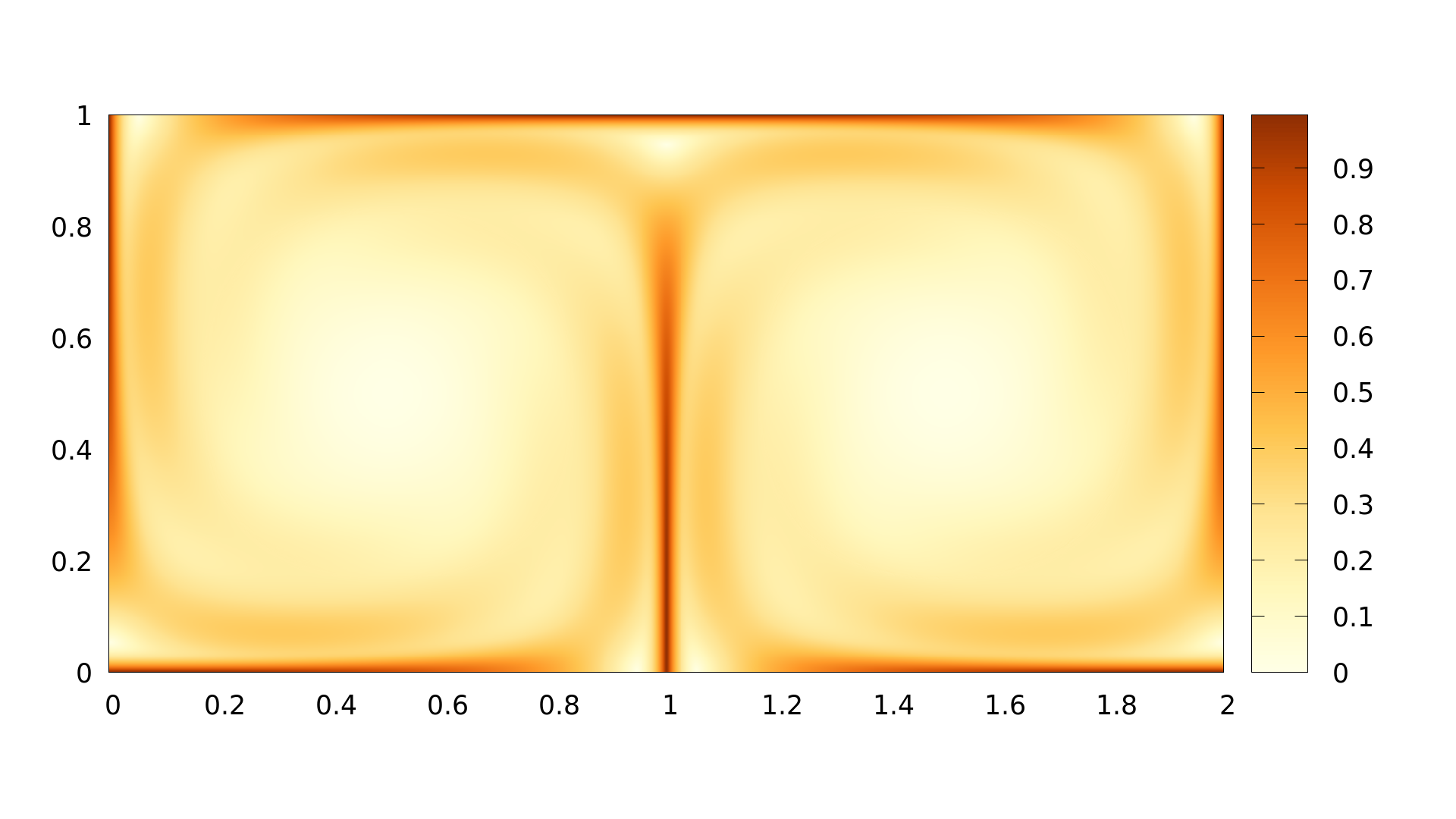}
  \label{fig:sdg_3_ftle_1}
  \end{subfigure}
 \hspace{-0.38cm}
   \begin{subfigure}{0.25\textwidth}
    \includegraphics[width= \textwidth]{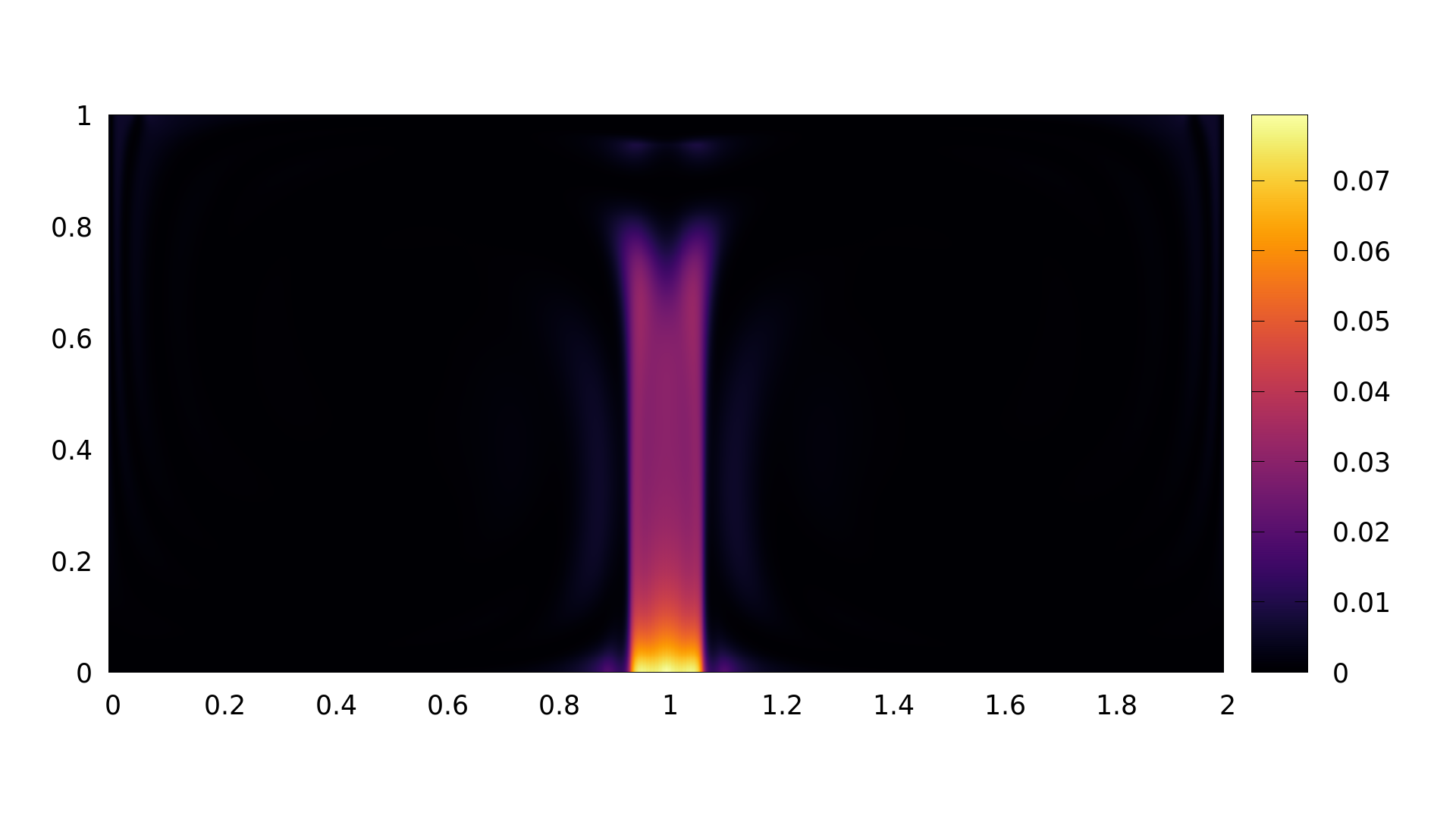}
  \label{fig:sdg_3_variance_d-ftle}
  \end{subfigure}
 \vspace{-0.9cm}
   \begin{subfigure}{0.25\textwidth}
    \includegraphics[width= \textwidth]{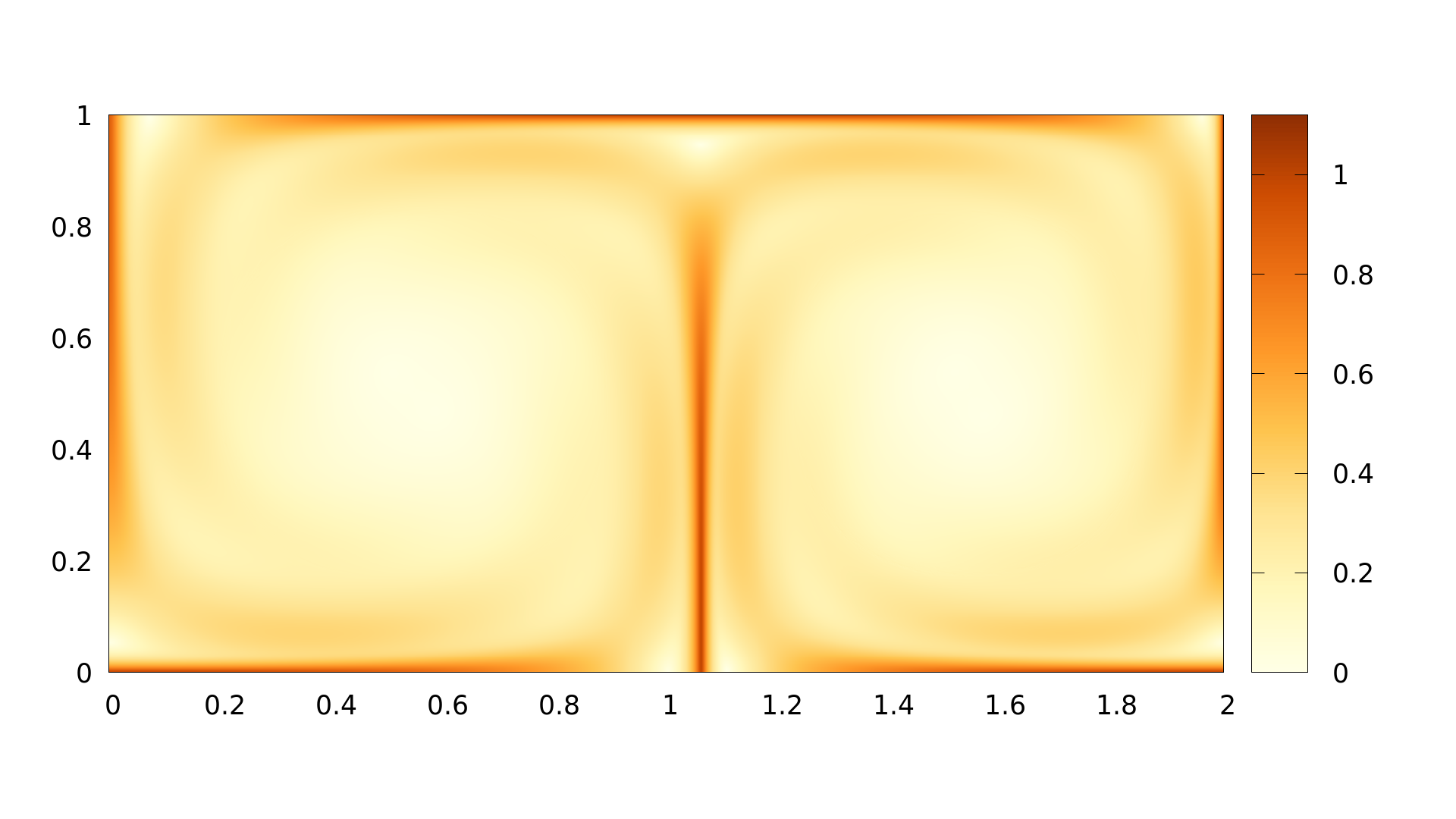}
  \label{fig:sdg_3_ftle_2}
  \end{subfigure}
 \hspace{-0.38cm}
   \begin{subfigure}{0.25\textwidth}
    \includegraphics[width= \textwidth]{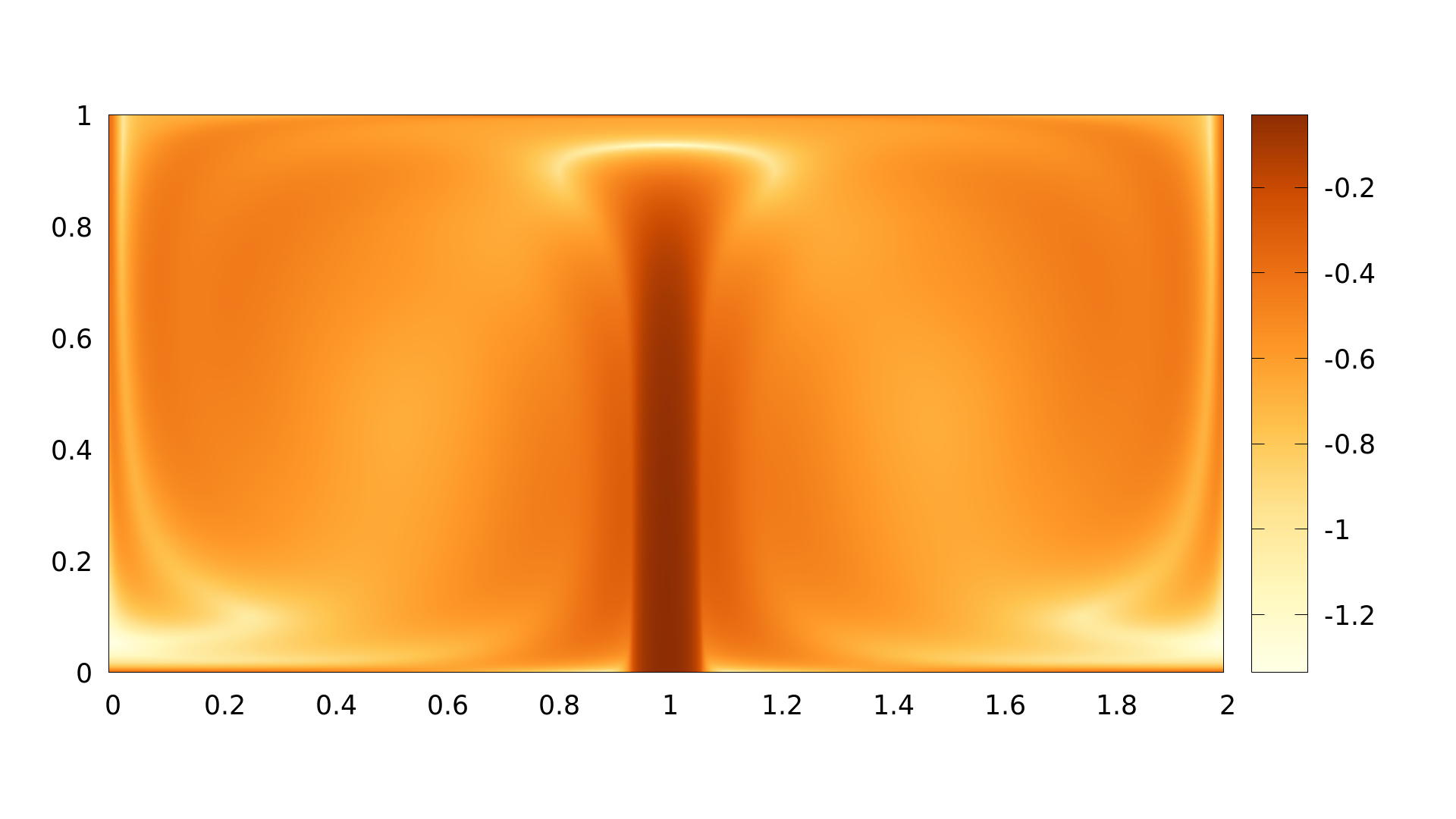}
  \label{fig:sdg_3_ftva}
  \end{subfigure}
    \caption{
    Ensemble $E_1$.
    Left column from top to bottom:
    FTLE of ensemble members $r=-0.4$, $r=0$, $r=0.4$. 
    Right column from top to bottom:
    D-FTLE, variance of D-FTLE, FTVA of the ensemble.
}
\label{fig:sdg_3}
 \end{figure}

The ensemble $E_1$ is illustrated in figure \ref{fig:sdg_3}. In the left column we see the FTLE fields of the ensemble members, while the right column gives a collection of standard uncertain FTLE measures (from top to down:   D-FTLE, variance of D-FTLE, FTVA. They show a common behavior: a rather blurry ridge near the line $x=1$.
Due to the 50 ensembles with a slight difference in the ridge position, there is no possibility to detect the source of the blurred ridge in the D-FTLE image. The variance does not help as it could also indicate variance in ridge strength.

\begin{figure}[h]
  \centering
  \begin{subfigure}{0.25\textwidth}
    \includegraphics[width= \textwidth]{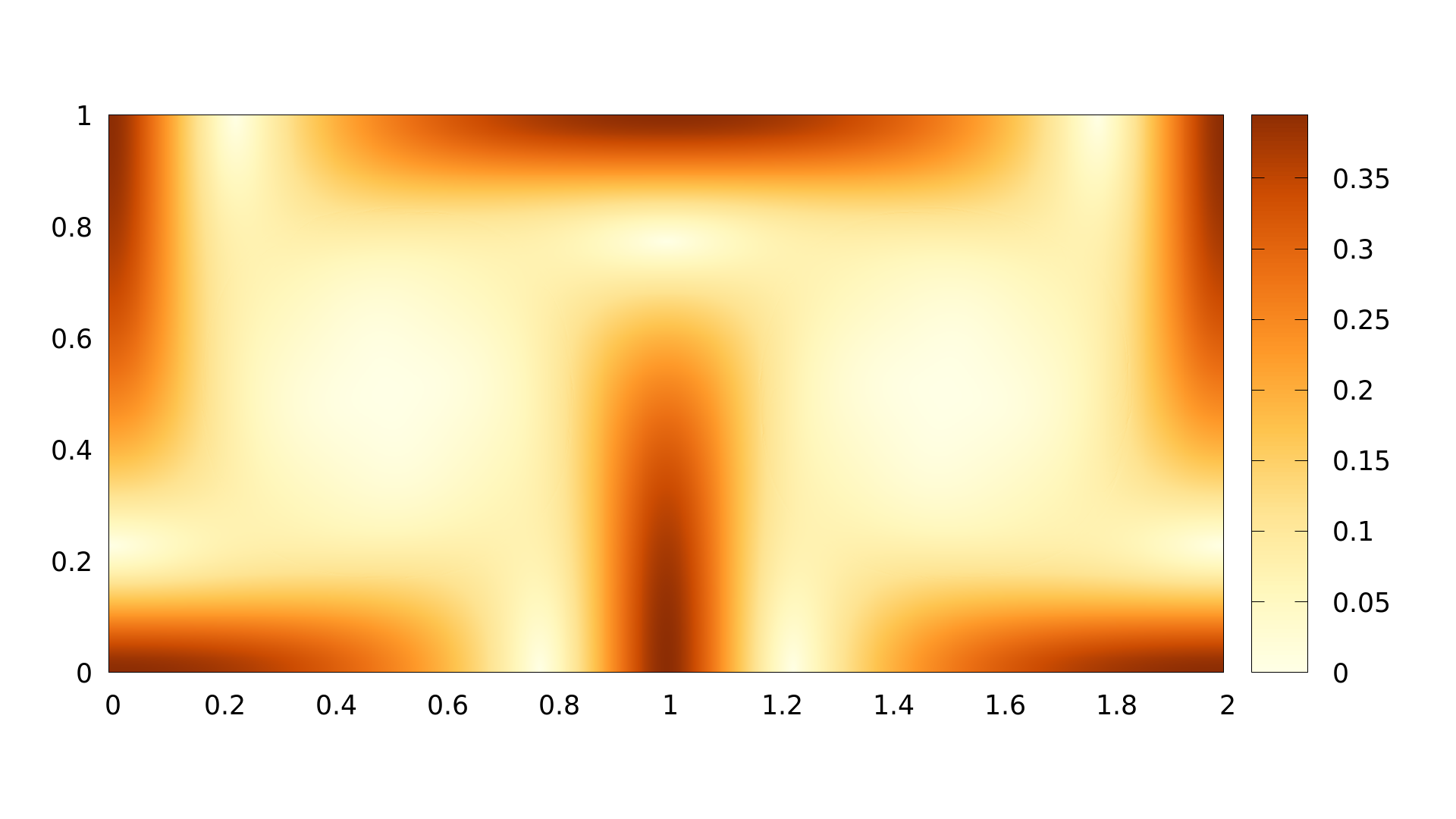}
  \label{fig:sdg_2_ftle_0}
  \end{subfigure}
 \hspace{-0.38cm}
 \vspace{-0.9cm}
      \begin{subfigure}{0.25\textwidth}
    \includegraphics[width= \textwidth]{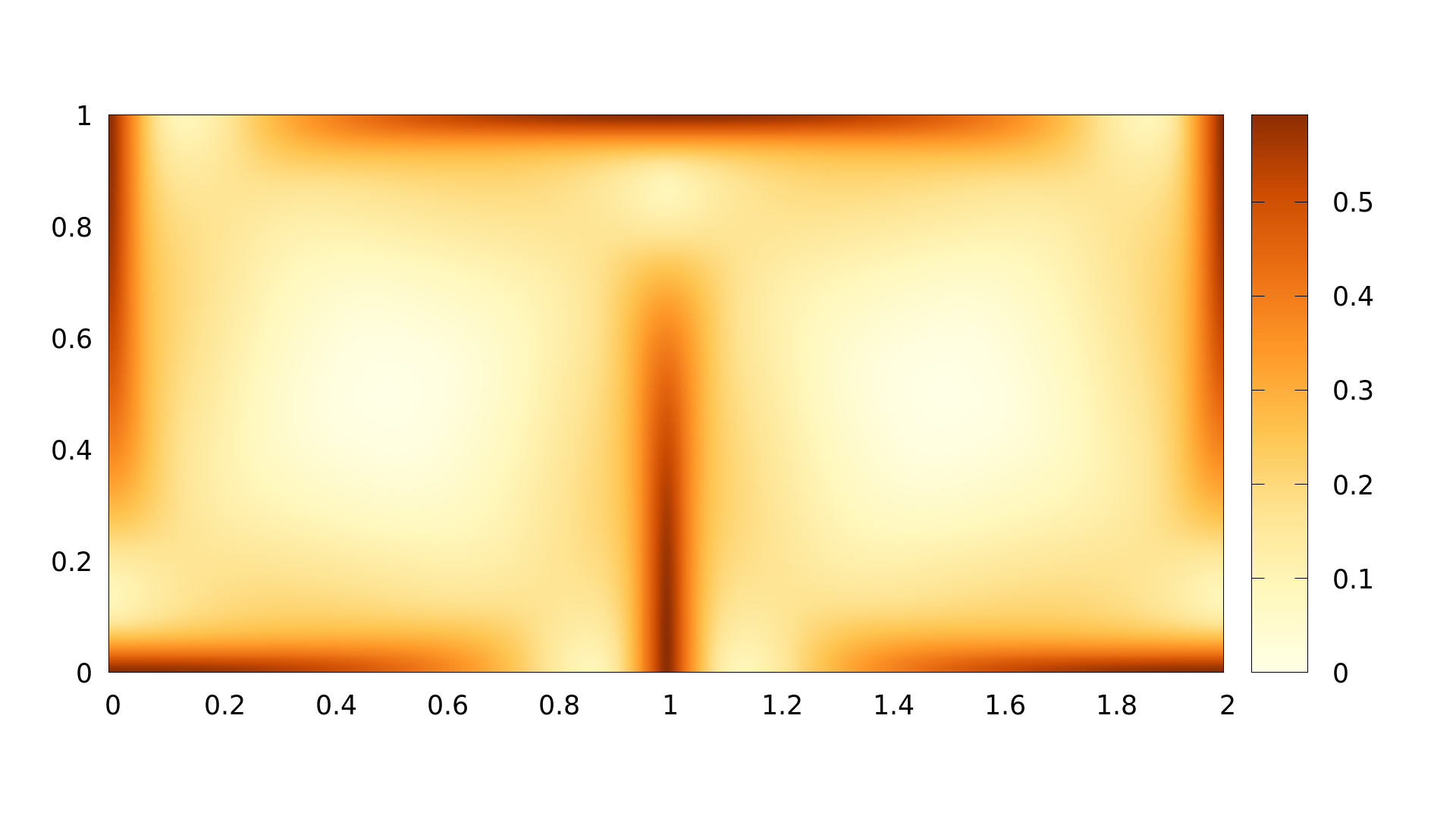}
  \label{fig:sdg_2_d-ftle}
  \end{subfigure}
 \vspace{-0.9cm}
    \begin{subfigure}{0.25\textwidth}
    \includegraphics[width= \textwidth]{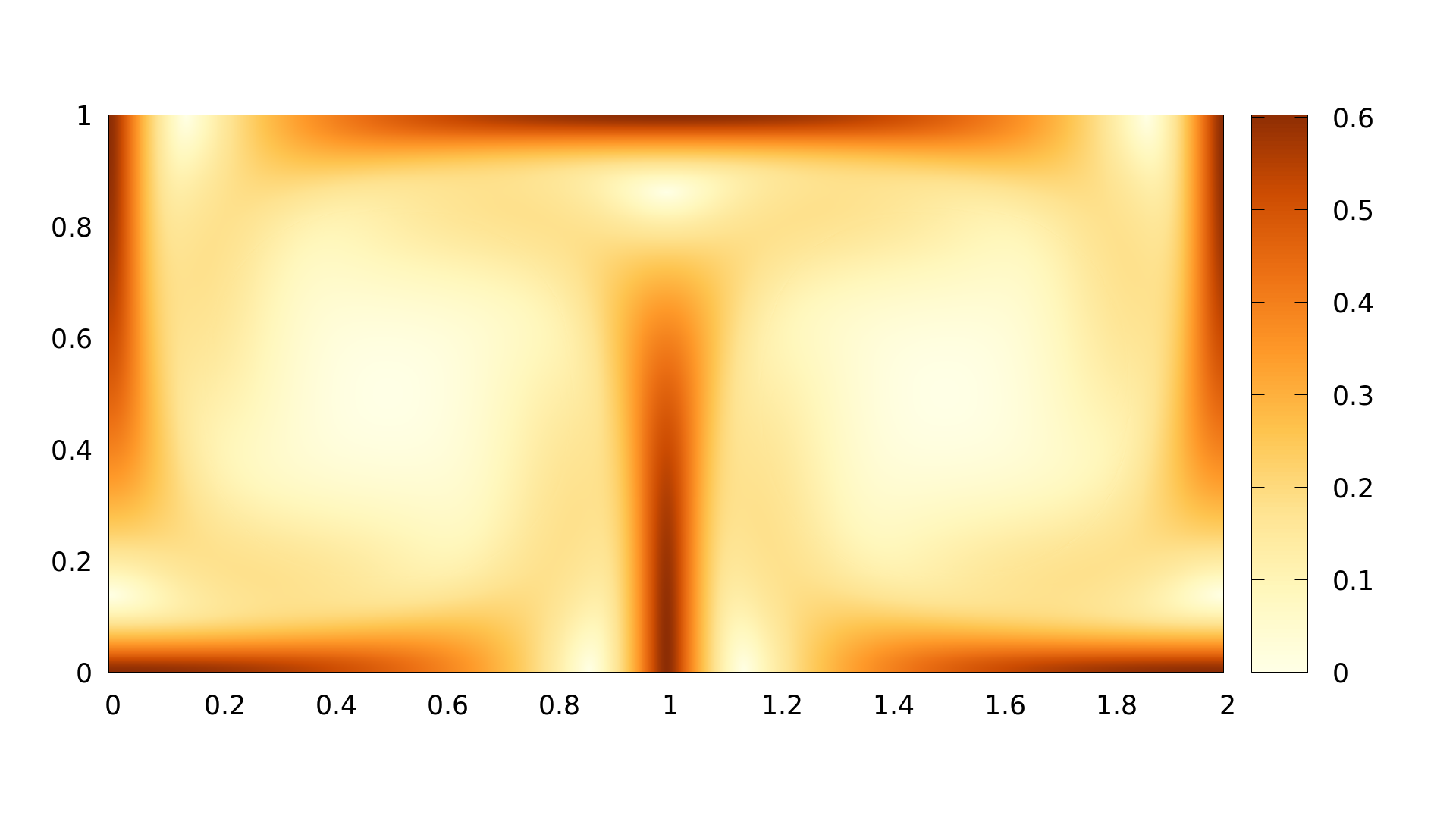}
  \label{fig:sdg_2_ftle_1}
  \end{subfigure}
 \hspace{-0.38cm}
   \begin{subfigure}{0.25\textwidth}
    \includegraphics[width= \textwidth]{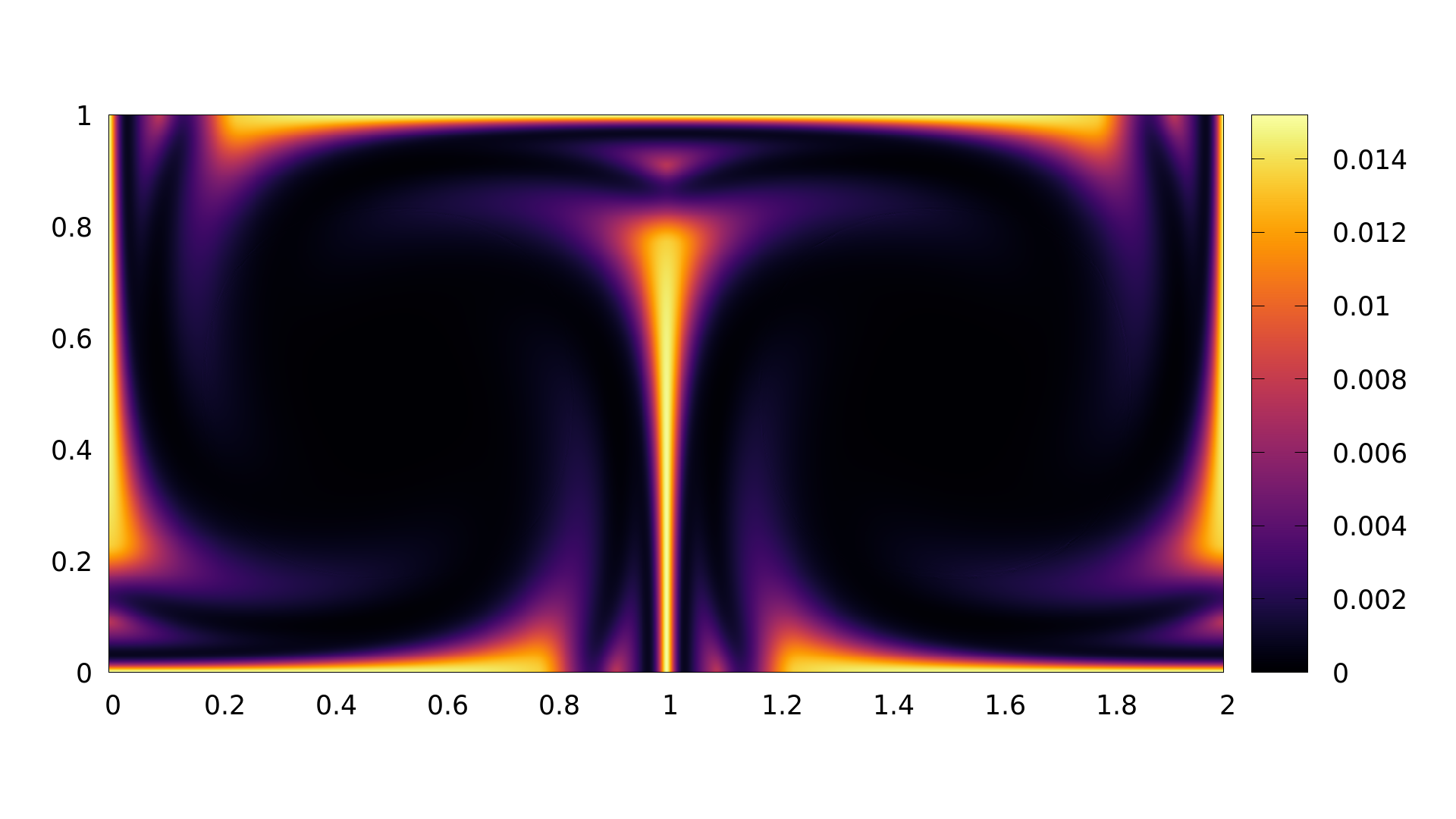}
  \label{fig:sdg_2_variance_d-ftle}
  \end{subfigure}
 \vspace{-0.9cm}
   \begin{subfigure}{0.25\textwidth}
    \includegraphics[width= \textwidth]{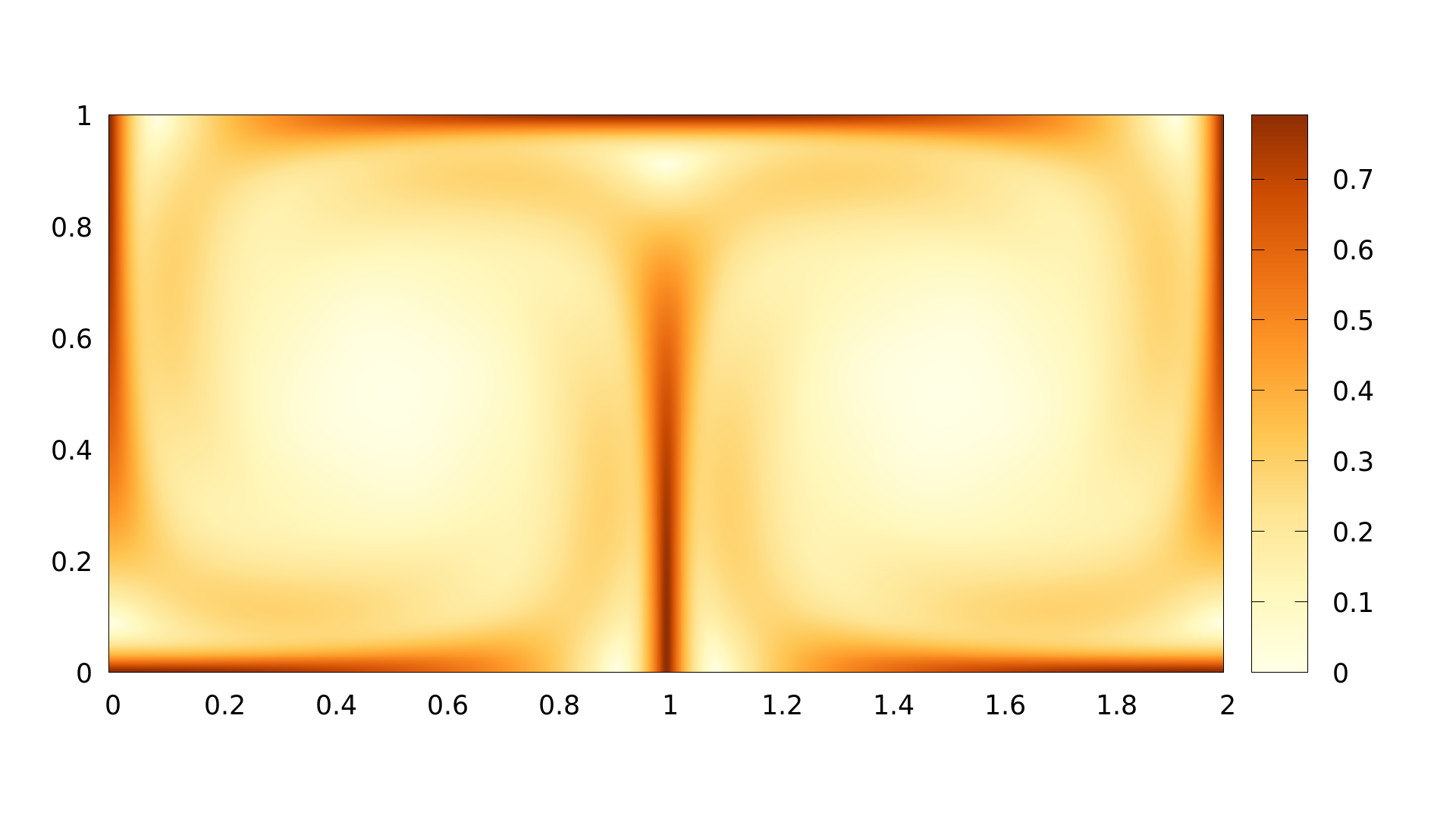}
  \label{fig:sdg_2_ftle_2}
  \end{subfigure}
 \hspace{-0.38cm}
   \begin{subfigure}{0.25\textwidth}
    \includegraphics[width= \textwidth]{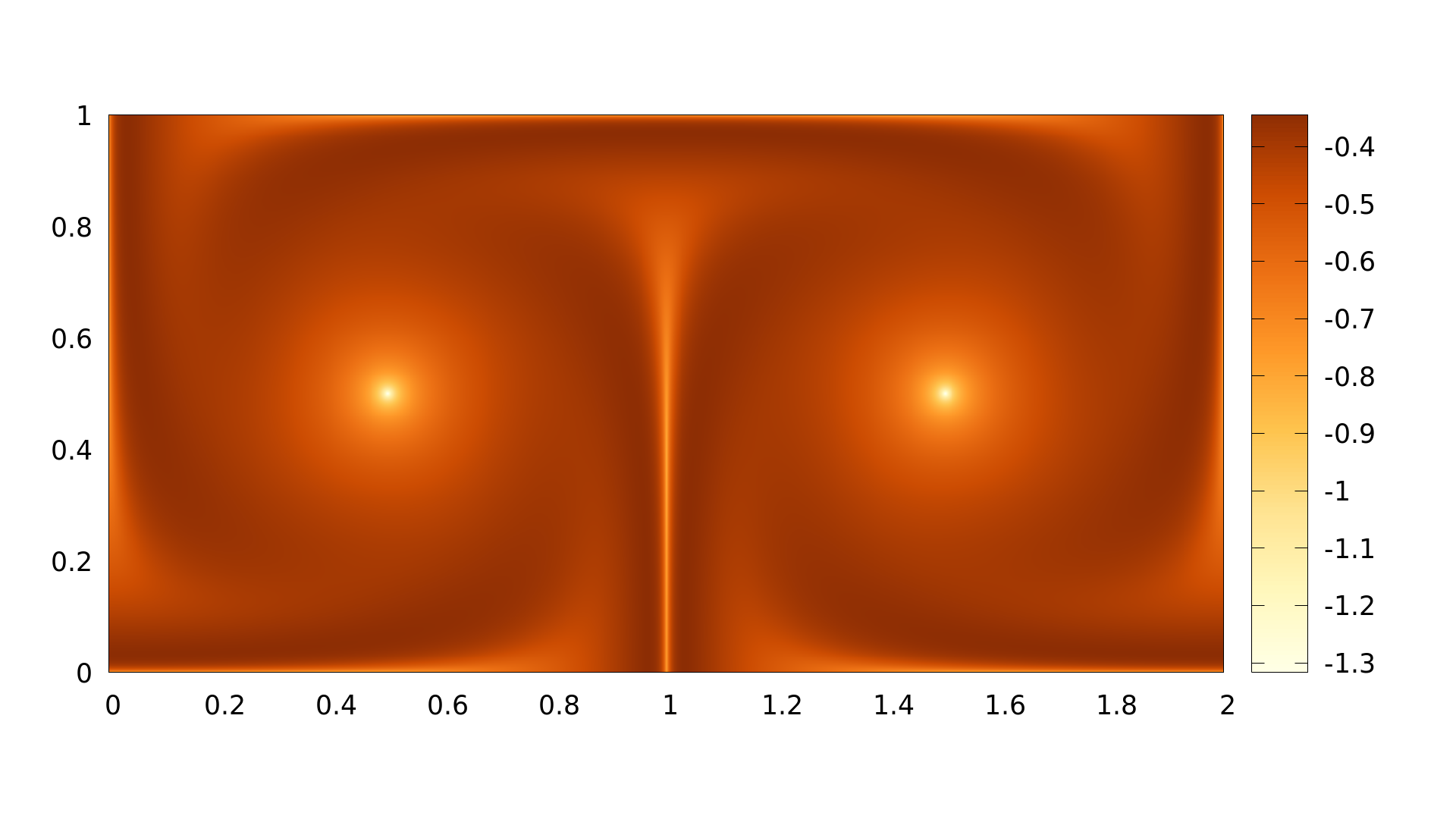}
  \label{fig:sdg_2_ftva}
  \end{subfigure}
    \caption{
    Ensemble $E_2$.
    Left row from top to bottom:
    FTLE of ensemble runs with $k=2$, $k=3$, $k=4$,. \\
    Right row from top to bottom:
    D-FTLE, variance of D-FTLE, FTVA.
}
  \label{fig:sdg_2}
 \end{figure}

 Ensemble $E_2$ is illustrated figure \ref{fig:sdg_2}.
In the left column we see the FTLE fields of the ensemble members,
the right column gives a collection of standard uncertain FTLE measures (from top to down:   D-FTLE, variance of D-FTLE, FTVA.
They also show a rather blurry ridge around the line $x=1$.

\begin{figure}[h!]
  \centering
  \begin{subfigure}{0.25\textwidth}
    \includegraphics[width= \textwidth]{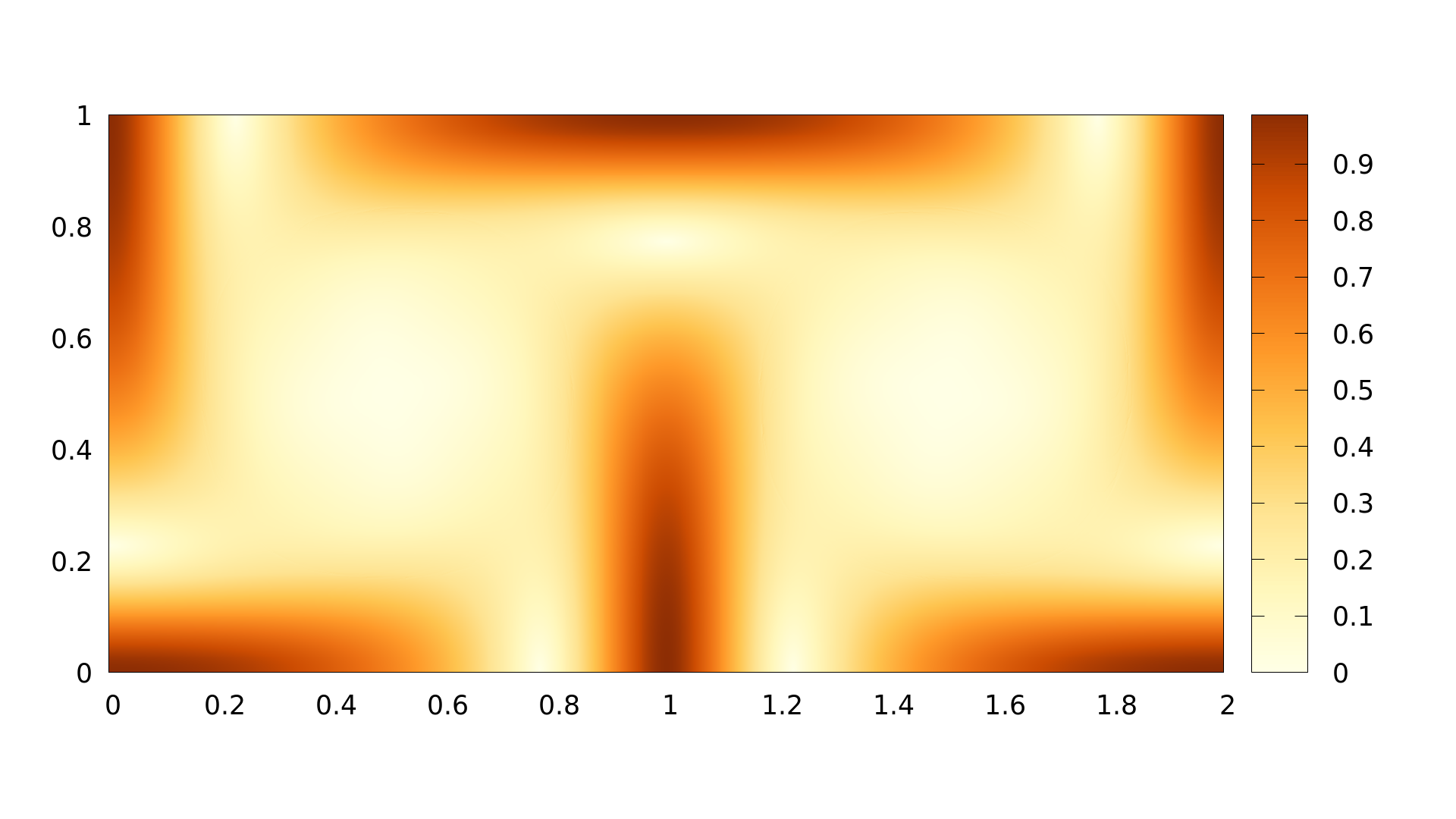}
  \label{fig:sdg_1_ftle_0}
  \end{subfigure}
 \hspace{-0.38cm}
 \vspace{-0.9cm}
      \begin{subfigure}{0.25\textwidth}
    \includegraphics[width= \textwidth]{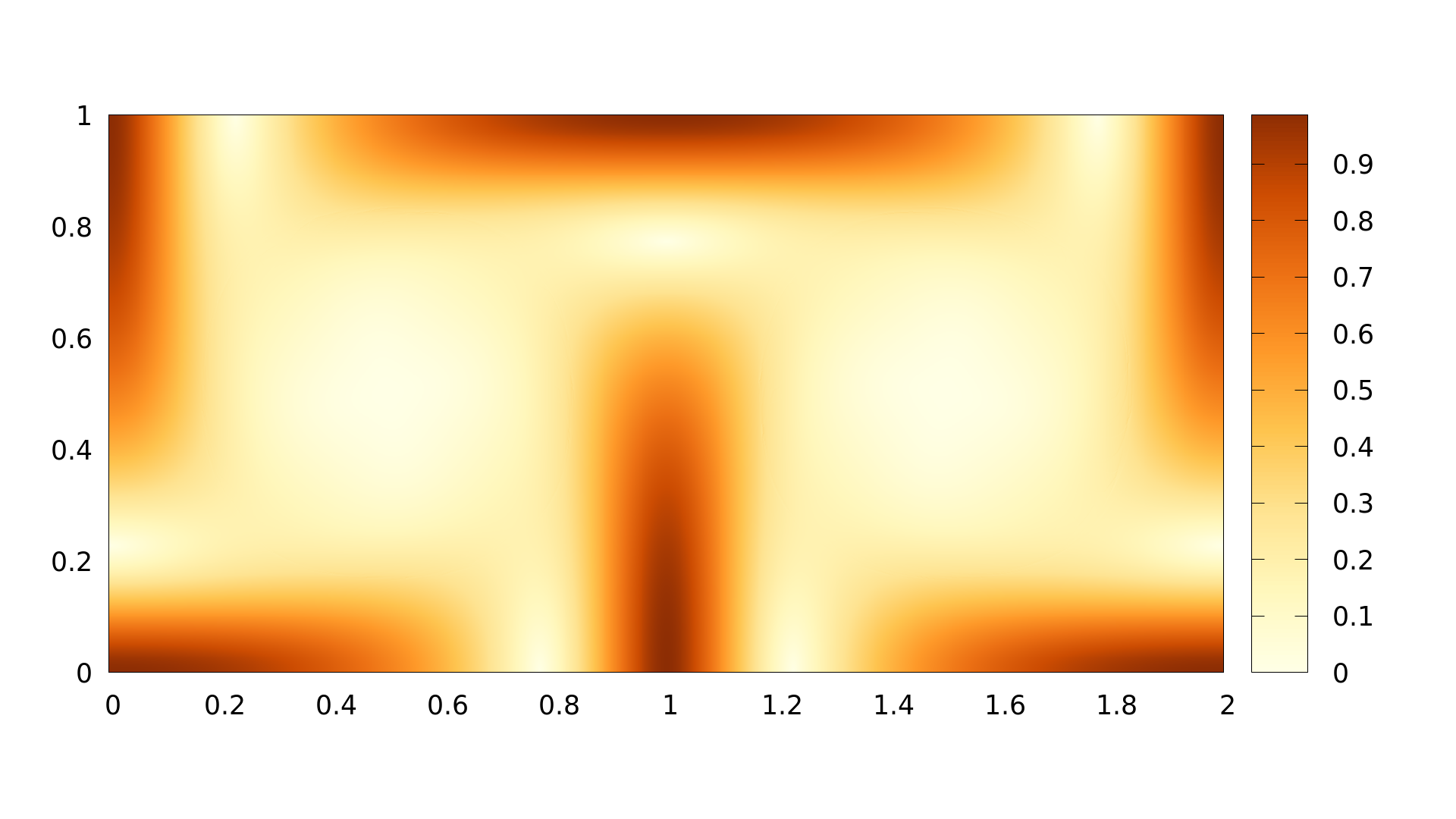}
  \label{fig:sdg_1_d-ftle}
  \end{subfigure}
 \vspace{-0.9cm}
    \begin{subfigure}{0.25\textwidth}
    \includegraphics[width= \textwidth]{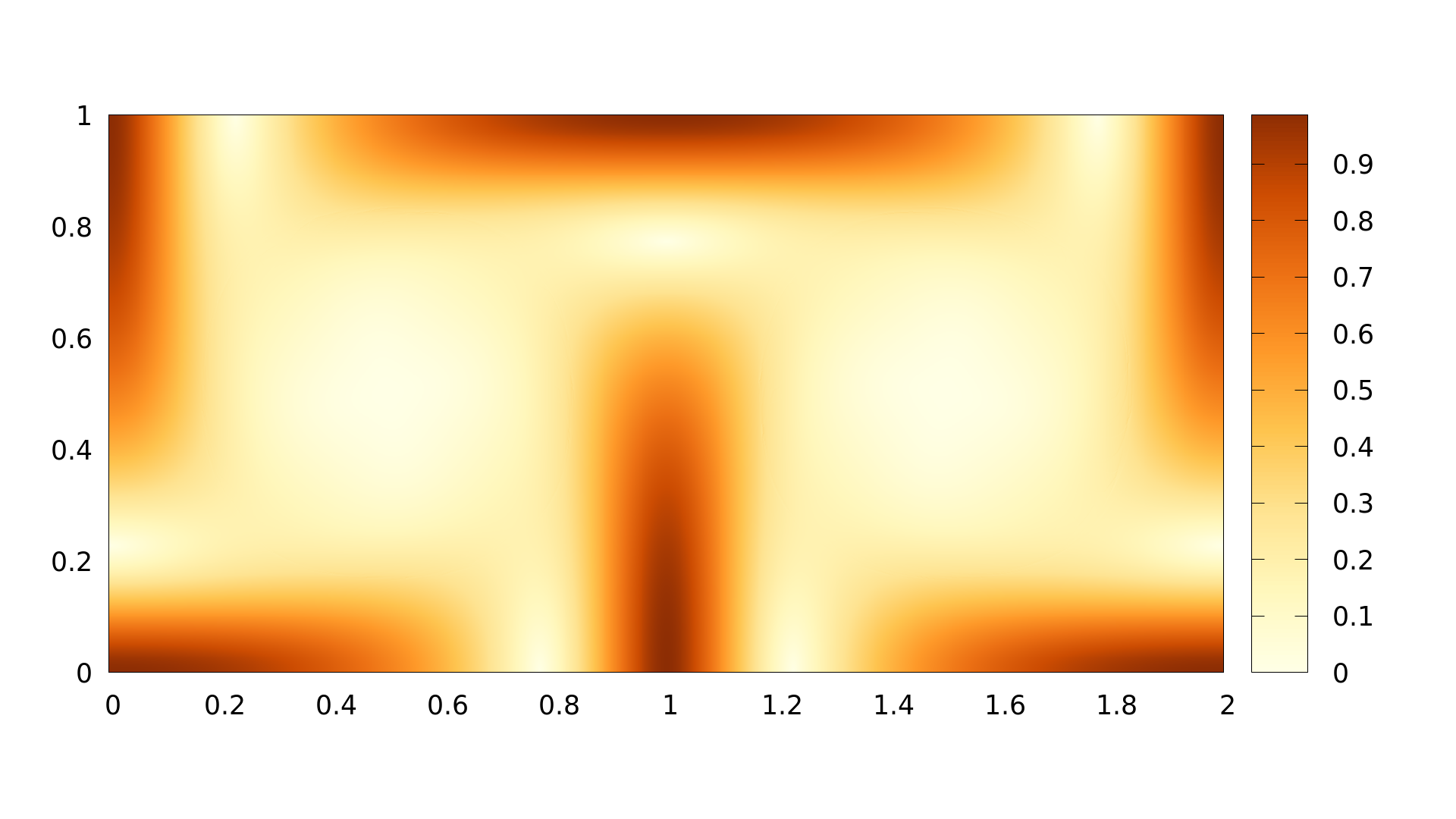}
  \label{fig:sdg_1_ftle_1}
  \end{subfigure}
 \hspace{-0.38cm}
   \begin{subfigure}{0.25\textwidth}
    \includegraphics[width= \textwidth]{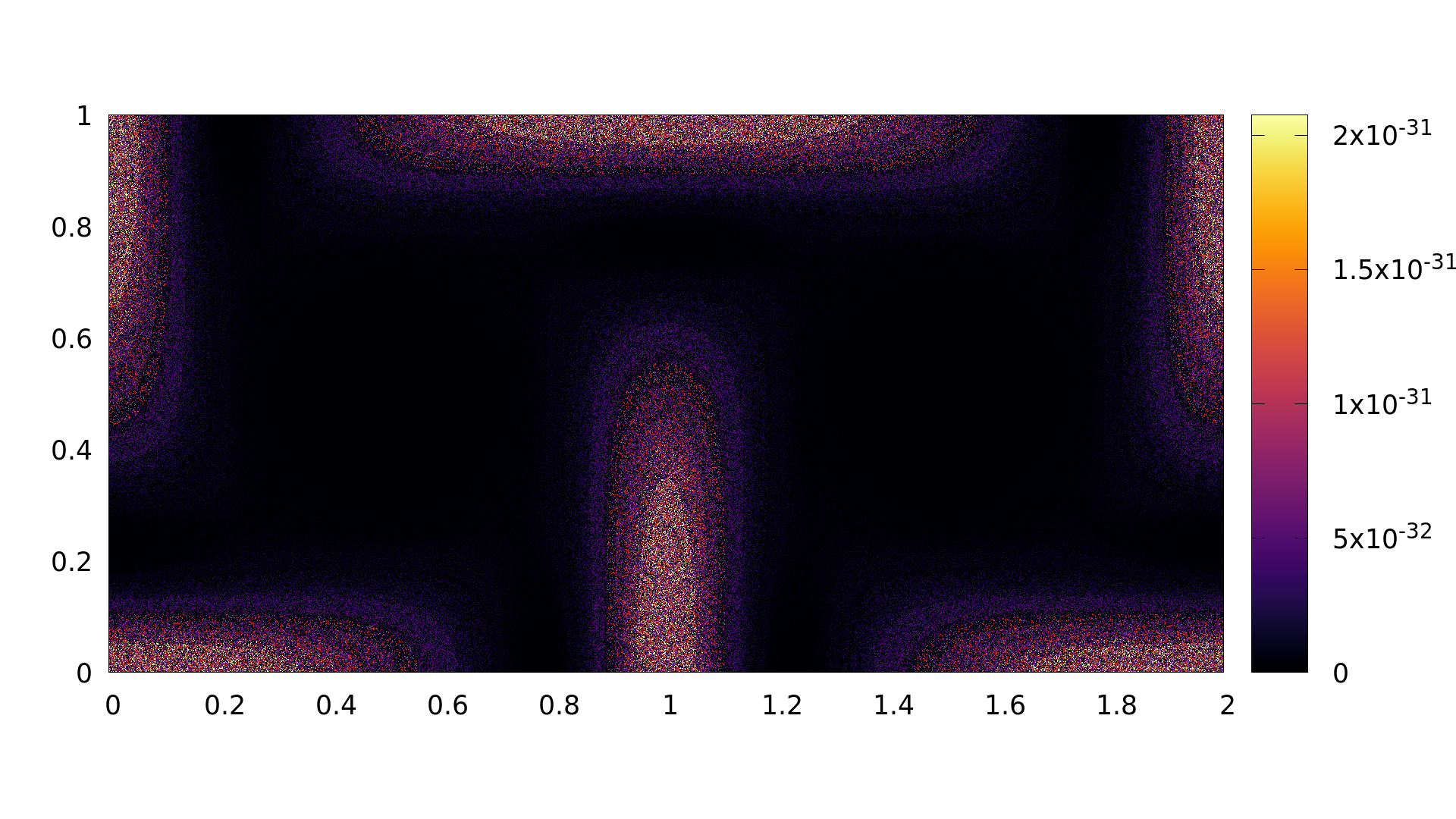}
  \label{fig:sdg_1_variance_d-ftle}
  \end{subfigure}
 \vspace{-0.9cm}
   \begin{subfigure}{0.25\textwidth}
    \includegraphics[width= \textwidth]{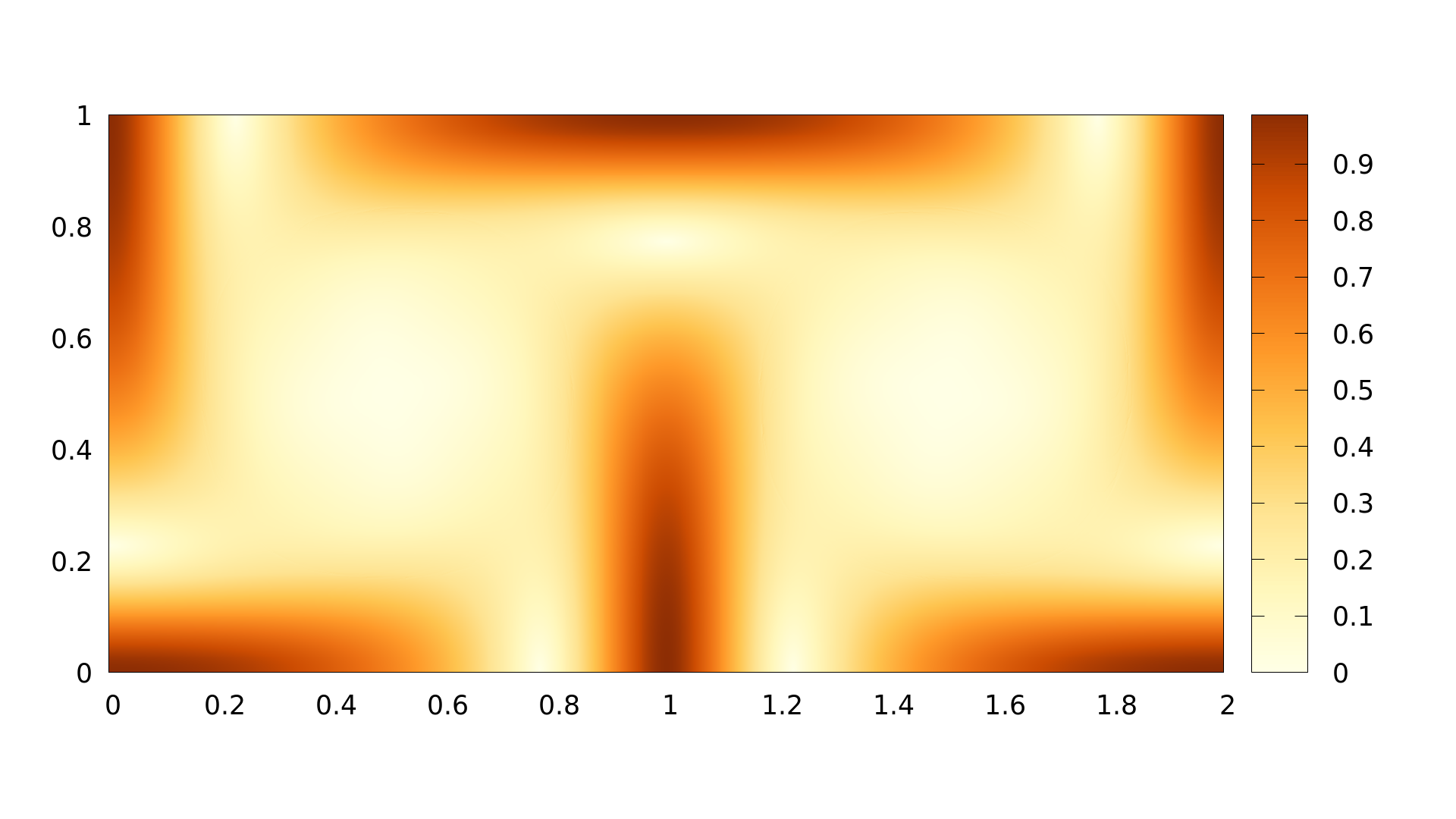}
  \label{fig:sdg_1_ftle_2}
  \end{subfigure}
 \hspace{-0.38cm}
   \begin{subfigure}{0.25\textwidth}
    \includegraphics[width= \textwidth]{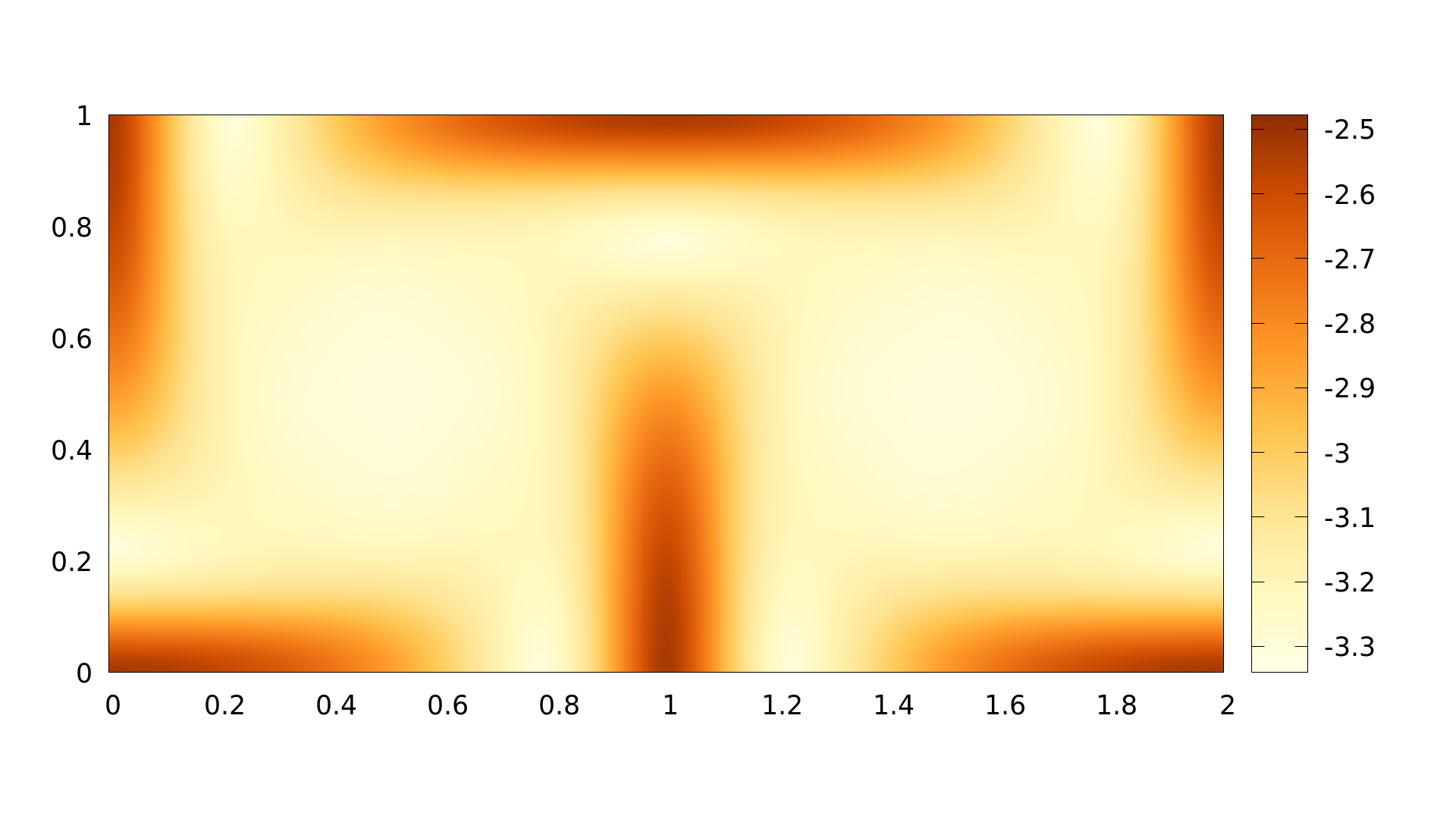}
  \label{fig:sdg_1_ftva}
  \end{subfigure}
    \caption{
    Ensemble $E_3$.
    Left row from top to bottom:
    FTLE of ensemble runs (all identical). \\
    Right row from top to bottom:
    D-FTLE, variance of D-FTLE, FTVA.
      \label{fig:sdg_1}
}
 \end{figure}

Finally, figure  \ref{fig:sdg_1} gives an illustration of ensemble $E_3$: in the left column we see the FTLE fields of the ensemble members, while right column gives a collection of standard uncertain FTLE measures (from top to bottom:   D-FTLE, variance of D-FTLE,s FTVA.
They also show a rather blurry ridge around the line $x=1$. In fact, here we have identical ensemble members having a weak separation.
The variance of D-FTLE is $0$.

Figures  \ref{fig:sdg_3}--\ref{fig:sdg_1} show that standard uncertain FTLE concepts do not give sufficient information to distinguish different sources of uncertainty of FTLE ridges. This is the motivation for introducing our new approach.

\section{Overview of our method}
Given an ensemble of velocity fields $\vv_1,...,\vv_n$ with the
corresponding flow maps $\phi_1,...,\phi_n$, our method consists of
the following algorithmic steps:
\begin{enumerate}
\item
  \textbf{\emph{Median flow}\/.}
  We find the ensemble index $m \in \{ 1,\dots,n \}$ such that
  $\phi_m$ is the best representative of the flows
  $\phi_1,\dots,\phi_n$.
  We select $\phi_m$ with the lowest sum of all squared point-wise distances to
  every other flow map. 
  We call $\phi_m$ the \emph{median flow}.
  \\

\item
  \textbf{Computation of domain displacements}.
  We compute a domain displacement for each pair of
  the median and the ensemble member.
  \[
    \vp_{m,i}: D \times T \times T ~\to~ \RRSet^q
  \]
  that maps points $(\vx,t)$ to displaced points $(\vx', t)$ with
  \[
    \vx'~=~\vx + \vp_{m,i}(\vx,t,\tau)~\in~ D~.
  \]
  Note that the displaced points $\vx'$ stay within the domain. \\
   \[
  \psi_i = \phi_i(\vx', t, \tau)
    \]
  We expect the displacement $\vp_{m,i}$ to be of small magnitude and smooth, and
  it should align the FTLE ridges of $\phi_m(\vx, t, \tau)$ as much as
  possible to the FTLE ridges of $\phi_i(\vx', t, \tau)$.
  \\

\item
  \textbf{Computation of displaced flow maps and displaced FTLE.}
  We apply the domain displacements on the corresponding flow maps
  such that they align with the median flow map.
  The flow map $\phi_i$ is displaced to ``match'' the median flow.
  We denote this displaced flow map $\psi_i$.
  We can compute gradients and FTLE from the displaced flow maps
  $\psi_i$.
  Due to the construction of the domain displacements, we expect the
  ``displaced'' FTLE ridges of $\psi_i$ to be aligned as much as
  possible with the FTLE ridges of the median
  $\psi_m=\phi_m$.
  \\
\item
  \textbf{Visual analysis.}
  The visual analysis consists of two parts.
  Firstly, a visualization of the distribution of displaced flow maps
  $\psi_i$, their gradients and their FTLE fields using existing
  approaches from the literature to visualize the distributions in
  $(\ref{eq_existingapproaches})$.
  Secondly, we conduct a visual analysis of the distribution of the
  domain $\vp_{m,1},\dots,\vp_{m,n}$, which captures information about
  the uncertainty of both the location and strength of the FTLE
  ridges.
\end{enumerate}

\section{Details of our method}
In the following, we present the steps of our method in detail.
For a concise notation, we may omit the arguments $t, \tau$ whenever
they can be regarded as constants.

\subsection{Selection of the median flow $\phi_m$}
We select the index $m$ of the median flow $\phi_m$ as the minimizer
of the sums
\[
  m = \underset{i}{\argmin} \left\{ \sum_{j=1}^n \int _D \invn{\phi_i -  \phi_j}^2 \diff \vx~|~ i=1,\dots,n \right\}
\]
This means that we choose $\phi_m$ so that the sum of squared distances for
all flows to $\phi_m$ becomes minimal.

\subsection[Computation of the displacement ??]{Computation of the displacement $\vp_{m,i}$}
\label{sec:displacement}
The objective of the domain displacement function $\vp_{m,i}$ is
aligning the FTLE ridges of $\phi_m$ as much as possible with the
ridges of the flow $\phi_i$.
To characterize the best alignment of $\phi_m$ and $\phi_i$, we define the
energy
\begin{equation*}
  A(\vp) = \frac{1}{\mbox{area}(D)} \int_D \invn{\phi_m(\vx + \vp(\vx)) - \phi_i(\vx)}^2_2 \; d \vx~,
\end{equation*}
which measures the deviation of the displaced flow map $\phi_m$ from
$\phi_i$ for varying a displacement function $\vp$.
In addition, the transformation $(\vx + \vp_{m,i})$ should be as rigid
as possible: %
$(\vx + \vp_{m,i})$ is rigid if
\begin{equation*}
  (\nabla(\vx + \vp_{m,i}))^T \; (\nabla(\vx + \vp_{m,i})) = \mI~,
\end{equation*}
which requires (due to $\nabla \vx \neq \vNull$)
\begin{equation*}
  \nabla\vp_{m,i}^T \; \nabla\vp_{m,i} +  \nabla\vp_{m,i}^T  + \nabla\vp_{m,i} = \mNull.
\end{equation*}
We capture this in a second energy term
\begin{equation*}
  B(\vp)= \frac{1}{\mbox{area}(D)} \int_D \invn{ \nabla\vp^T \; \nabla\vp +  \nabla\vp^T  + \nabla\vp}^2_F \; d \vx~,
\end{equation*}
Where $\invn{\cdot}_F$ denotes the Frobenius norm of a matrix.
Depending on the application, we may also consider the additional
boundary condition
\begin{equation}
  \label{eq_boundarycondition}
  \delta D + \vp_{m,i}(\delta D ) = \delta D
\end{equation}
where $\delta D$ denotes the boundary of $D$.
\eqref{eq_boundarycondition} ensures that a point on the boundary of
$D$ is mapped to (another) boundary point of $D$.

We minimize a blend of the alignment term $A$ and the ``rigidity''
term $B$ to determine the optimal domain displacement as
\begin{equation}
  \label{eq_finalenergy}
  \vp_{m,i} = \underset{\vp}{\argmin}~ (1-\rho) \; A(\vp) + \rho \; B(\vp)
\end{equation}
for a weight $\rho \in [0,1]$.
Note that for the extreme cases $\rho=0$ and $\rho=1$,
\eqref{eq_finalenergy} has a closed-form minimizer:
For $\rho=1$ the trivial minimizer is
\begin{equation}
  \label{eq_closedsolution2}
  \vp_{m,i}=\vNull
\end{equation}
For $\rho=0$, the function
\begin{equation}
  \label{eq_closedsolution1}
  \vp_{m,i} = \phi_m\left(\phi_i(\vx,\,t,\,\tau),\, t+\tau,\, -\tau\right) - \vx
\end{equation}
yields $A(\vp_{m,i})=0$ and this minimizes \eqref{eq_finalenergy}).
This minimizer \eqref{eq_closedsolution1} is obtained as follows:
for each location and time $(\vx,t)$, we first integrate $\vv_j$ over
a period $\tau$.
From the resulting point $\phi_i(\vx,t,\tau)$, we apply a backward
integration of $\vv_m$ over the negative period $-\tau$.

Note that neither \eqref{eq_closedsolution1} nor
\eqref{eq_closedsolution2} are useful solutions:
(\ref{eq_closedsolution1}) is perfectly rigid but has no effect, and
while (\ref{eq_closedsolution1}) maximizes the alignment of $\phi_m$
and $\phi_i$, it also tends to have strong gradients (which grow
exponentially in $\tau$) and can therefore neither expected to be
small nor smooth.
Instead, we search for a solution of \eqref{eq_finalenergy} for a
parameter $0<\rho<1$ that keeps $\vp_{m,i}$ small and smooth while
simultaneously aligning the ridges.
The influence of the parameter $\rho$ is discussed in
section \ref{sec:discussion} and figure \ref{fig:dg_rho}.

Minimizing \eqref{eq_finalenergy} is a nonlinear optimization problem.
We describe the discretization and the numerical solution in
section~\ref{sec:optimization}

 \begin{figure}[H]
  \centering
  \begin{subfigure}{0.164\textwidth}
    \includegraphics[width= \textwidth]{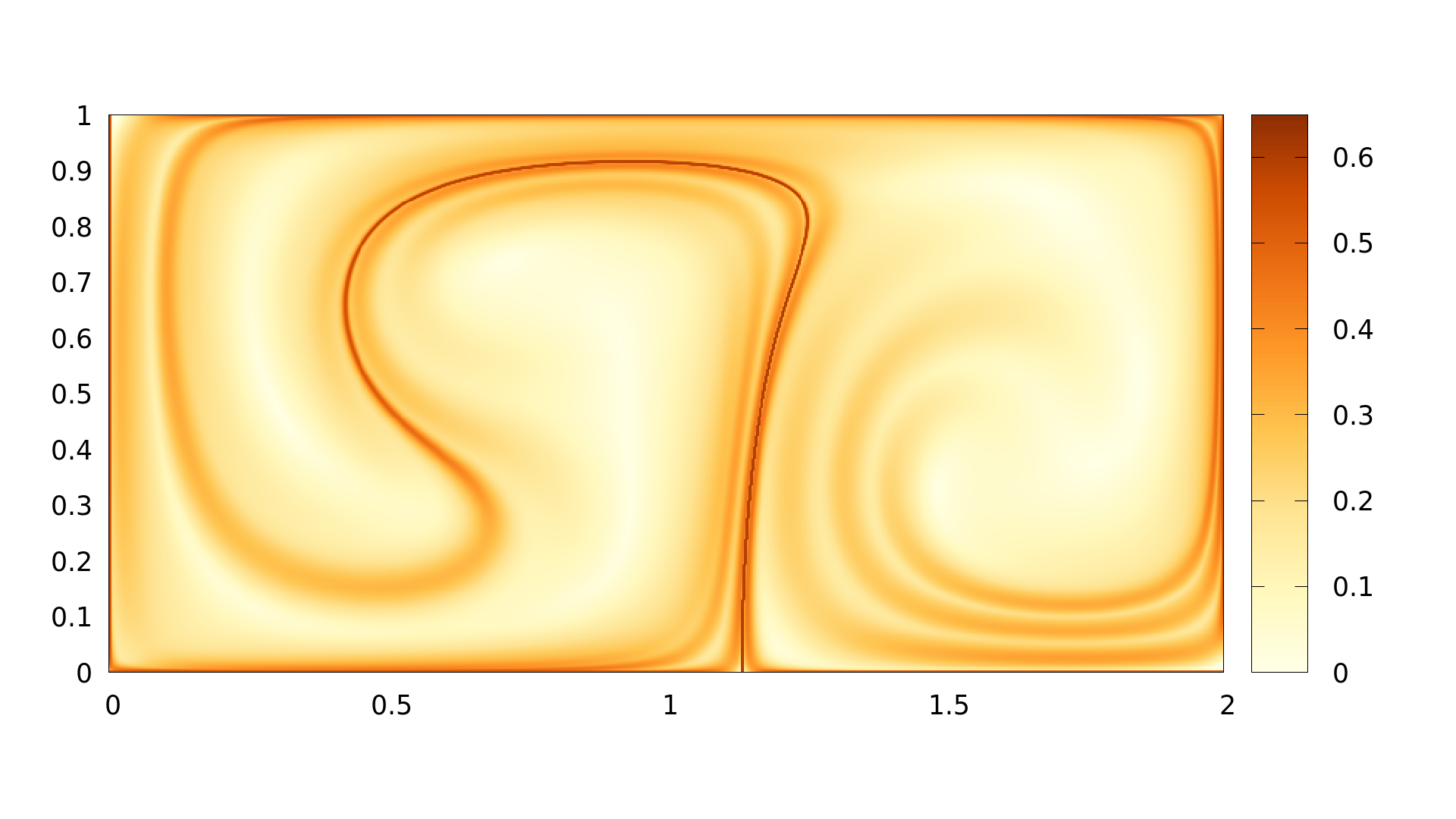}
  \label{fig:dg_opt_ftle_0_h}
  \end{subfigure}
 \hspace{-0.3cm}
  \vspace{-0.5cm}
      \begin{subfigure}{0.164\textwidth}
    \includegraphics[width= \textwidth]{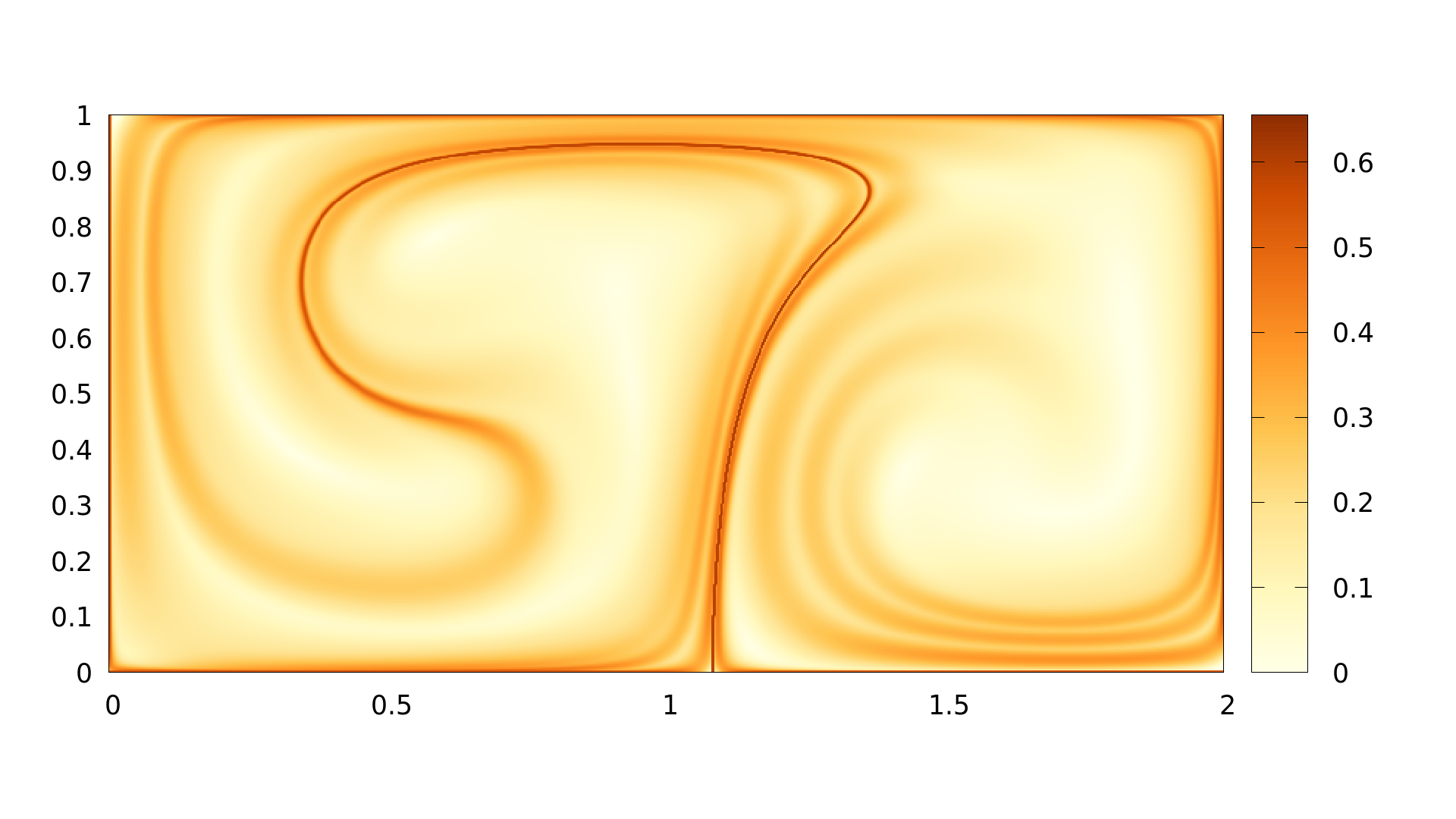}
  \label{fig:dg_opt_iter_0_ftle}
  \end{subfigure}
   \hspace{-0.3cm}
    \begin{subfigure}{0.164\textwidth}
    \includegraphics[width= \textwidth]{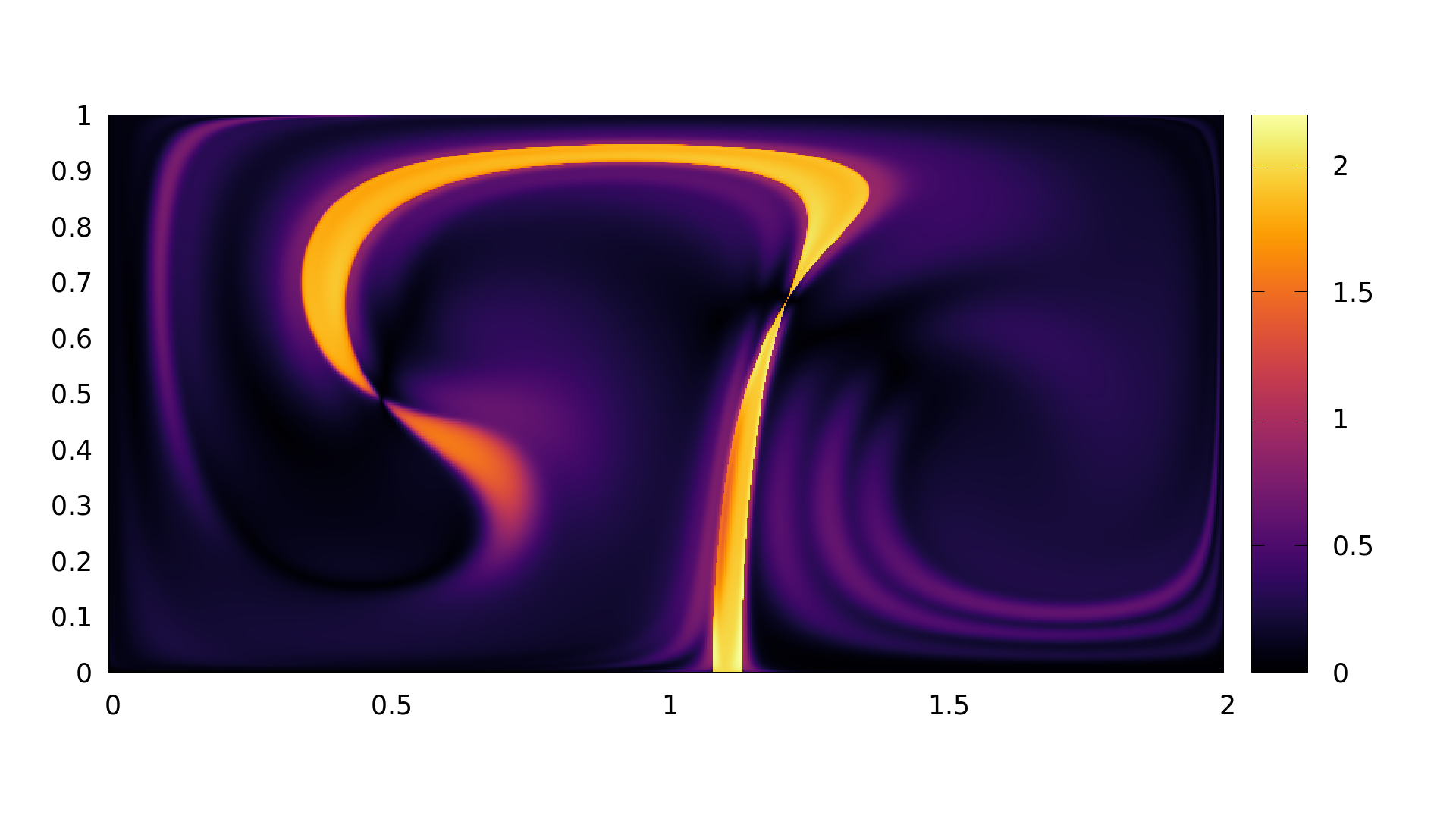}
  \label{fig:dg_opt_iter_0_error}
  \end{subfigure}

    \begin{subfigure}{0.164\textwidth}
    \includegraphics[width= \textwidth]{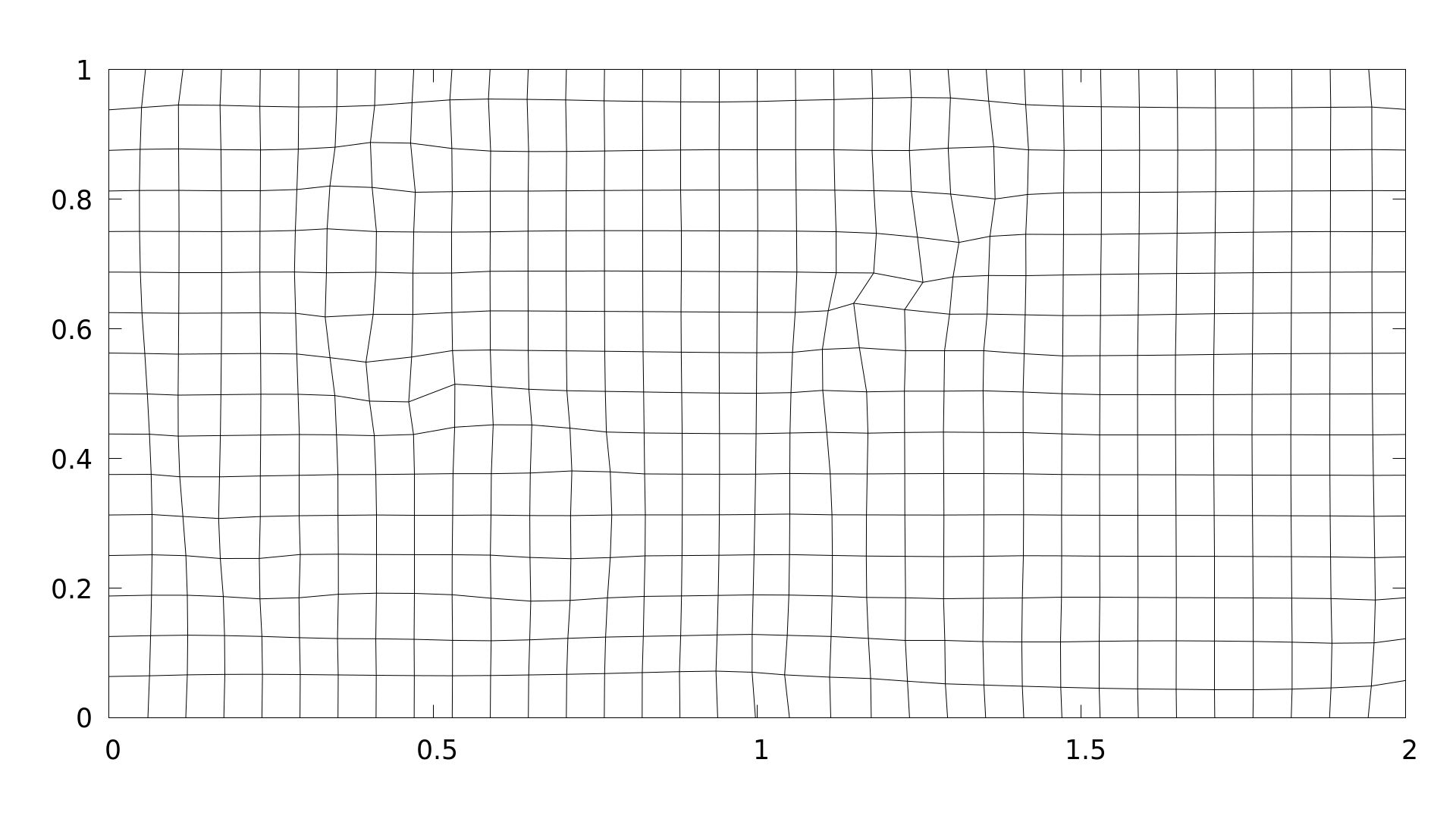}
  \label{fig:dg_opt_iter_2_h}
  \end{subfigure}
 \hspace{-0.3cm}
  \vspace{-0.5cm}
      \begin{subfigure}{0.164\textwidth}
    \includegraphics[width= \textwidth]{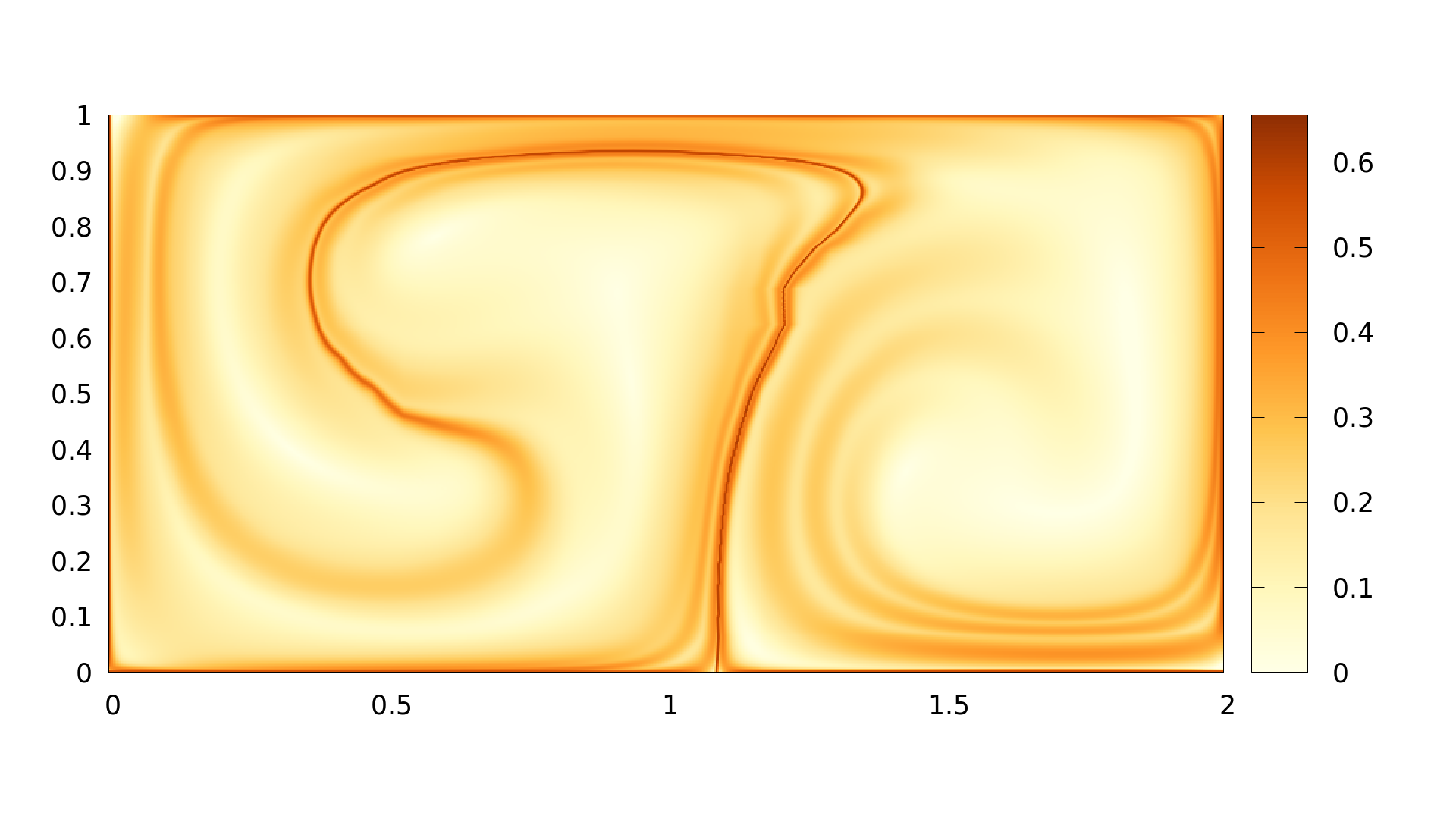}
  \label{fig:dg_opt_iter_2_ftle}
  \end{subfigure}
   \hspace{-0.3cm}
    \begin{subfigure}{0.164\textwidth}
    \includegraphics[width= \textwidth]{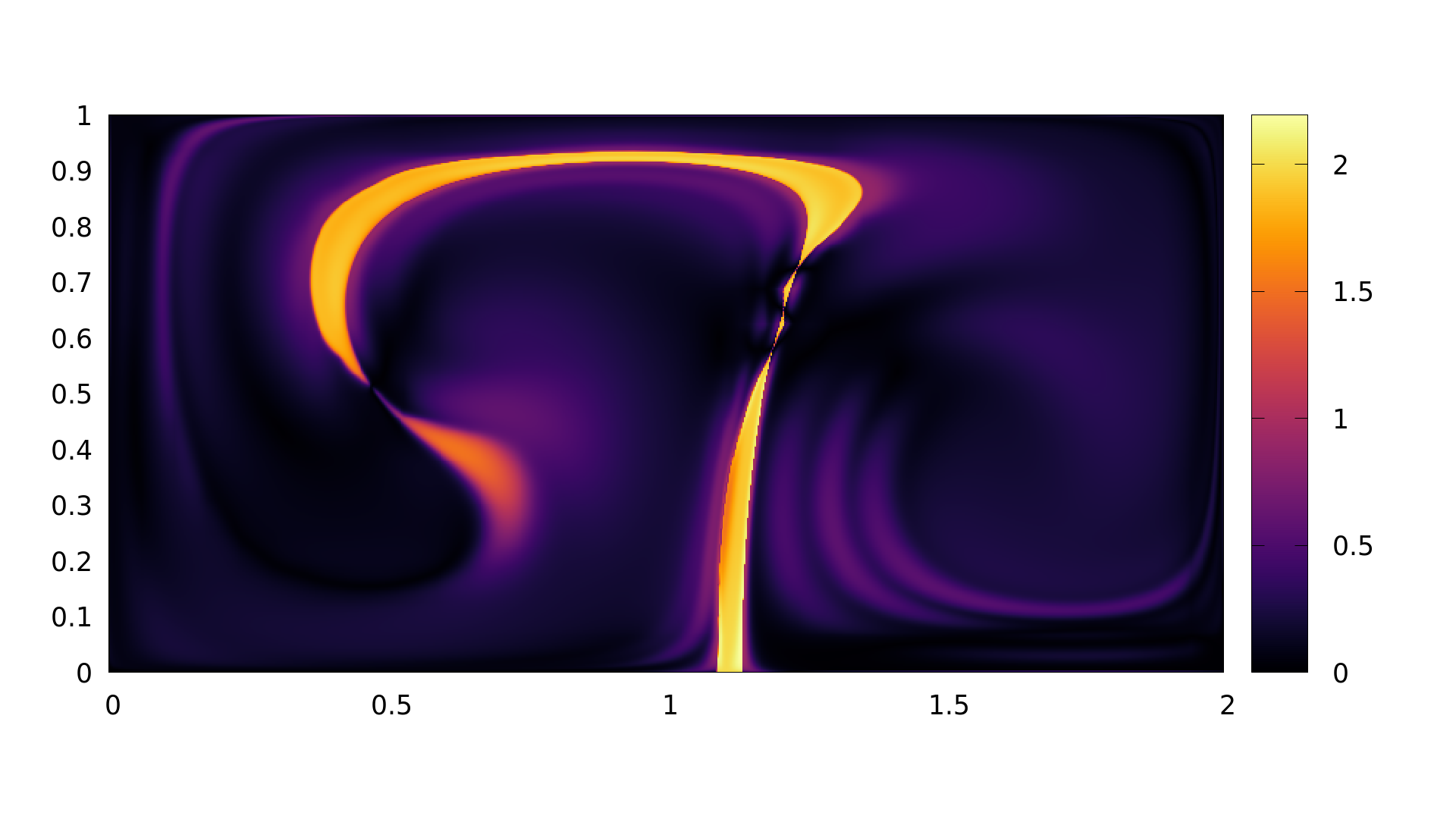}
  \label{fig:dg_opt_iter_2_error}
  \end{subfigure}

      \begin{subfigure}{0.164\textwidth}
    \includegraphics[width= \textwidth]{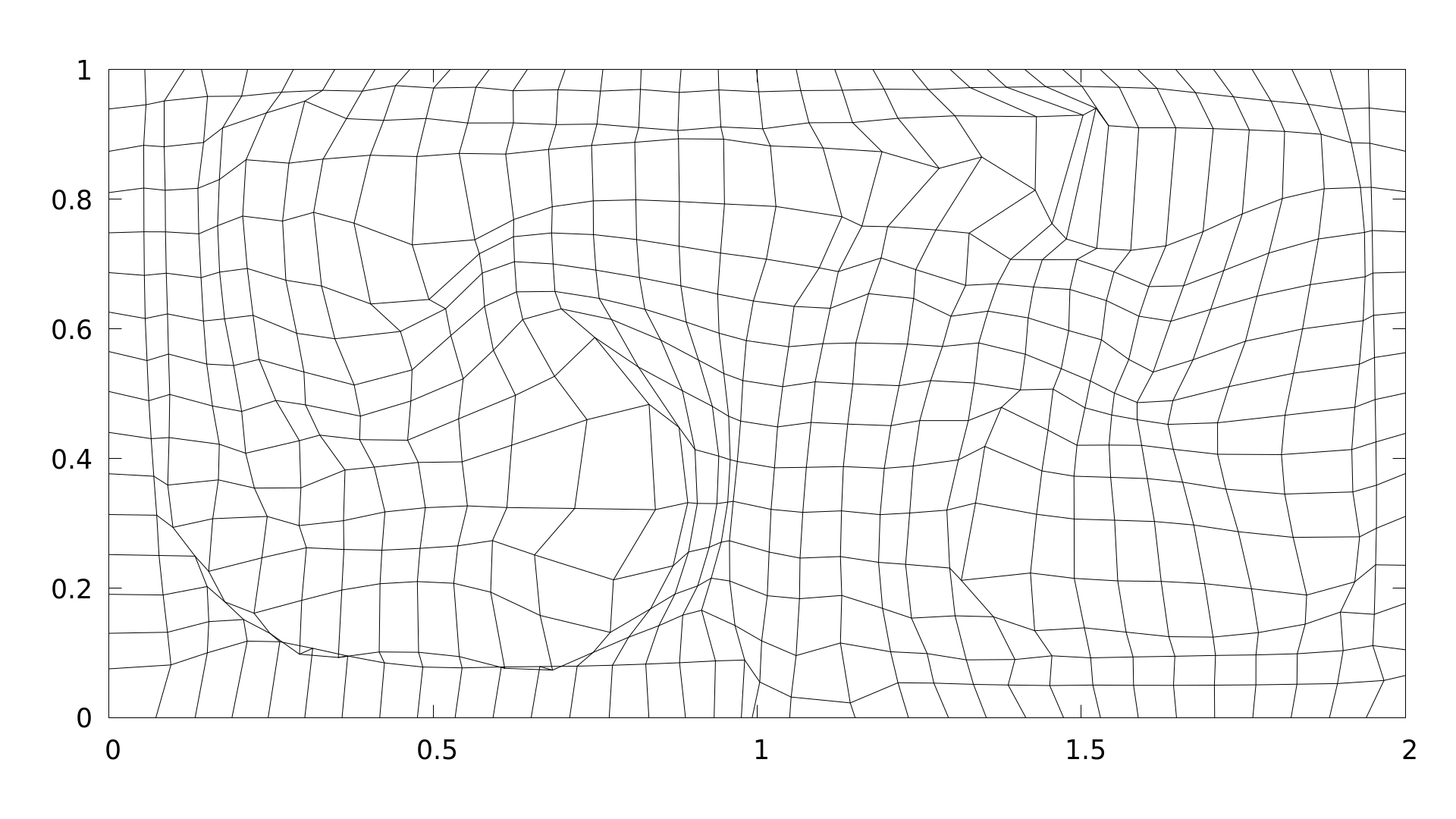}
  \label{fig:dg_opt_iter_10_h}
  \end{subfigure}
 \hspace{-0.3cm}
  \vspace{-0.5cm}
      \begin{subfigure}{0.164\textwidth}
    \includegraphics[width= \textwidth]{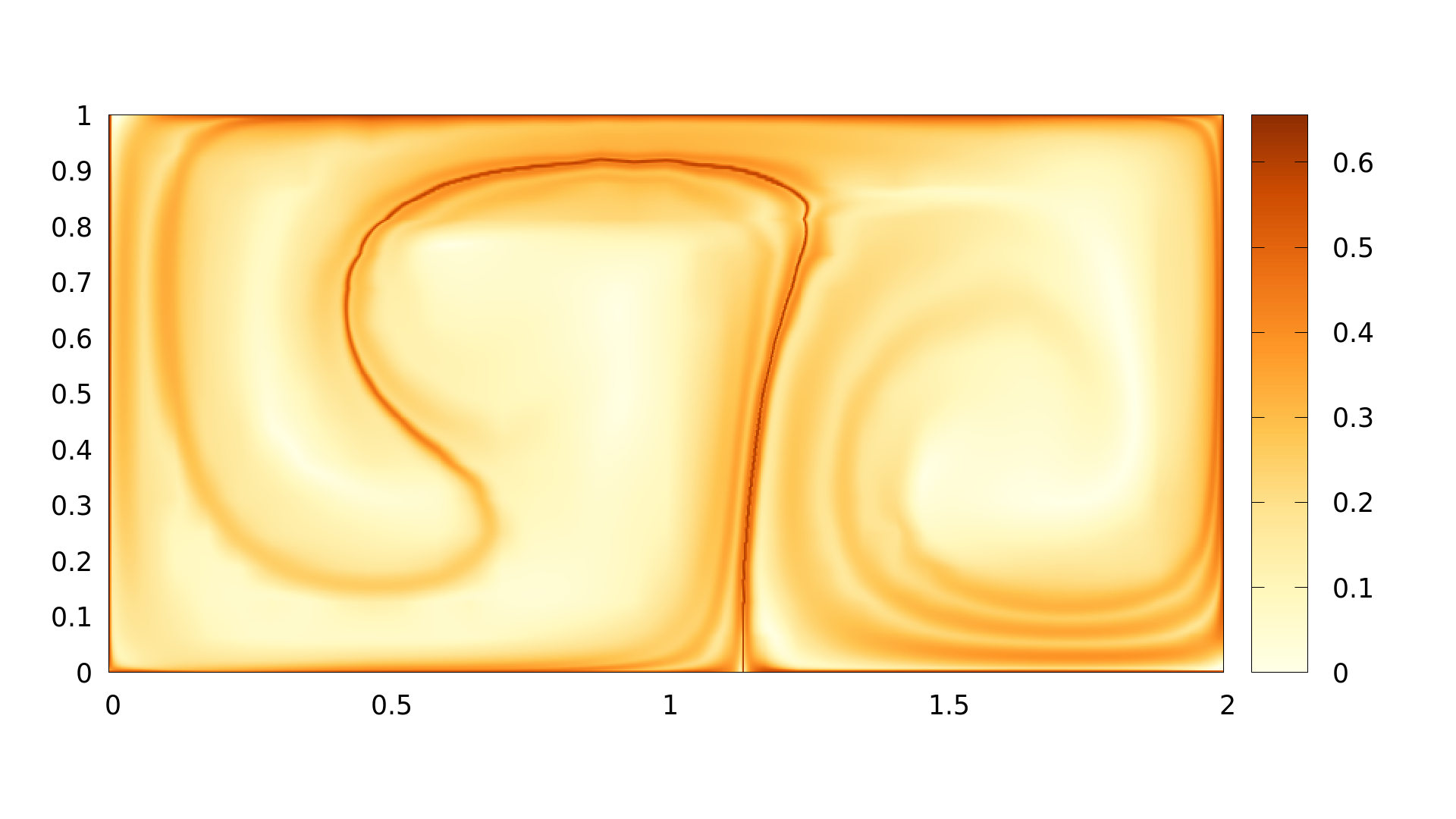}
  \label{fig:dg_opt_iter_10_ftle}
  \end{subfigure}
   \hspace{-0.3cm}
    \begin{subfigure}{0.164\textwidth}
    \includegraphics[width= \textwidth]{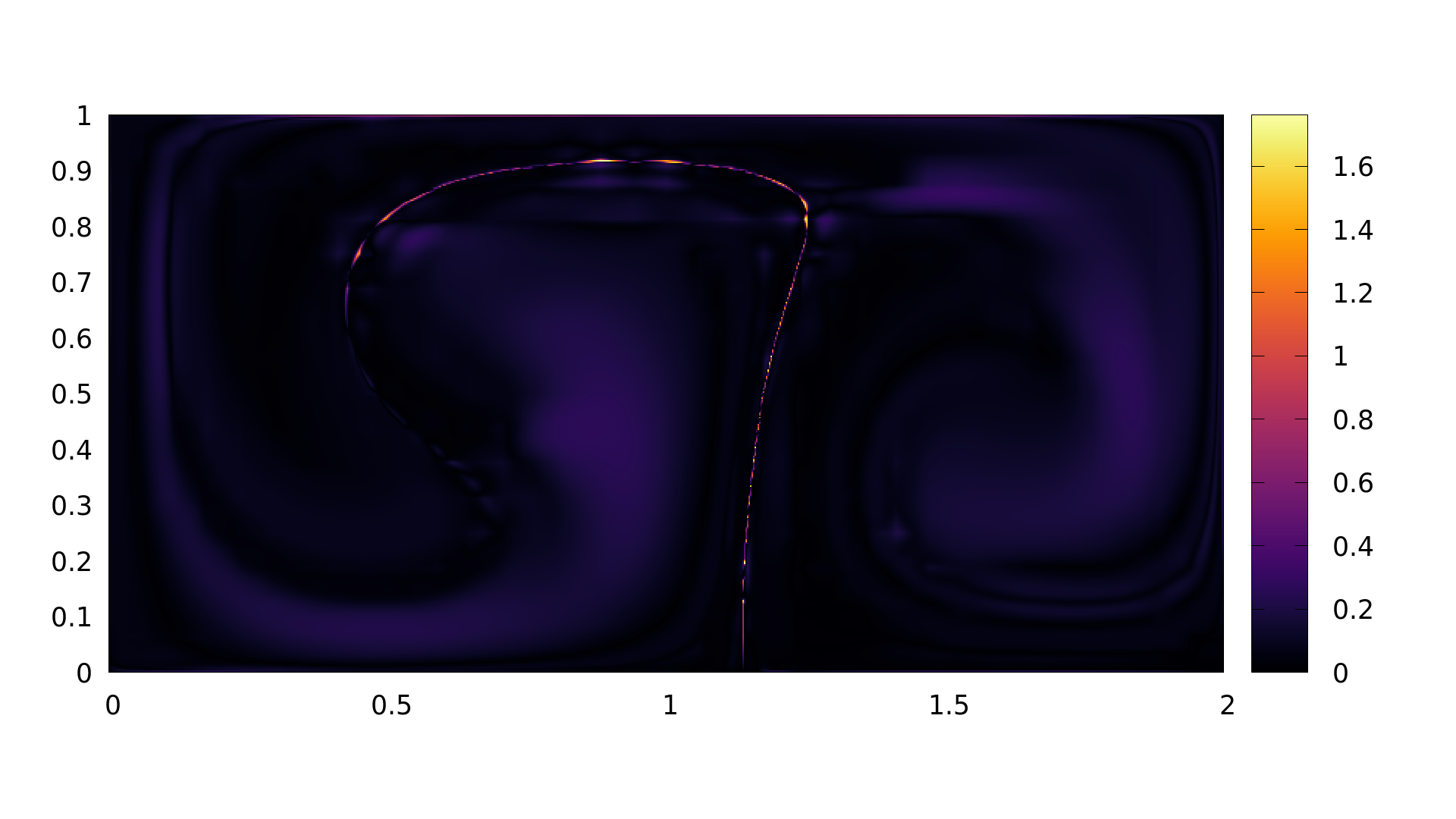}
  \label{fig:dg_opt_iter_10_error}
  \end{subfigure}

      \begin{subfigure}{0.164\textwidth}
    \includegraphics[width= \textwidth]{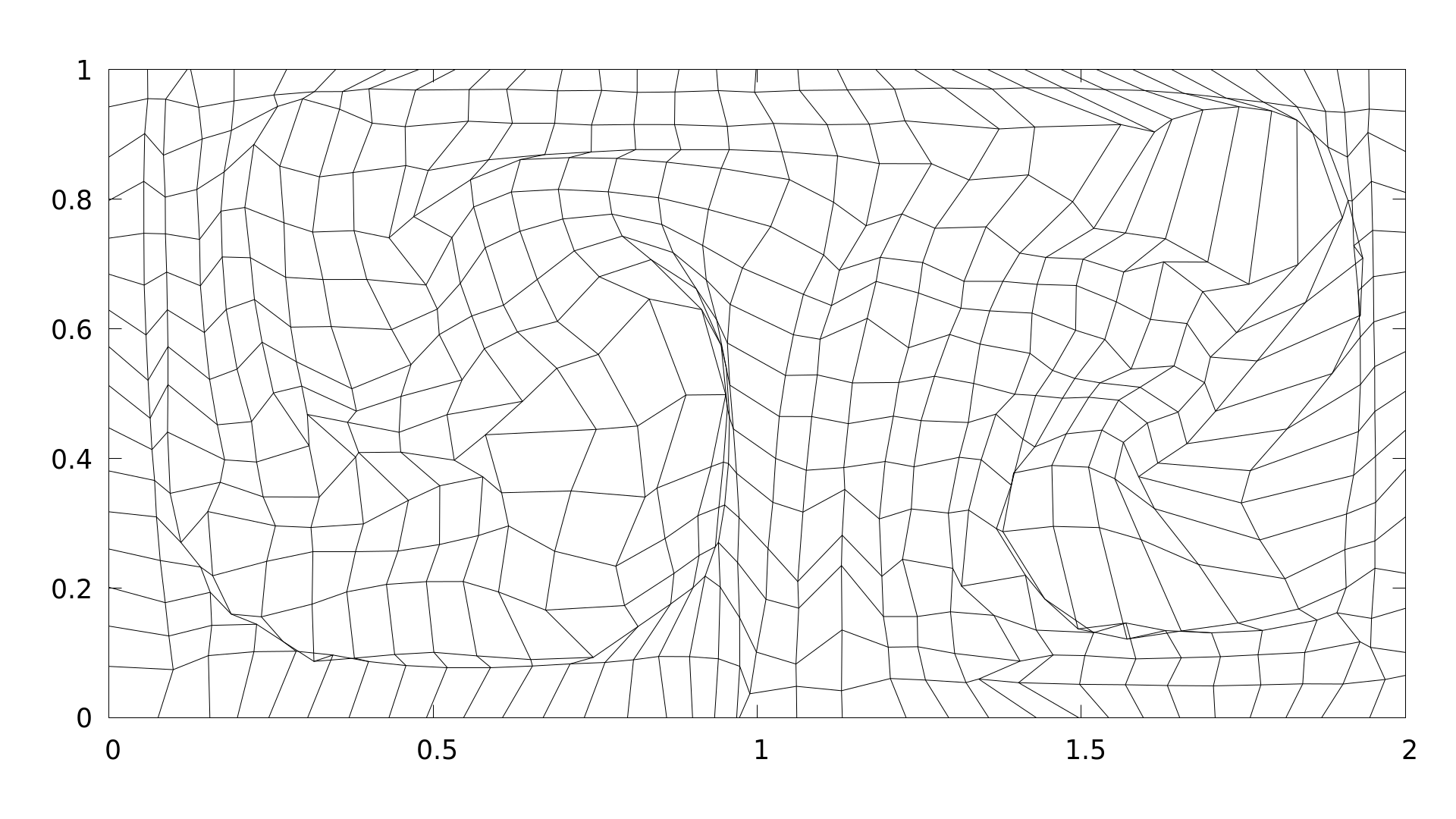}
  \label{fig:dg_opt_iter_30_h}
  \end{subfigure}
 \hspace{-0.3cm}
  \vspace{-0.5cm}
      \begin{subfigure}{0.164\textwidth}
    \includegraphics[width= \textwidth]{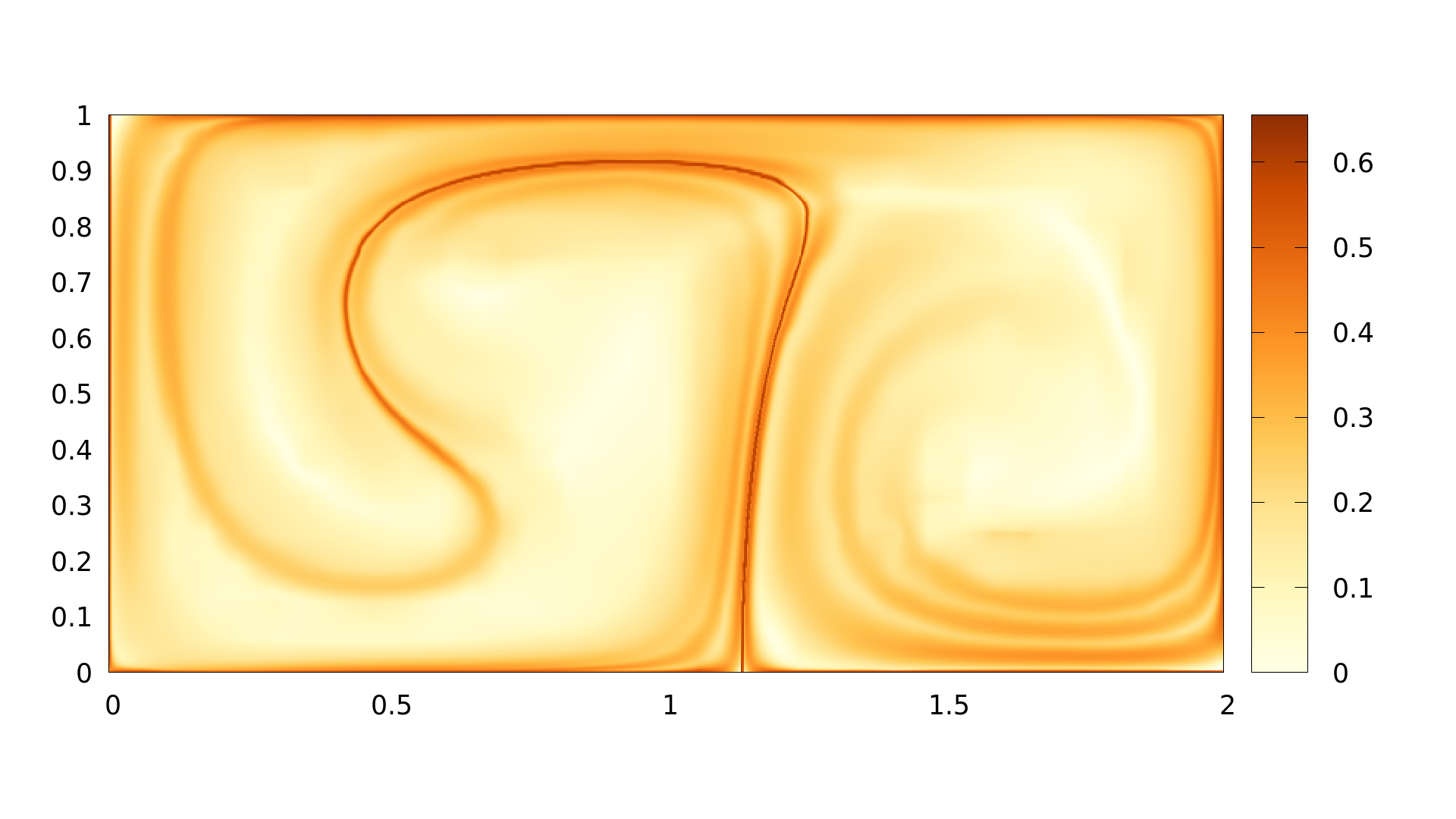}
  \label{fig:dg_opt_iter_30_ftle}
  \end{subfigure}
   \hspace{-0.3cm}
    \begin{subfigure}{0.164\textwidth}
    \includegraphics[width= \textwidth]{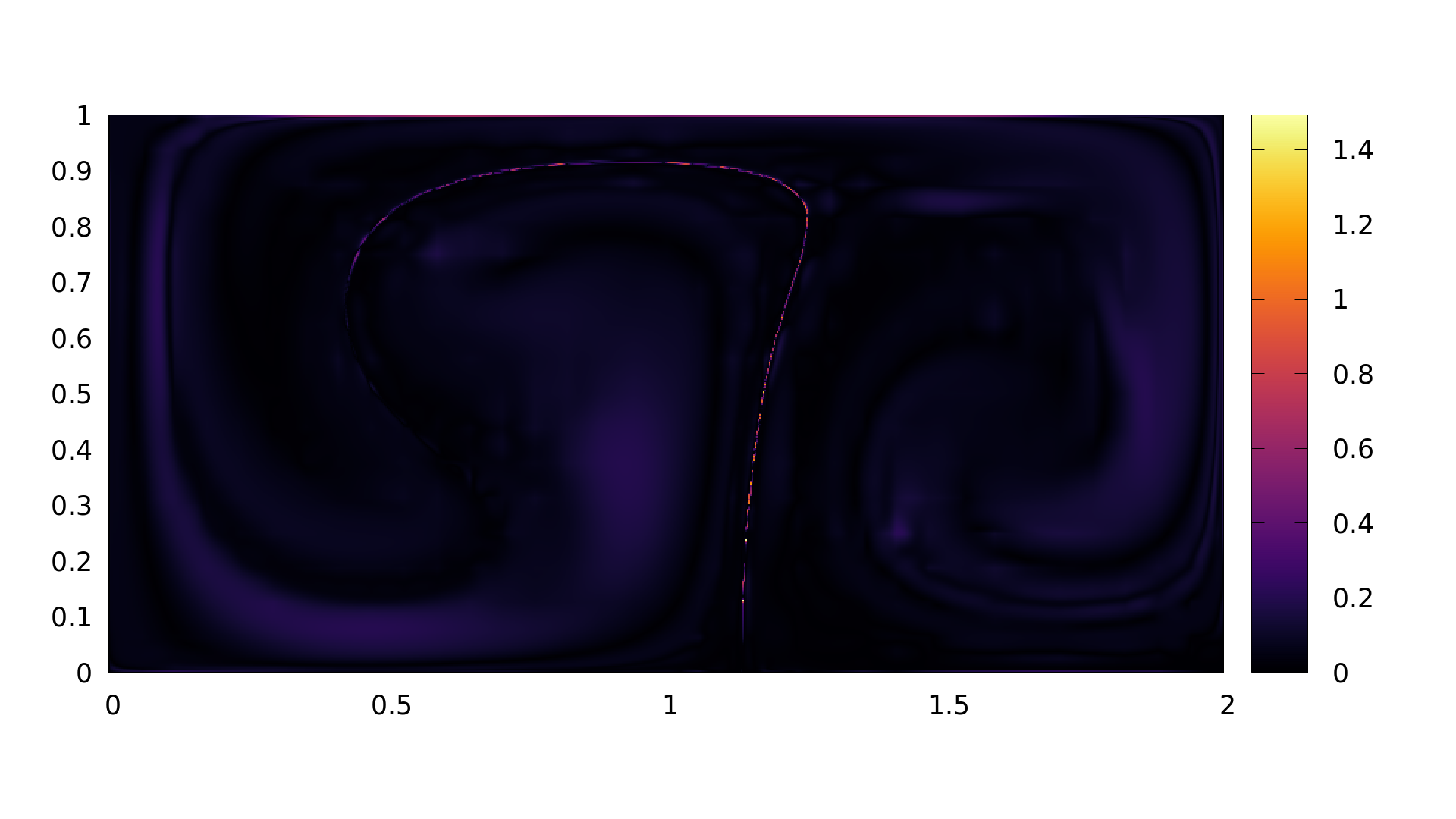}
  \label{fig:dg_opt_iter_30_error}
  \end{subfigure}

        \begin{subfigure}{0.164\textwidth}
    \includegraphics[width= \textwidth]{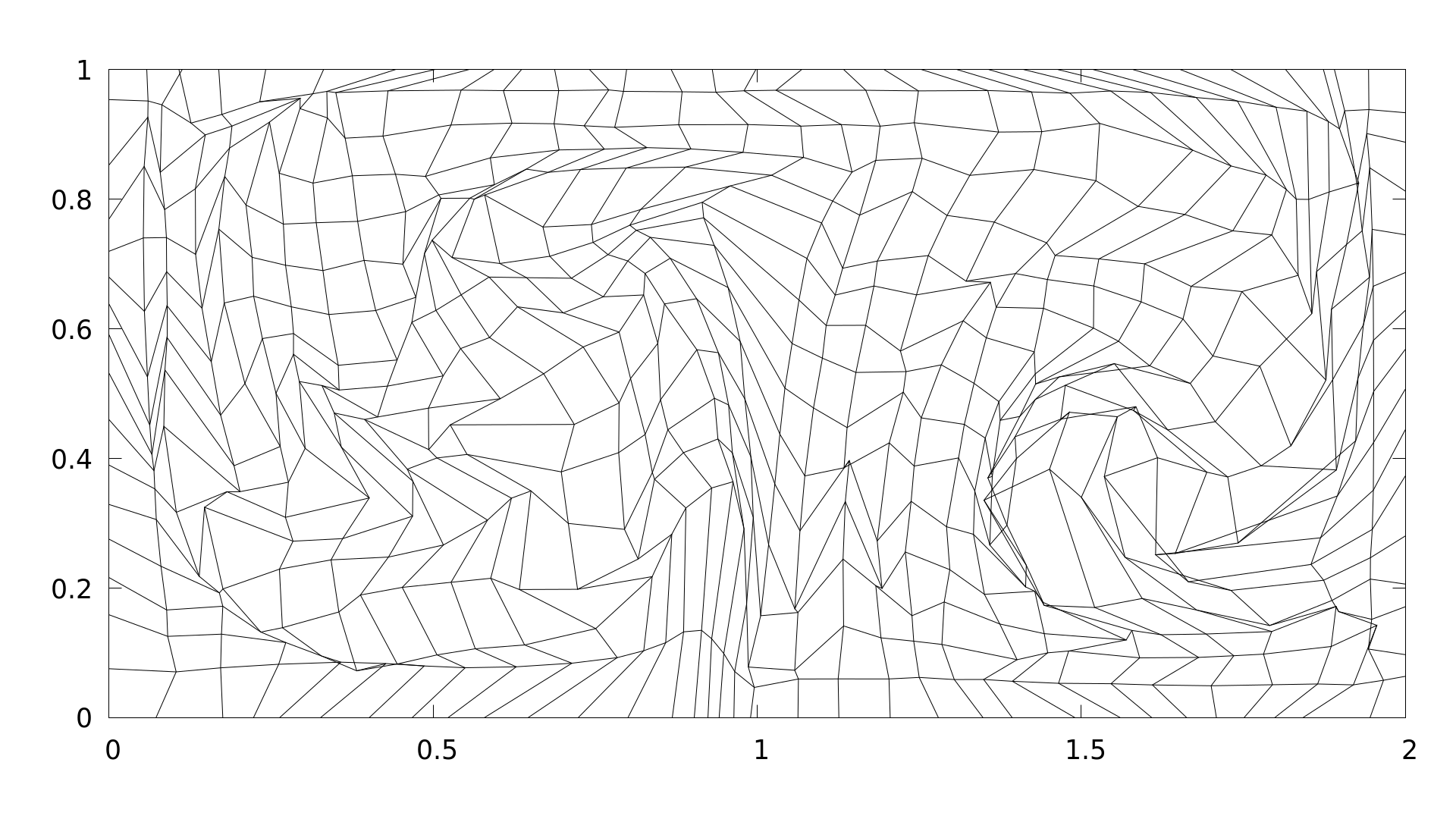}
  \label{fig:dg_opt_iter_last_h}
  \end{subfigure}
 \hspace{-0.3cm}
  \vspace{-0.5cm}
      \begin{subfigure}{0.164\textwidth}
    \includegraphics[width= \textwidth]{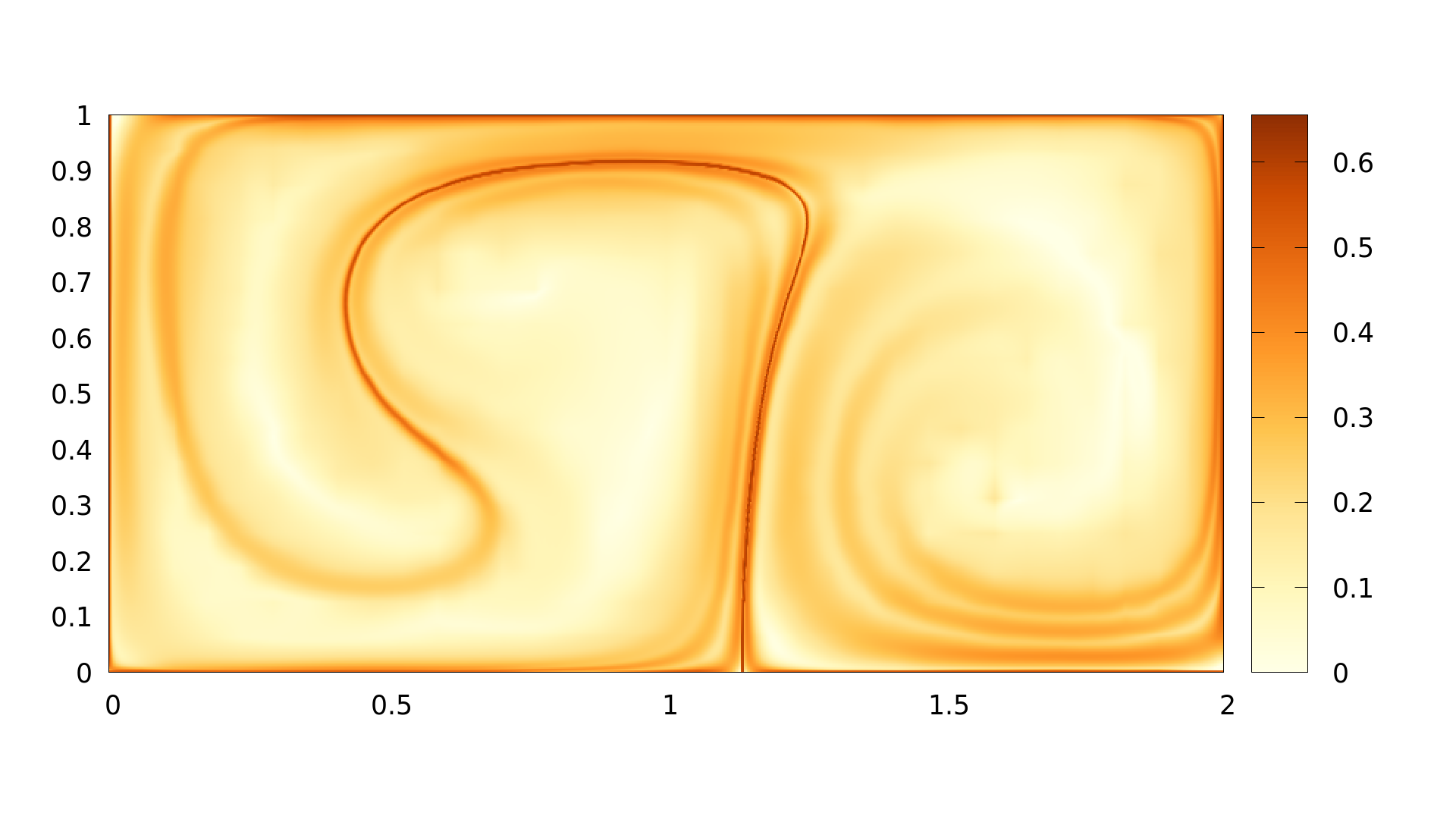}
  \label{fig:dg_opt_iter_last_ftle}
  \end{subfigure}
   \hspace{-0.3cm}
    \begin{subfigure}{0.164\textwidth}
    \includegraphics[width= \textwidth]{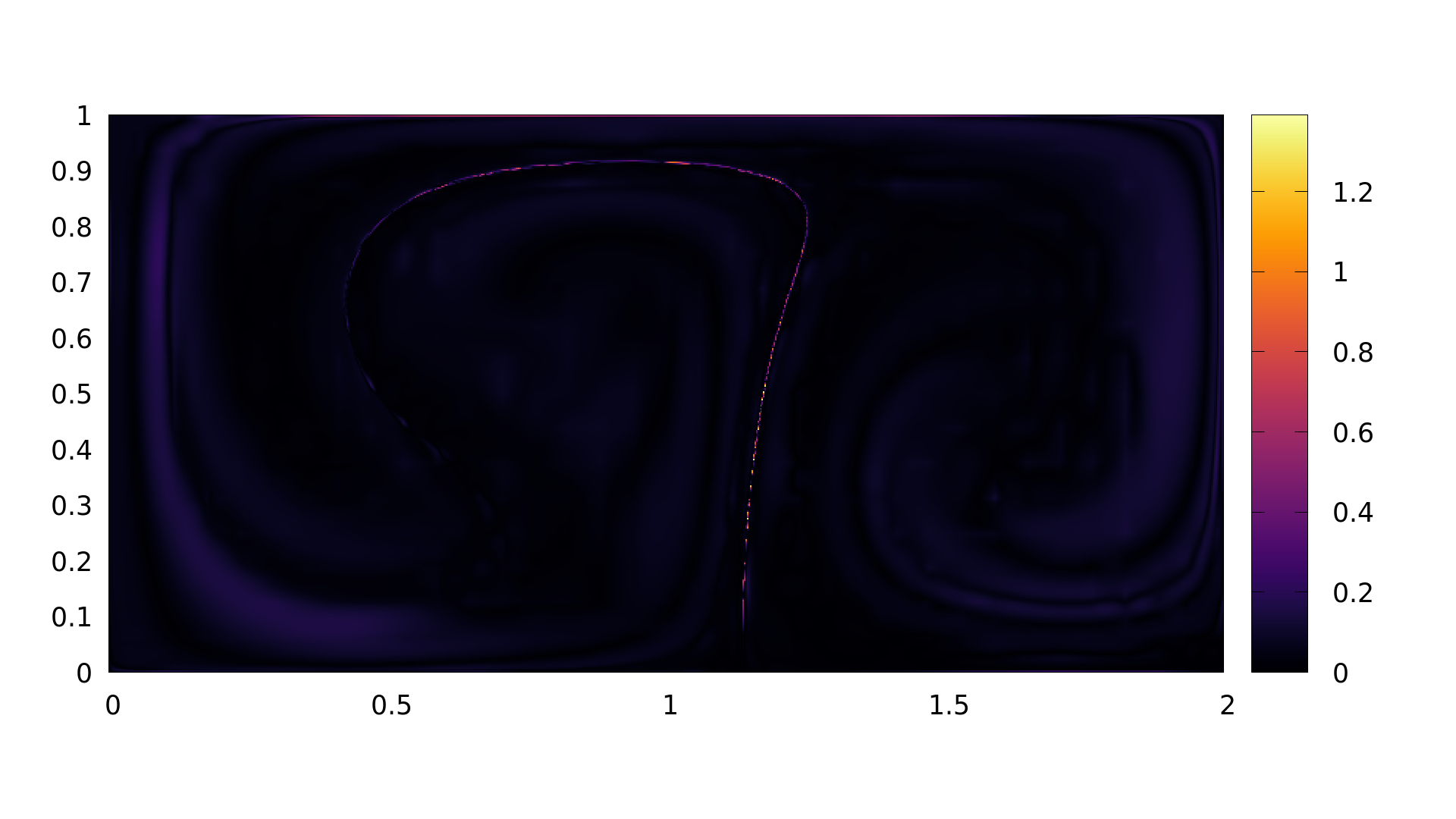}
  \label{fig:dg_opt_iter_last_error}
  \end{subfigure}

    \caption{
	Optimization for alignment of 2 Double Gyre ensembles.
	Top row (from left): two ensemble members and energy $A(\vp)$ to be minimized.\\
	Rows 2--5 show the state of optimization after 7, 46, 99 and 151  iterations.
        Each row shows (from left):
	$\vp$, $\overline{\mbox{FTLE}}$, $A(\vp)$. \\
    \label{fig:dg_opt}
  }
 \end{figure}

\subsection{Displaced flow maps and displaced FTLE.}
For each ensemble member $i=1\dots,n$, we compute the \emph{displaced
  flow map}\/
\begin{equation}
  \label{eq:definepsi}
  \psi_i(\vx,t,\tau) = \phi_i(\vx + \vp_{m,i}(\vx,t,\tau)  ,t,\tau)~.
\end{equation}
We can compute the (spatial) gradient
\begin{equation}
\label{eq:definenablapsi}
  \nabla \psi_i(\vx,t,\tau) = %
  \nabla \phi_i(\vx + \vp_{m,i}(\vx,t,\tau)  ,t,\tau) \; %
  \left( \mI + \nabla \vp_{m,i}(\vx,t,\tau) \right)
\end{equation}
where $\mI$ denotes the identity.
Based on $\nabla \psi_i$, we define \emph{displaced FTLE}\/
\begin{equation}
  \overline{\mbox{FTLE}}_i(\xx,t,\tau) = \frac{1}{|\tau|} \ln \sqrt{\lambda_{\max}\left( (\nabla \psi_i)^T \, \nabla \psi_i\right)}~.
\end{equation}
Note again that the idea behind the displacement $\vp_{m,i}$ is to align
the ridges of $\overline{\mbox{FTLE}}_i$ with the
ridges of $\overline{\mbox{FTLE}}_m$ as close as possible ,  i.e., the ridges of the median flow.
Remarks:
\begin{itemize}
\item
 (\ref{eq:definepsi}) shows that $\psi_i$ is not obtained by integrating a new velocity field. Instead, $\psi_i$ is obtained by a deformation of the precalculated flow map $\phi_i$ with the domain displacement $\vp_{m,i}$.
\item
(\ref{eq:definenablapsi}) shows the relation between  the gradients of $\phi_i$ and $\psi_i$. In particular, it shows that $\nabla \phi_i$ only becomes extremal if either $\nabla \psi_i$ or  $\nabla \vp_{m,i}$ become extremal. Since the condition $B(\vp_{m,i})$ prevents $\nabla \vp_{m,i}$ from becoming extremal and keeping in mind that FTLE is based on considering the extremals in  $\phi_i$ and $\psi_i$ respectively, there is a one-to-one relation between the ridges of $\nabla \phi_i$ and $\nabla \psi_i$. Therefore, our alignment approach cannot produce new FTLE ridges or make existing ones disappear. It can only transform and align their locations.
\end{itemize}

\subsection{Visual analysis and interpretation}
\label{sec:analysis}
Visual analysis can be based on either of
\begin{equation*}
  \psi_1,\dots,\psi_n\quad\textbf{or}\quad
  \nabla \psi_1,\dots, \nabla \psi_n \quad\textbf{or}\quad\overline{\mbox{FTLE}}_1,\dots,\overline{\mbox{FTLE}}_n~.
\end{equation*}
For the visualization of $\psi_1,\dots,\psi_n$, we apply existing
techniques for uncertain FTLE visualization:
D-FTLE and FTVA.

In addition, we can visually analyze the distribution of the domain
displacements $\vp_{m, i}$.
This provides information about the uncertainty of both, the location
and the strength of FTLE ridges.
Assuming a normal distribution of all $\vp_{m, i}$, we analyze the
covariance
\begin{equation}
  \label{eq:cov}
  \mC(\vx,t,\tau) =  \frac{1}{n}  \sum_{i=1}^n \, \vp_{m,i}(\vx,t,\tau) \,\, \vp_{m,i}(\vx,t,\tau)^T~.
\end{equation}
Assume that the integration time $t$ and period $\tau$ are fixed.
Then the tensor field $\mC$ assigns a symmetric positive definite matrix
to every domain point $\vx$ .
For a given direction vector $\vr$ with $\invn{\vr}=1$, the quadratic
form $\vr^T \, \mC \, \vr$ measures the variance of the domain
displacements in this direction.
We are interested in the variances near FTLE ridges:
high variance in a direction \emph{perpendicular}\/ to a ridge
indicates a high uncertainty in \emph{ridge location}\/, whereas high
variance in a direction \emph{parallel}\/ to a ridge indicates high
uncertainty in the strength of \emph{separation}\/.
The case $\mC=\vNull$ indicates that the ridges of all ensemble
members are perfectly aligned and have similar strength.

This tool complements the analysis of the distribution of
$\overline{\mbox{FTLE}}$:
Visualizing $\overline{\mbox{FTLE}}_1,\dots,\overline{\mbox{FTLE}}_n$
gives information about the average strength of separation, while the
analysis of $\mC$ reveals the uncertainty of ridge location and
strength.

\begin{figure}[h!]
  \centering
  \includegraphics[width= 0.17\textwidth]{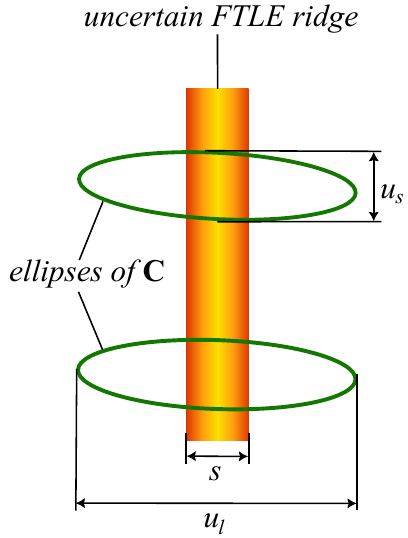}
  \caption{
    Sources of uncertainty encoded by our approach using tensors
    $\mC$ (see \eqref{eq:cov}):
    The width $s$ of a ridge corresponds to the strength of
    separation.
    The expansion $u_s$ of the ellipse parallel to the ridge encodes
    uncertainty of the strength of separation.
    Expansion $u_l$ perpendicular to the ridge encodes uncertainty of
    ridge location.
  }
  \label{fig_illustrationSourcesUncertainty}
\end{figure}

We visualize the tensor field $\mC$ by placing non-overlapping ellipses.
We are therefore computing the corresponding ellipse for every FTLE value
above a certain FTLE value.
As higher FTLE values tend to be close to each other, ellipses would overlap.
We therefore prioritize ellipses with corresponding higher FTLE values in a case of overlapping.
In order to prevent occluding the LCS structures, the ellipses are rendered semi-transparent.
This reduces clutter while preserving to render ellipses on the more important regions.
Ellipses below a certain diameter will also be discarded to prevent the clutter of dots on ridges with a high certainty.
High FTLE values which are not covered by ellipses therefore have a high certainty.
The scale of the ellipses is chosen as the 95 percentile of the normal
distribution.
Figures \ref{fig:sdg_3_dd} and \ref{fig:sdg_2_dd} show examples.
Based on this, we observe the extent and orientation of the ellipses
along the ridges of the underlying uncertain FTLE fields, which gives
the following interpretation:
\begin{itemize}
\item
  The width and sharpness of a ridge indicate the strength of
  \emph{separation}\/:
  the sharper the ridge, the stronger the flow separation.
\item
  The \emph{uncertainty in the strength of separation}\/ is encoded in
  expansion of the ellipses in the direction of the ridge:
  large expansion in ridge direction corresponds to high uncertainty
  in the strength of separation along the ridge.
\item
  The \emph{uncertainty in the ridge location}\/ is encoded in the
  ellipse expansion in the direction perpendicular to the ridges:
  the larger the ellipse expansion perpendicular to the ridge is, the
  more uncertain the ridge locations.
\end{itemize}
Figure ~\ref{fig_illustrationSourcesUncertainty} illustrates the
interpretation of uncertain ridges.

\subsection{Numerical optimization}
\label{sec:optimization}
We sample flow maps $\phi_i$ for all ensemble members and compute
pairwise distances to select the median flow map.
As the flow maps are sampled on a regular grid and represented as
matrices, we measure the distance between two flow maps $i, j$ simply
as $invn{\phi_i-\phi_j}_F$.
Then the main task is finding an optimal displacement $p_{i,m}$ for
each flow field $\phi_i$ w.r.t. the median flow $\phi_m$ as outlined
in section~\ref{sec:displacement}.

We represent domain displacements as bilinear maps $D\to{}D$, which
are parameterized by a grid of nodes or control points in 2d.
For each map we can evaluate the cost function \eqref{eq_finalenergy}
by replacing the integrals in the terms $A$ and $B$ by discrete sums.
Note that this summation acts on the discrete flow map, i.e., in the
sampling resolution.
We use the L-BFGS algorithm, a quasi-Newton method, for minimizing the
cost function.
This requires also the evaluation of the gradient of the cost
function.
We use central differences for approximating the gradient an the
following observation for avoiding redundant computations:
The displacement map is defined using a bilinear basis functions with
compact support, i.e., each control points acts only on a neighborhood
w.r.t. to the control grid.
This is a well-known property: our setting is a simple case of a
tensor product B-splines.

For the estimation of a partial derivative w.r.t one node using a
finite difference, this means that the summation (i.e., the discrete
integration) can be restricted to a local neighborhood:
as all contributions away from this neighborhood stay constant, they
cancel in the finite difference.
Note that a similar argument applies in a continuous formulation.

As the gradient estimation dominates the computational cost for
minimization, this observation has a significant impact on
performance, because it ``decouples'' the resolution of the sampling
grid from the resolution of the grid of nodes controlling the
displacement:
The computation cost is de facto bound by the significantly lower
resolution of the control grid.
For additional speedup, we take advantage of parallel computation for
both, the summation over the whole domain for evaluation the cost
function and for the evaluation of partial derivatives.

We apply the following boundary conditions for the minimization:
all nodes must stay within the domain, and displacements -- i.e.,
nodes -- on the boundary must map to a boundary point, i.e., they have
only one degree of freedom along the boundary curve.
This ensures that the displaced points stay within the domain.
Displacements at the corners of the rectangular domain vanish: corners
remain corners.
Note that the latter condition is only meaningful if there are feature
points such as corners on the domain boundary curve (e.g., for a
rectangle).

Note that the term $B(\vp)$ that penalizes nonrigid behavior enforces
a smooth displacement and to some extent penalized non-bijective
displacements maps, i.e., ``flipped'' cells in the deformed grid.
We don't see any need to add further constraints that guarantee a
bijective deformation of the domain (see also section~\label{sec:discussion}).

Finally, the computations required for the visual analysis and
interpretation (section~\ref{sec:analysis}), are straightforward:
replace $\phi_i$ by $\psi_i$.
We sample $\psi_i$ for computing $\nabla\psi_i$ and
$\overline{\mbox{FTLE}}$ similarly as for computing $\nabla\phi_i$ and
$\mbox{FTLE}$ from sampled flow maps $\phi_i$.

\section{Results}

\subsection{Introductory example}

\begin{figure}[h]
  \centering
   \vspace{-1.cm}
  \begin{subfigure}{0.5\textwidth}
    \includegraphics[width= \textwidth]{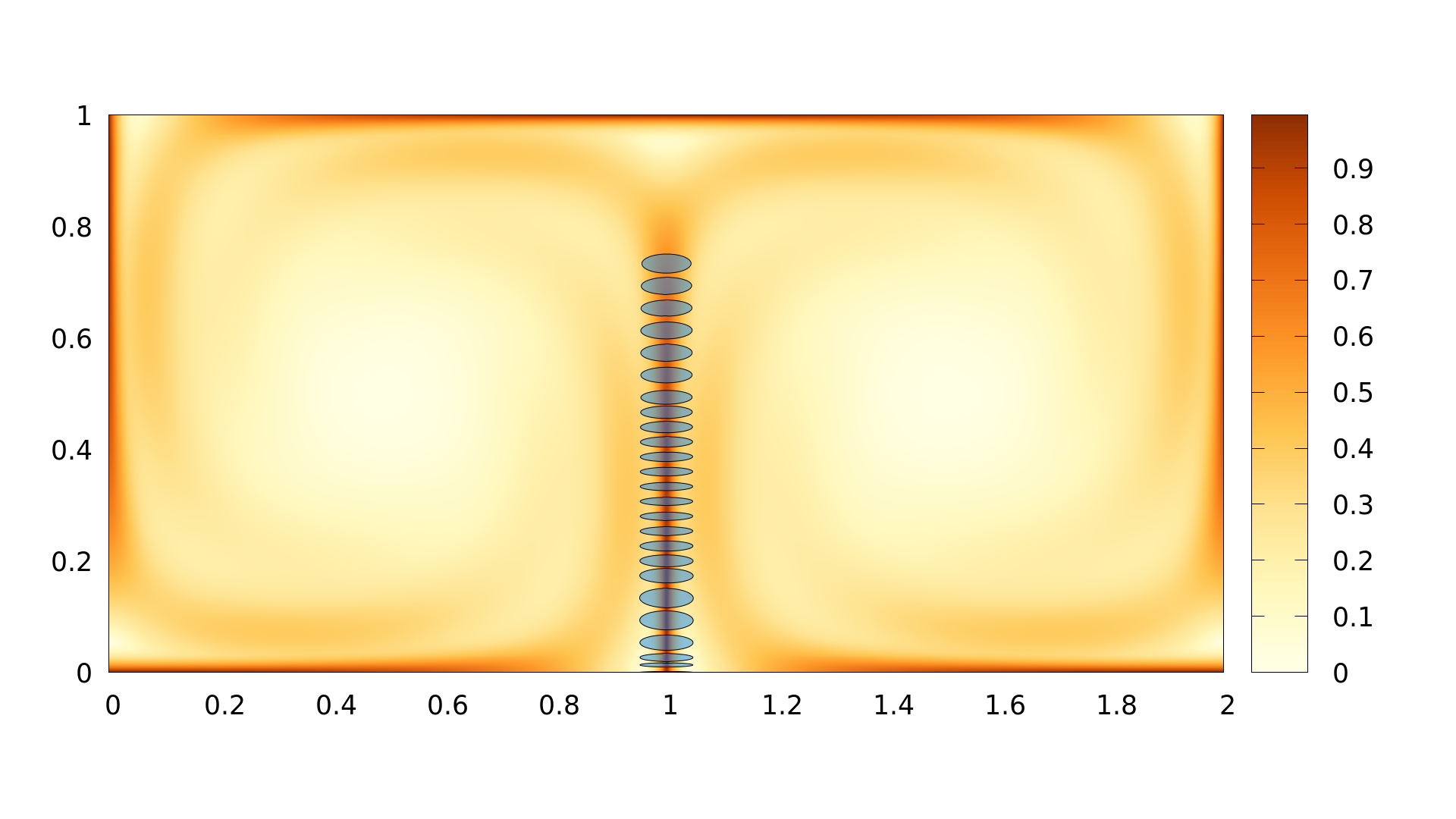}
  \label{fig:sdg_3_dd_d-ftle}
  \end{subfigure}
  \vspace{-1.cm}
      \begin{subfigure}{0.5\textwidth}
      \vspace{-1.5cm}
    \includegraphics[width= \textwidth]{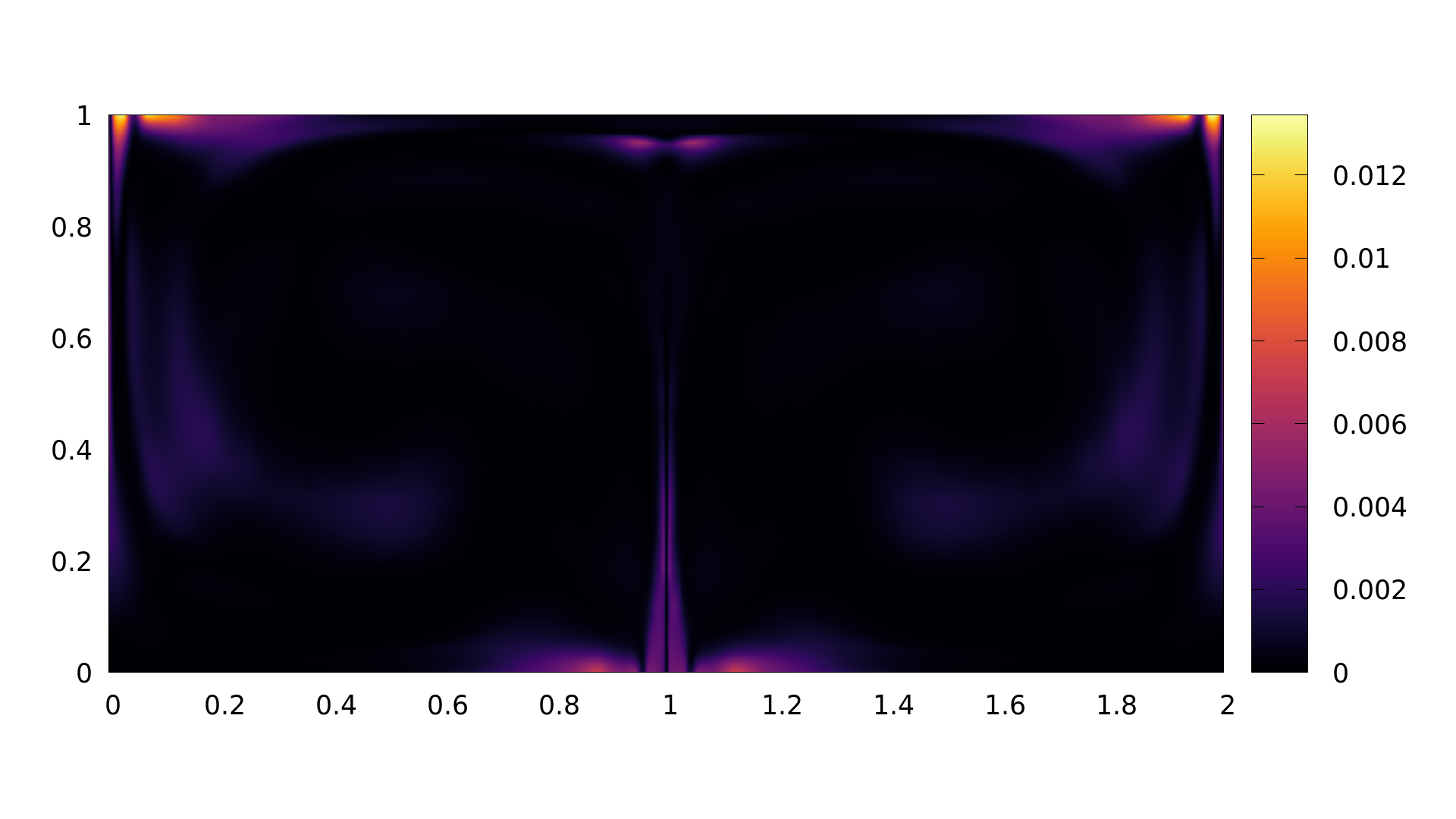}
  \label{fig:sdg_3_variance_dd-dftle}
  \end{subfigure}
     \caption{
	The optimal domain displaced D-FTLE and its variance for  ensemble $E_1$.
      \label{fig:sdg_3_dd}
      }
  \end{figure}

\begin{figure}[H]
  \centering
  \vspace{-1.2cm}
  \begin{subfigure}{0.5\textwidth}
    \includegraphics[width= \textwidth]{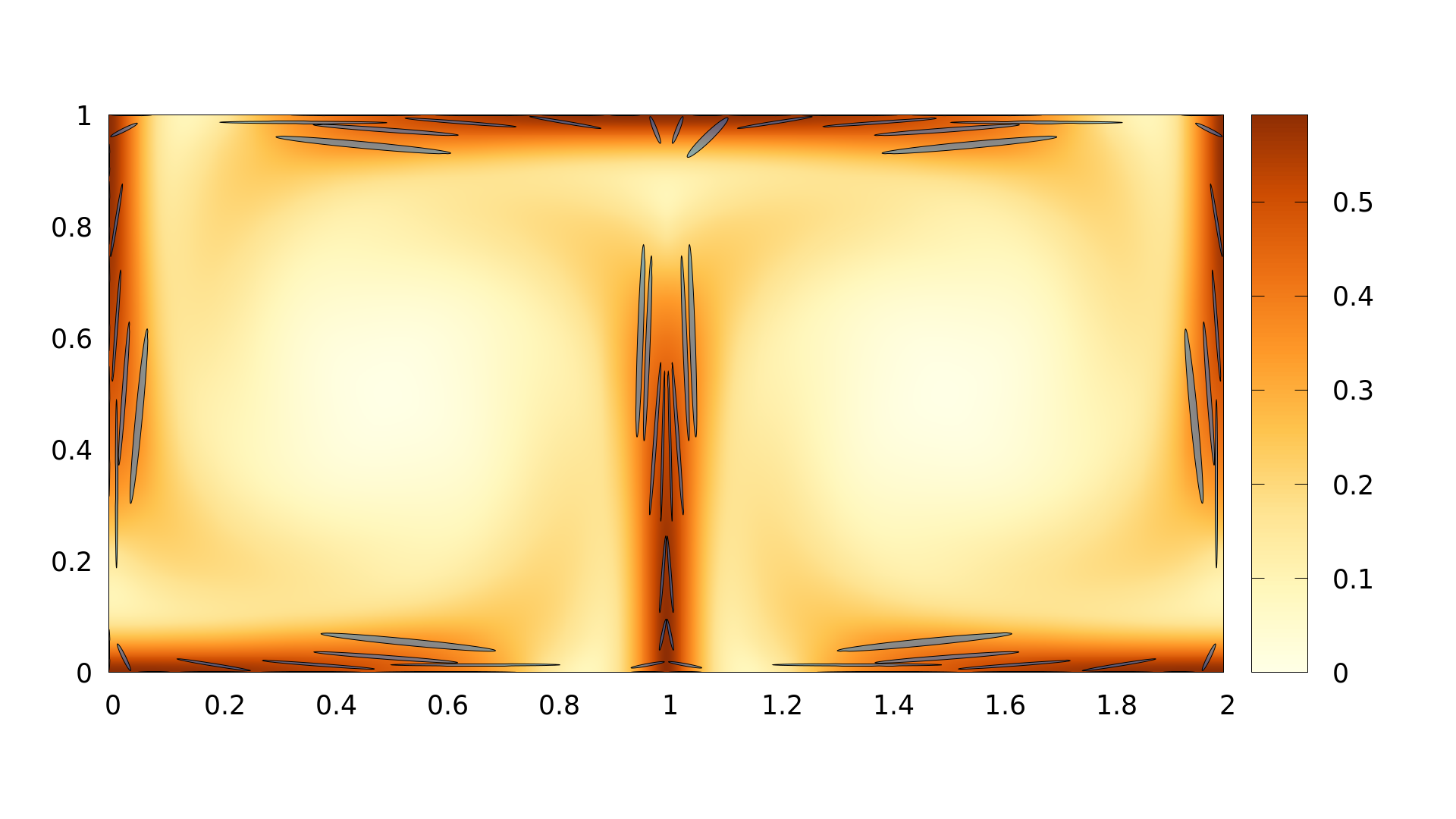}
  \label{fig:sdg_2_dd_d-ftle}
  \end{subfigure}
  \vspace{-1.2cm}
      \begin{subfigure}{0.5\textwidth}
   \vspace{-1.5cm}
    \includegraphics[width= \textwidth]{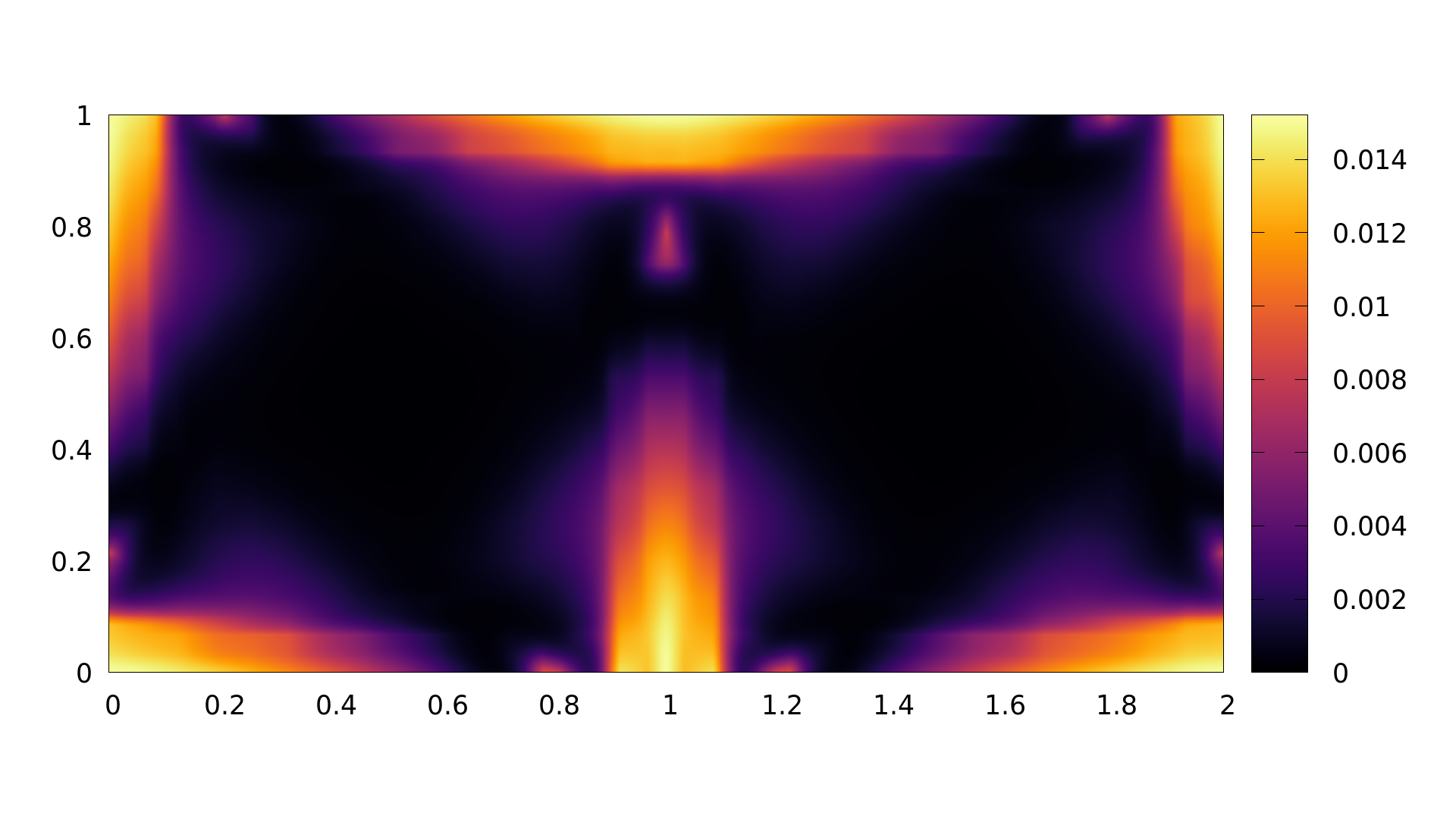}
  \label{fig:sdg_2_variance_dd-dftle}
  \end{subfigure}
  \hfill
     \caption{
	The optimal domain displaced D-FTLE and its variance for the ensemble $E_2$.
      \label{fig:sdg_2_dd}
      }
  \end{figure}

\begin{figure}[H]
  \centering
    \vspace{-1.cm}
  \begin{subfigure}{0.25\textwidth}
    \includegraphics[width= \textwidth]{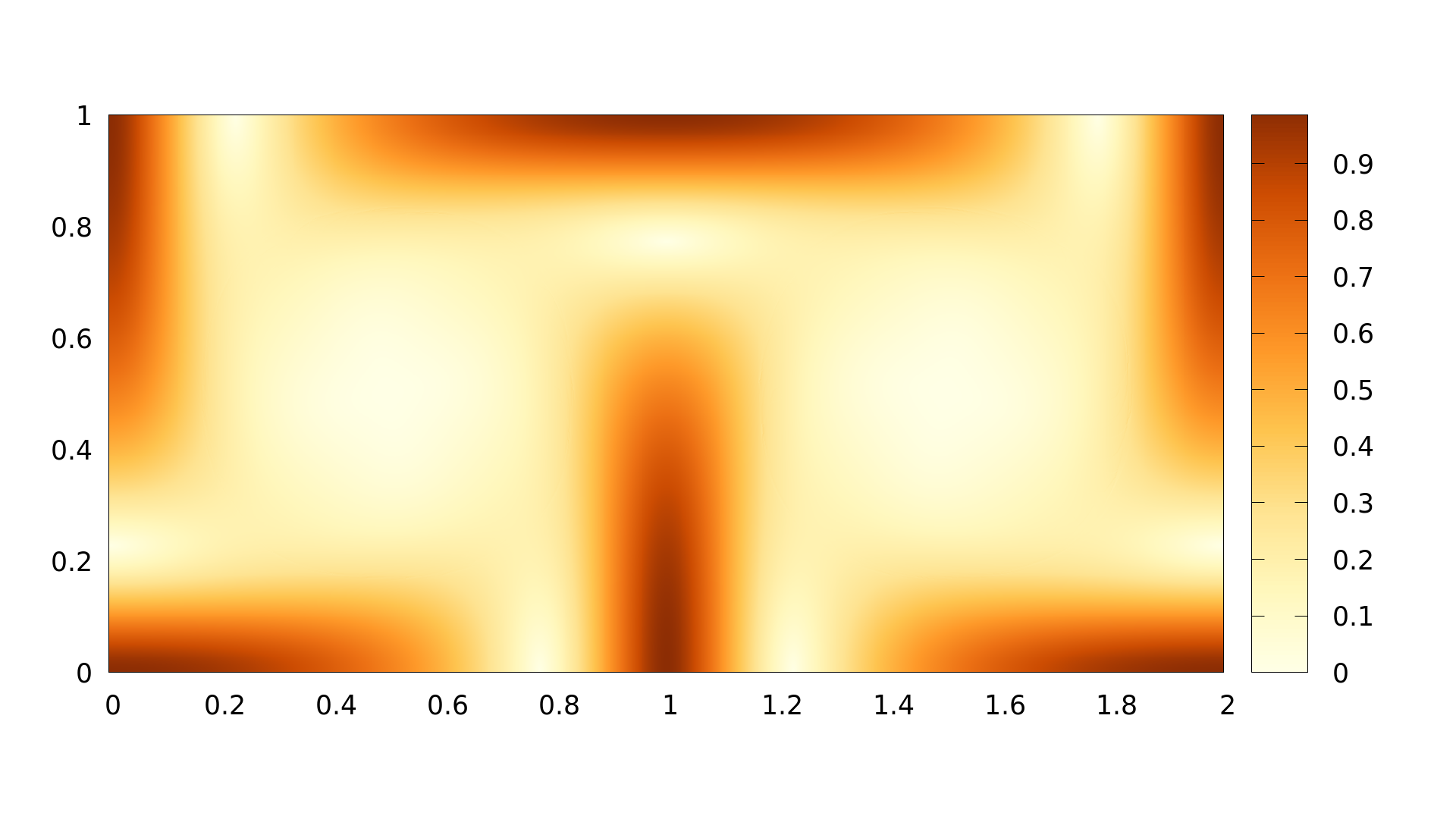}
  \label{fig:sdg_1_dd_d-ftle}
  \end{subfigure}
  \vspace{-1.0cm}
  \hspace{-0.38cm}
      \begin{subfigure}{0.25\textwidth}
    \includegraphics[width= \textwidth]{Steady_Double_Gyre_1_variance_d-ftle.png}
  \label{fig:sdg_1_variance_dd-dftle}
  \end{subfigure}
    \hfill
     \caption{
	The optimal domain displaced D-FTLE and its variance for the ensemble $E_3$.
      \label{fig:sdg_1_dd}
      }
  \end{figure}

We start the analysis of our method in comparison to the introductory example in section \ref{sec:example1}, previously illustrated in figures   \ref{fig:sdg_3}--\ref{fig:sdg_1}.
Figure  \ref{fig:sdg_3_dd} (top) shows D-FTLE of the optimal domain displaced flow ensemble $\psi_1,...,\psi_{50}$. We see a rather sharp ridge at the location $x=1$, encoding a strong separation. Further, we see that the ellipses of  $\mC$ on the ridges have a strong expansion in the direction perpendicular to the ridge direction, where the expansion parallel to the ridge expansion is low. This encodes the strong uncertainty of ridge location, and weak uncertainty of the strength of separation. Figure \ref{fig:sdg_3_dd} (bottom) shows the variance of domain displaced D-FTLE, in comparison to its non-displaced version in figure\ref{fig:sdg_3} (right column, second row).
Figure  \ref{fig:sdg_2_dd} (top) shows the result for the ensemble $E_2$. Here the ellipses of $\mC$ align with the   ridge direction, encoding a strong uncertainty of the separation strength but low uncertainty in the ridge location. For reference, compare the variance of the domain displaced D-FTLE figure
\ref{fig:sdg_2_dd} (bottom) with the variance of the original D-FTLE figure \ref{fig:sdg_2} (right column, second row).
Figure  \ref{fig:sdg_1_dd} shows D-FTLE and its variance for the ensemble $E_3$. Here, we see no ellipses of $\mC$ at all. In fact, all ellipses are vanishing, indicating no uncertainty in ridge location and strength of separation.

\subsection{Double gyre}

\begin{figure}[H]
  \centering
  \vspace{-0.5cm}
  \begin{subfigure}{0.25\textwidth}
    \includegraphics[width= \textwidth]{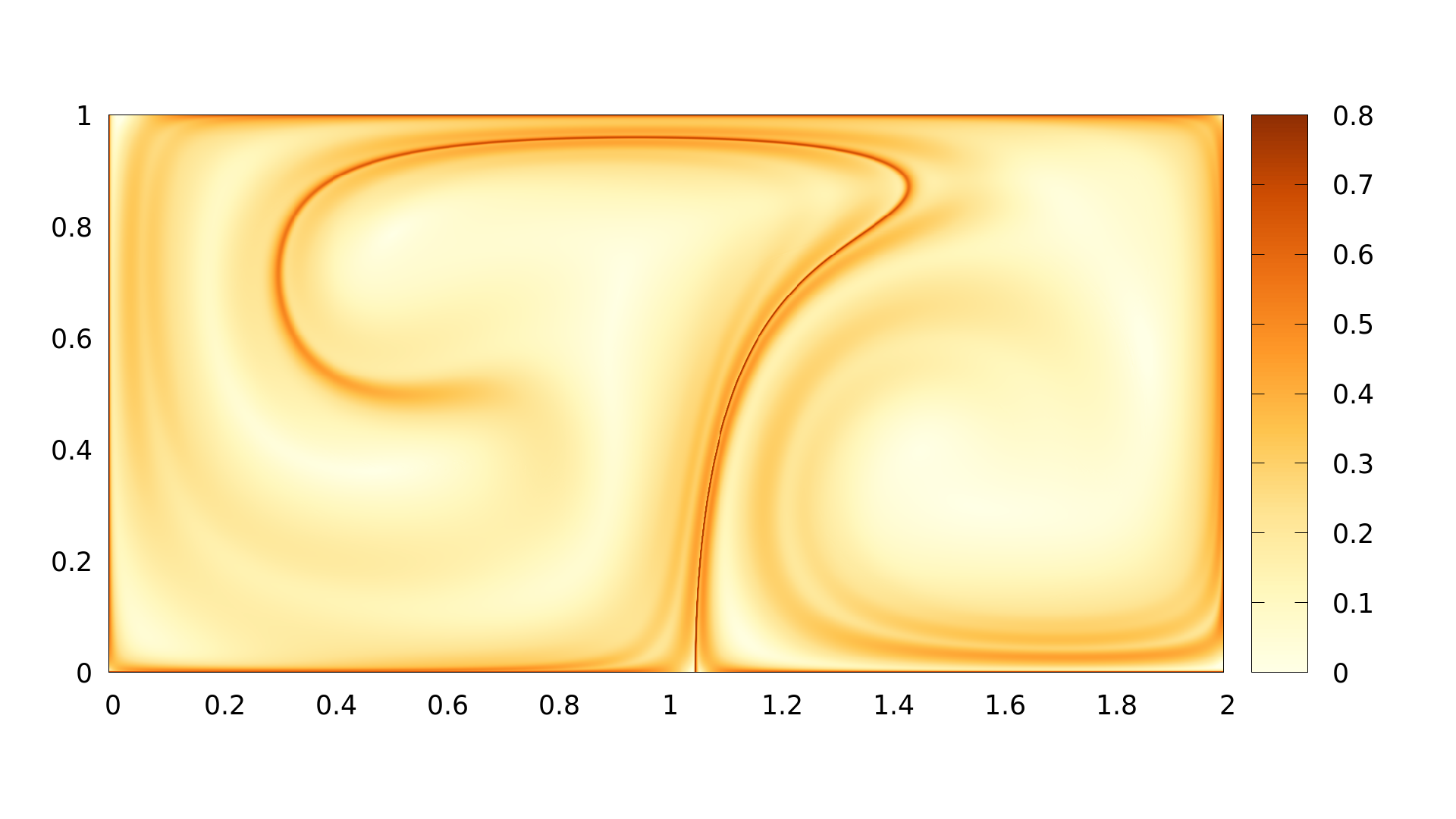}
  \label{fig:dg_2_ftle_0}
  \end{subfigure}
 \hspace{-0.38cm}
 \vspace{-0.8cm}
      \begin{subfigure}{0.25\textwidth}
    \includegraphics[width= \textwidth]{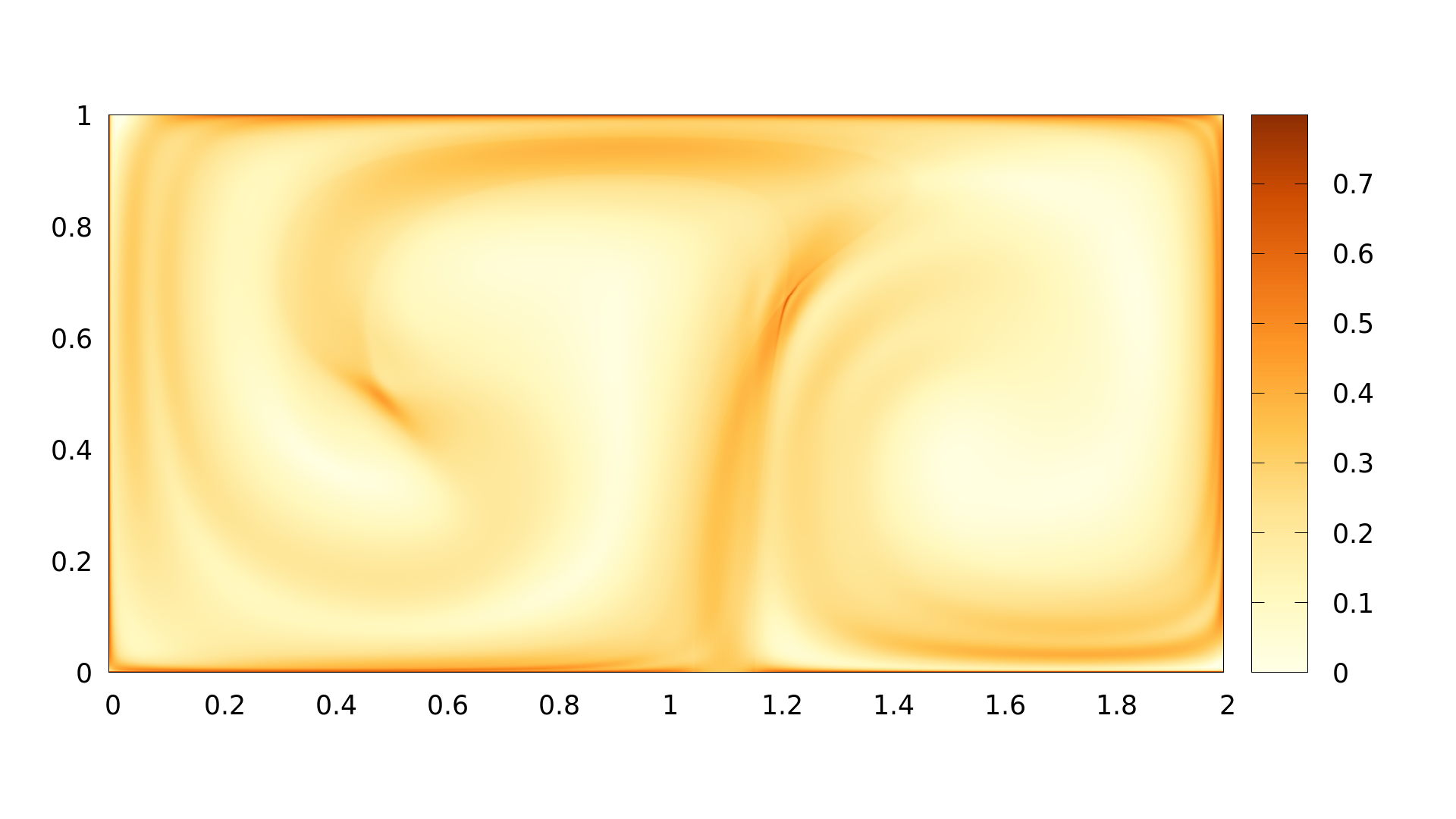}
  \label{fig:dg_2_d-ftle}
  \end{subfigure}
 \vspace{-0.8cm}
    \begin{subfigure}{0.25\textwidth}
    \includegraphics[width= \textwidth]{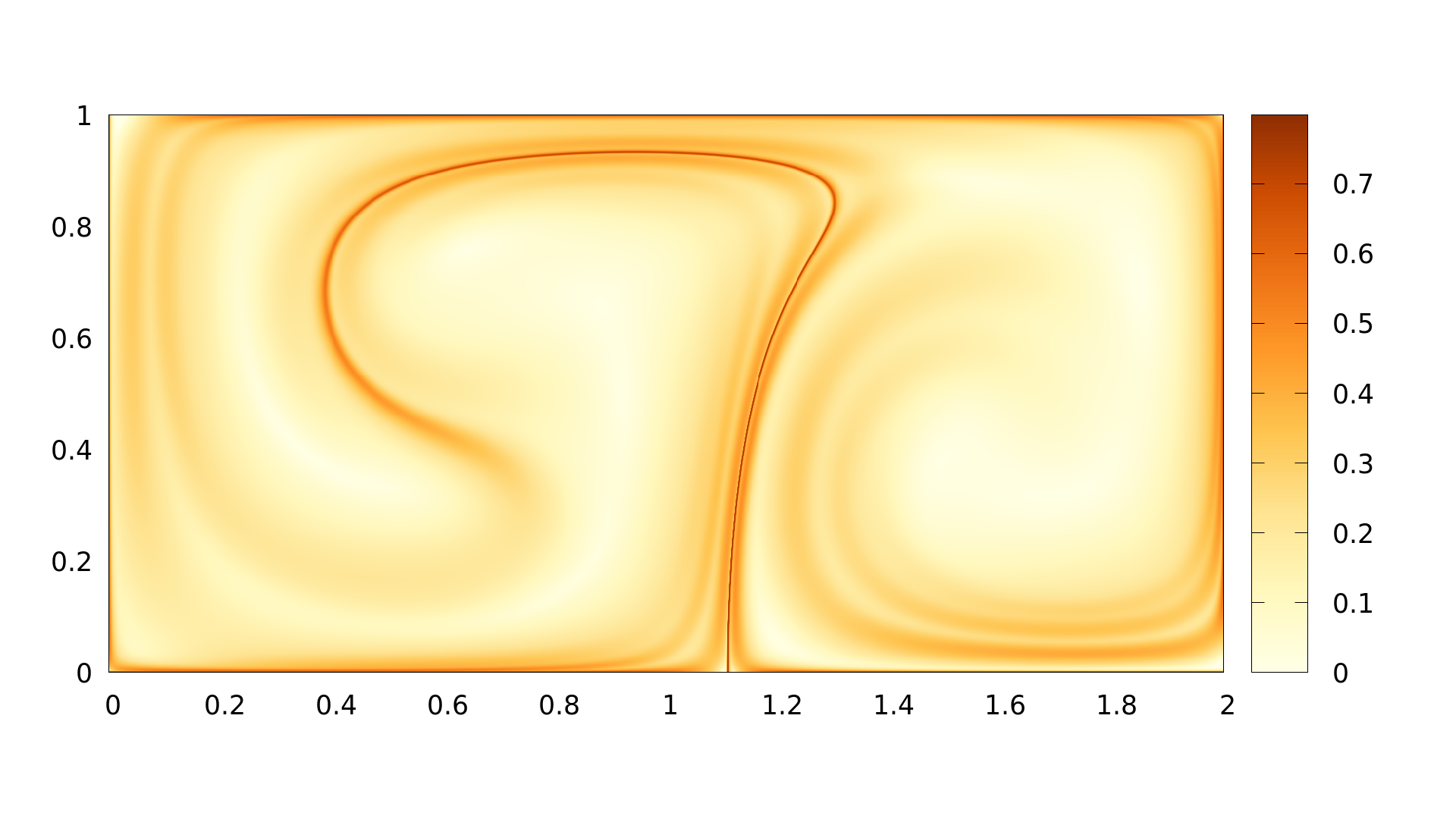}
  \label{fig:dg_2_ftle_1}
  \end{subfigure}
 \hspace{-0.38cm}
   \begin{subfigure}{0.25\textwidth}
    \includegraphics[width= \textwidth]{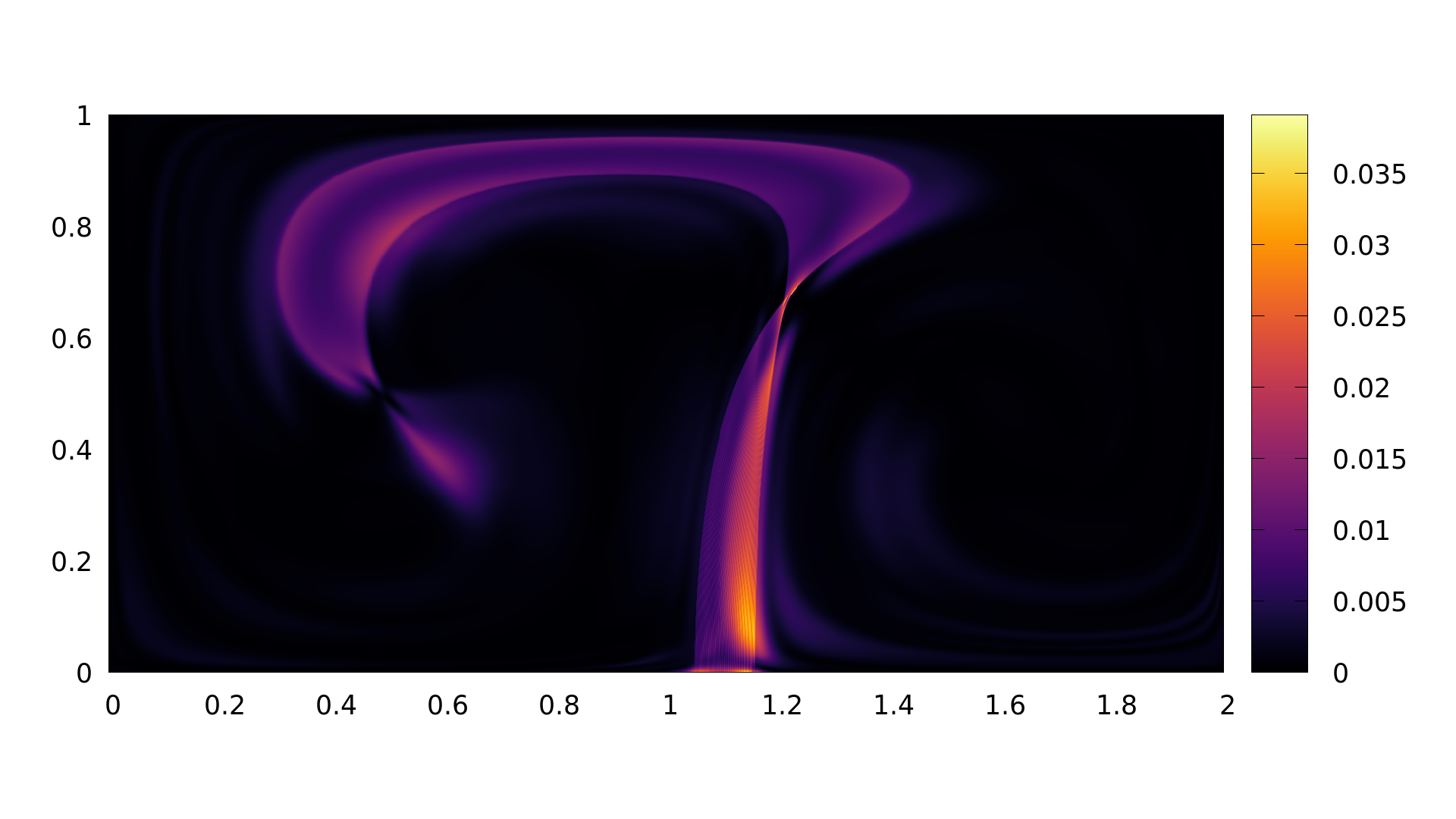}
  \label{fig:dg_2_variance_d-ftle}
  \end{subfigure}
 \vspace{-0.5cm}
   \begin{subfigure}{0.25\textwidth}
    \includegraphics[width= \textwidth]{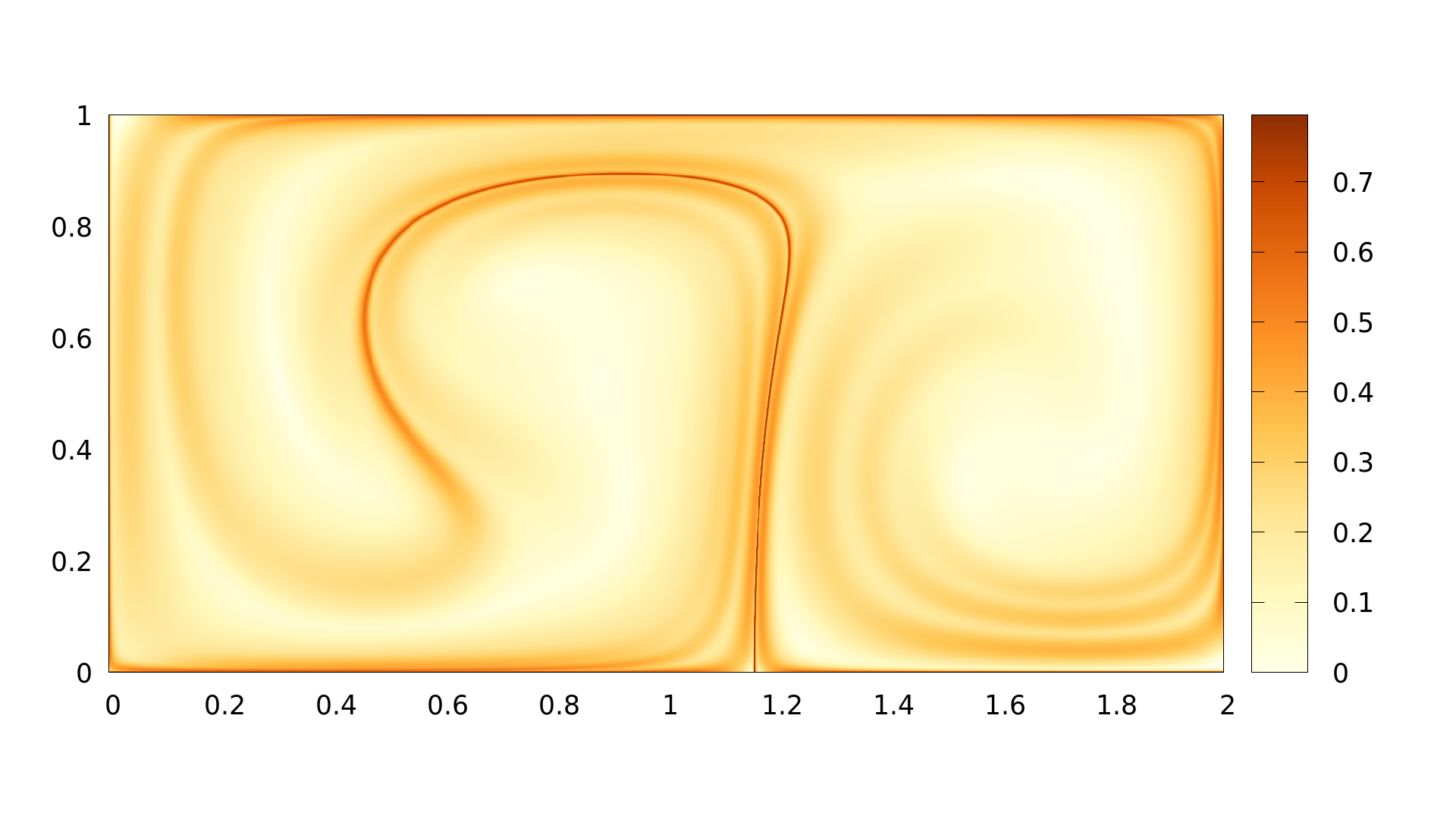}
  \label{fig:dg_2_ftle_2}
  \end{subfigure}
 \hspace{-0.38cm}
   \begin{subfigure}{0.25\textwidth}
    \includegraphics[width= \textwidth]{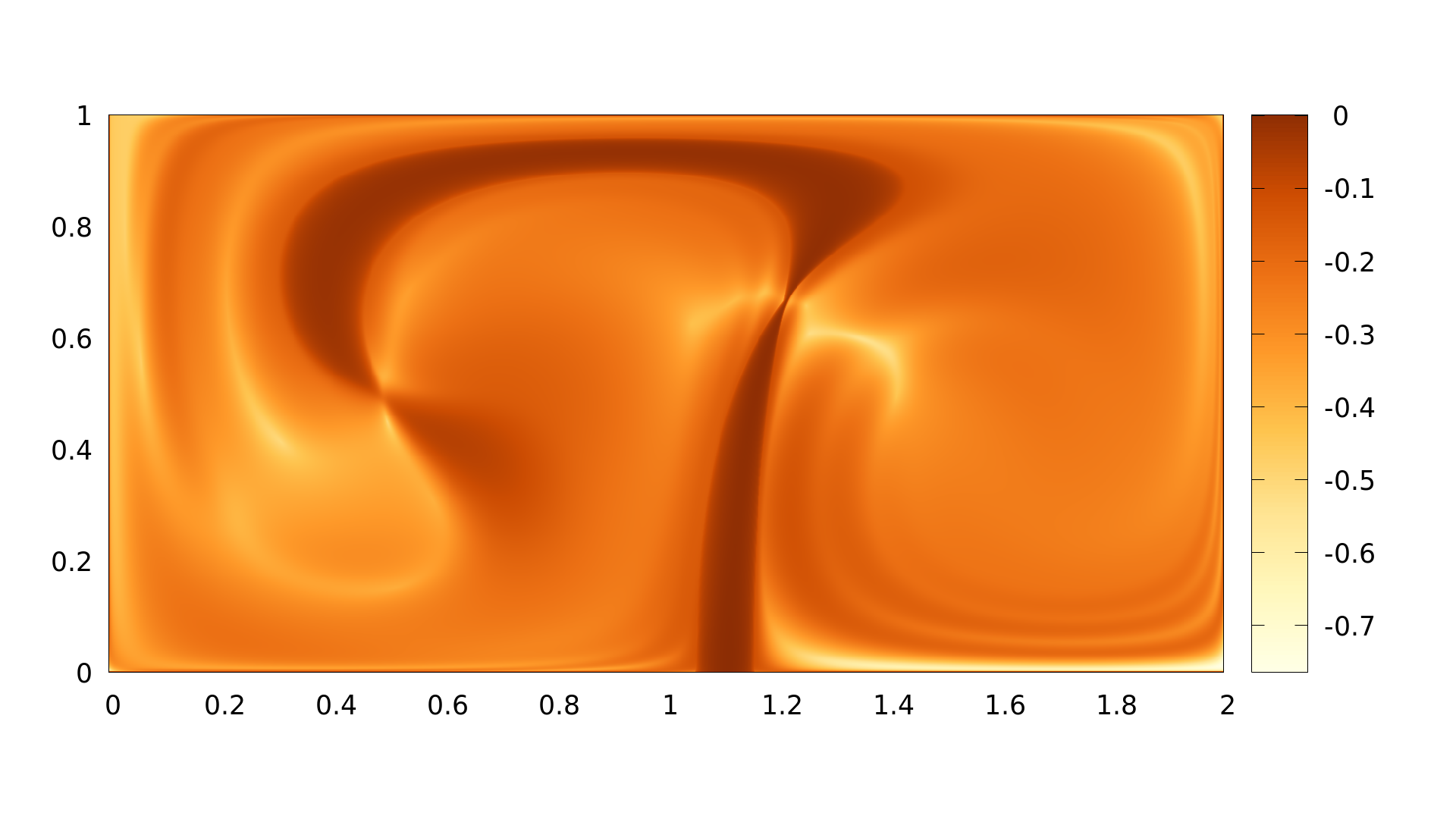}
  \label{fig:dg_2_ftva}
  \end{subfigure}
    \caption{
    Double Gyre ensemble with $\tau = 11$.
    Left column from top to bottom:
    FTLE of ensemble runs with $t=-0.5$, $t=0$, $t=0.5$,. \\
    Right column from top to bottom:
    D-FTLE, variance of D-FTLE, FTVA.
}
 \label{fig:dg_2}
 \end{figure}

We consider the double gyre data set from \cite{Shadden2005} that is defined as
\begin{eqnarray}
f(x,t) &=& a(t)x^2+b(t)x
\\
a(t) &=& \epsilon \, sin(\omega t)
\\
b(t) &=& 1- 2\epsilon \, sin(\omega t)
\\
\label{eq_doublegyre}
\vv_{DG}(\vx,t) = \vv_{DG}(x,y,t) &=& \begin{pmatrix}
\pi \, A \, sin(\pi f(x))cos(\pi y)\\
\pi \, A \, cos(\pi f(x))sin(\pi y)
\end{pmatrix}
\end{eqnarray}
We use $A=0.1$, $\epsilon=0.25$ and $\omega=\frac{2\pi}{10}$.
The data set is an interesting test case because we can create different ensemble members by varying $t$ and $\tau$.
In fact, varying $t$ results in ridges at different positions, as the Gyres oscillate.
Different integration times $\tau$ result in FTLE ridges of different strength and width.
Figure \ref{fig:dg_2} shows an ensemble of 50 members with times in the range of $t={-0.5, 0.5}$ and constant integration time $\tau = 11$ of the Double Gyre. \\

\begin{figure}[H]
  \centering
    \vspace{-1.cm}
  \begin{subfigure}{0.5\textwidth}
    \includegraphics[width= \textwidth]{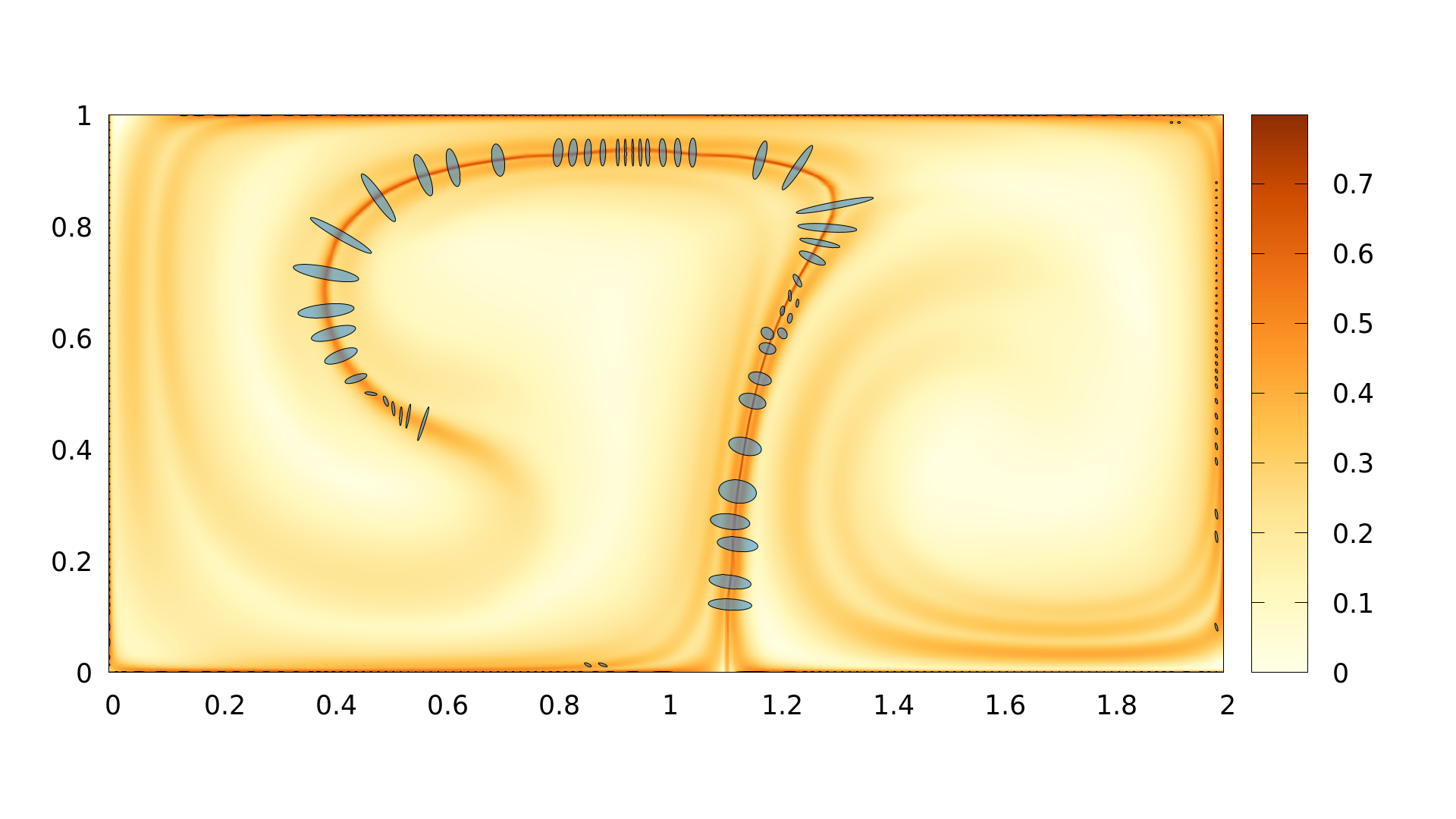}
  \label{fig:dg_2_dd_d-ftle}
  \end{subfigure}
  \hfill
  \vspace{-2.cm}
      \begin{subfigure}{0.5\textwidth}
    \includegraphics[width= \textwidth]{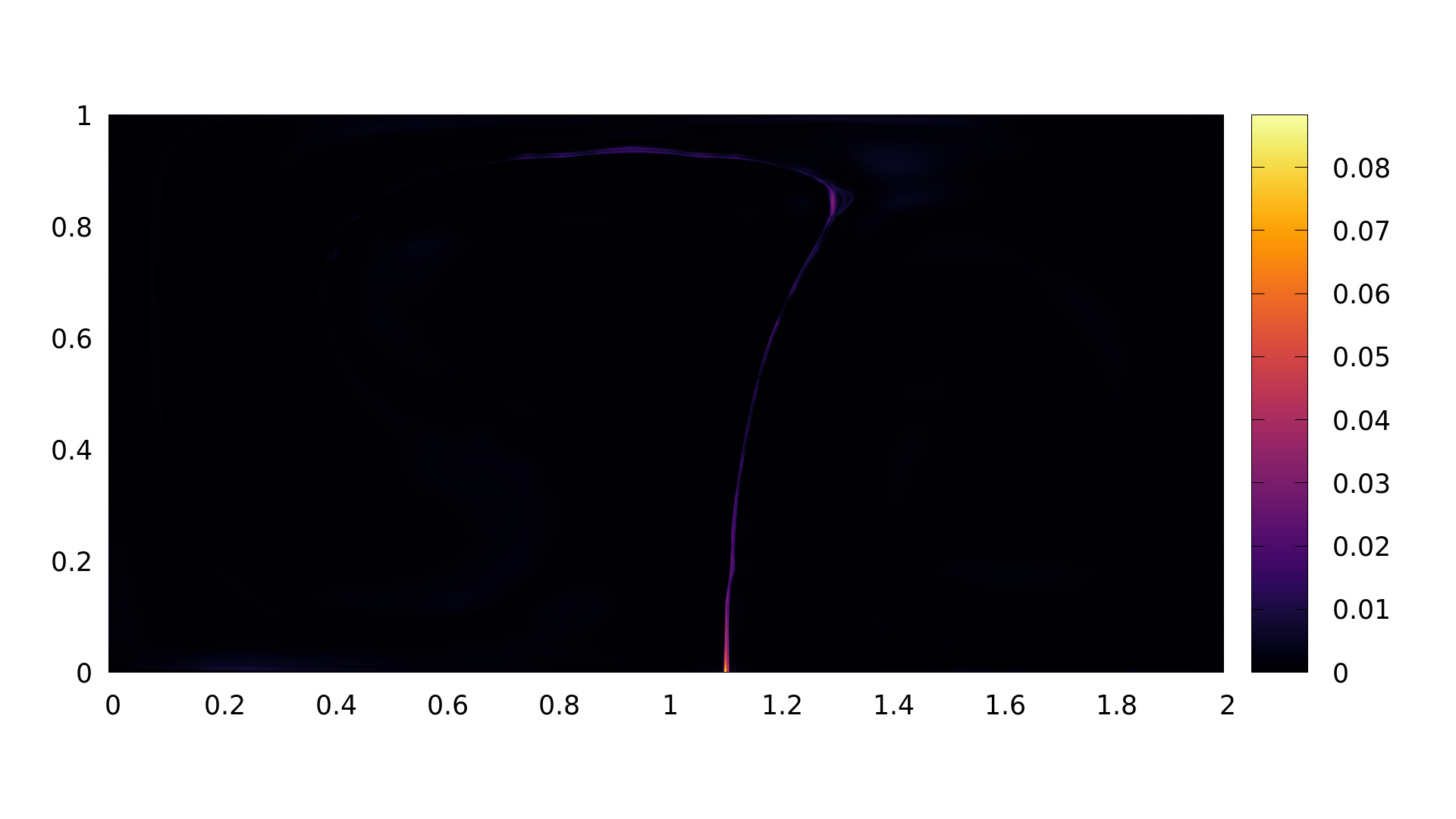}
  \label{fig:dg_2_variance_dd-dftle}
  \end{subfigure}
  \hfill
  \vspace{-1.2cm}
     \caption{
	The optimal domain displaced D-FTLE and its variance for the Double Gyre ensemble.
      \label{fig:dg_2_dd}
      }
  \end{figure}
Figure \ref{fig:dg_2}(left column) shows the FTLE fields of the ensemble members. The right-hand column of \ref{fig:dg_2} shows standard uncertain FTLE visualizations: D-FTLE, variance of D-FTLE, FTVA (from top to bottom). They show broad and unsharp ridges which are nearly vanishing. \\
 Figure \ref{fig:dg_2_dd} shows the result of our approach: D-FTLE and its variance for the domain displaced ensemble members. We see sharp ridges, encoding a strong separation. Along the ridges we observe different kinds of behavior of the ellipses of $\mC$. In most regions they are rather "thin" (i.e, having a strong aspect ratio). Further, in some regions they are aligned with the ridge directions, while in other parts orientation is perpendicular to the ridge direction. 

 \subsection{Displacement example}
\begin{figure}[H]
  \centering
  \begin{subfigure}[][][t]{0.22\textwidth}
    \includegraphics[width= \textwidth]{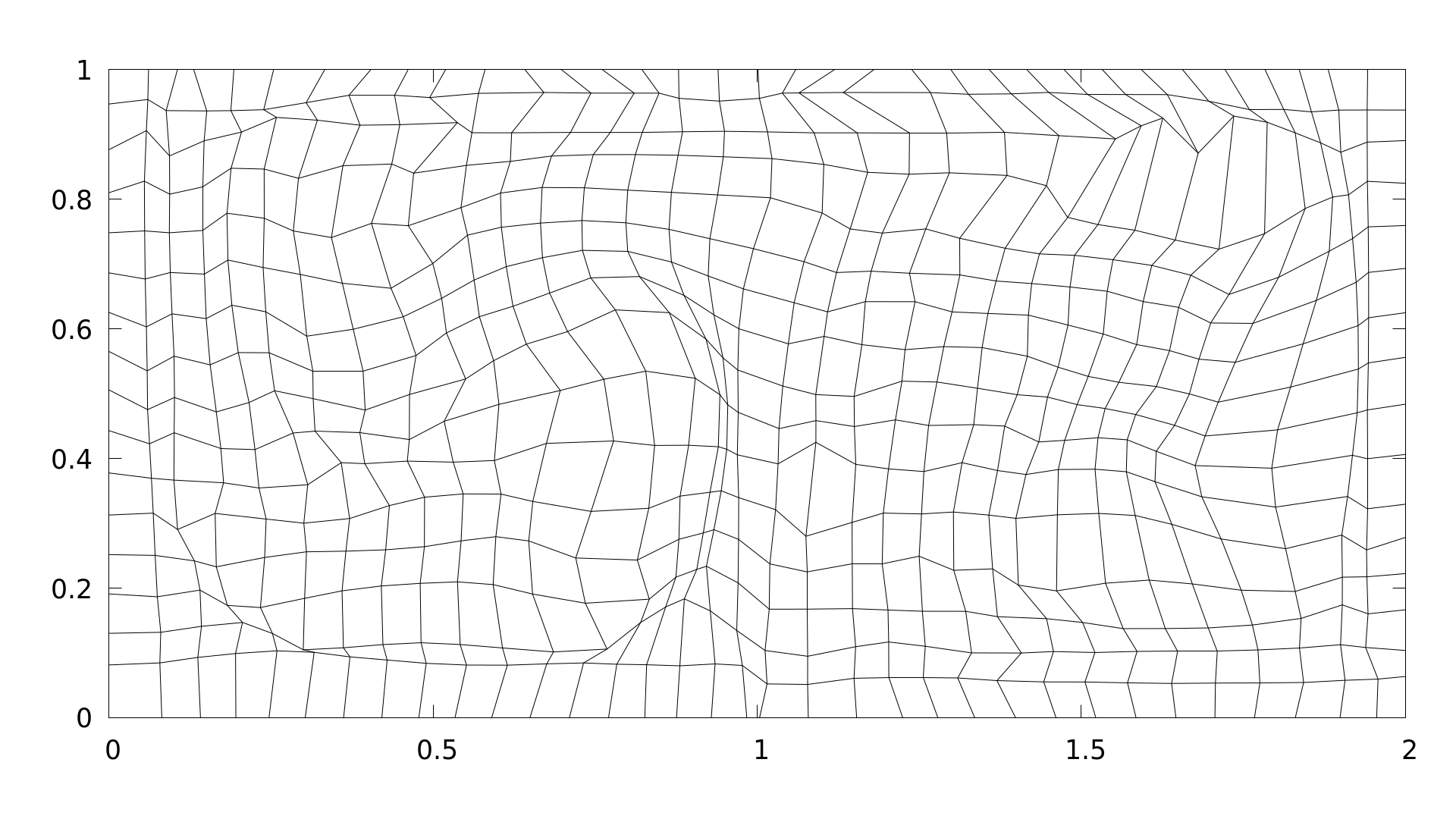}
  \label{fig:dg_h_grid_2_0}
  \end{subfigure}
 \hspace{-0.38cm}
 \vspace{-0.8cm}
  \begin{subfigure}[][][t]{0.26\textwidth}
    \includegraphics[width= \textwidth]{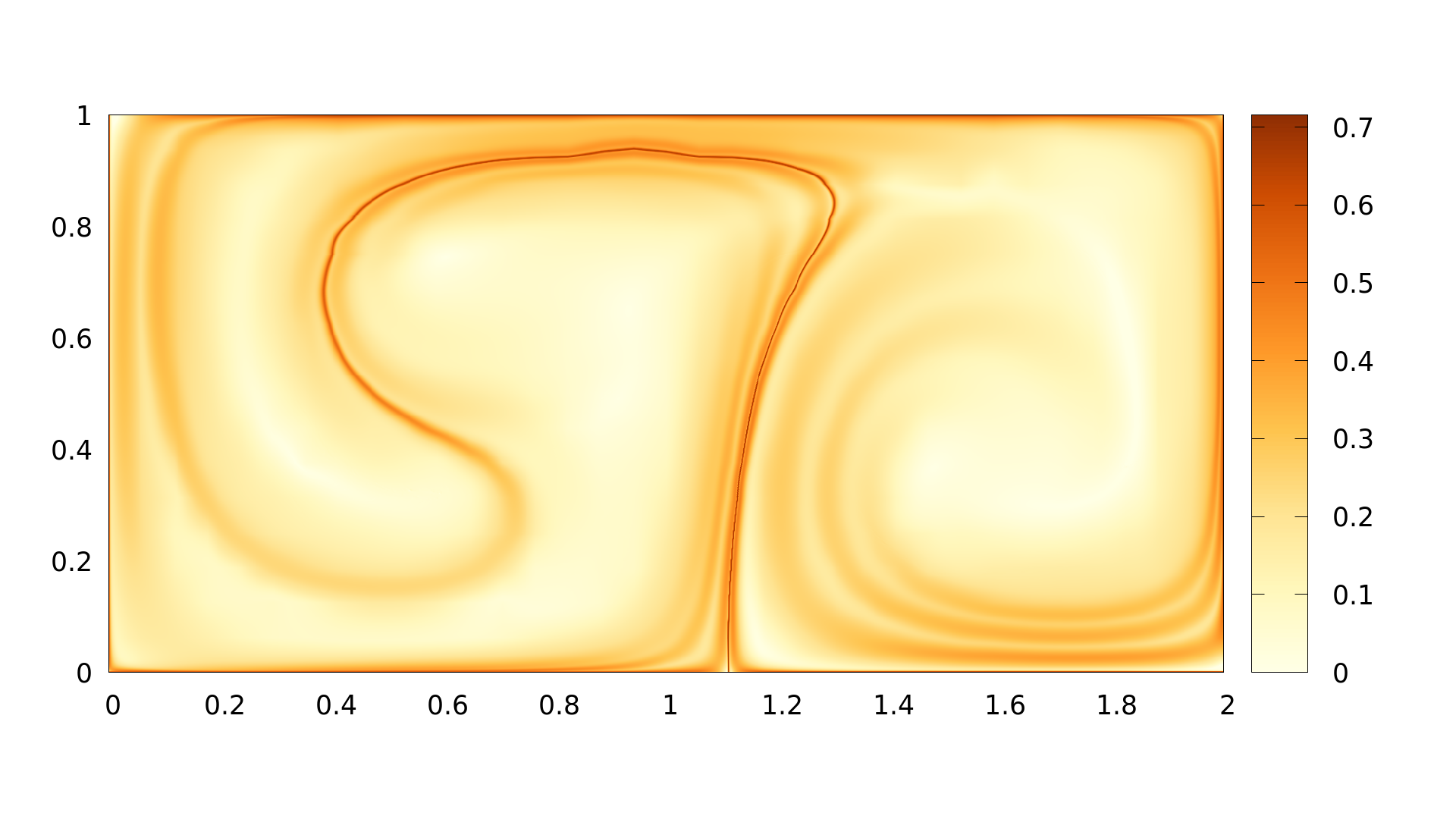}
  \label{fig:dg_h_ftle_2_0}
  \end{subfigure}
  \begin{subfigure}[][][t]{0.22\textwidth}
    \includegraphics[width= \textwidth]{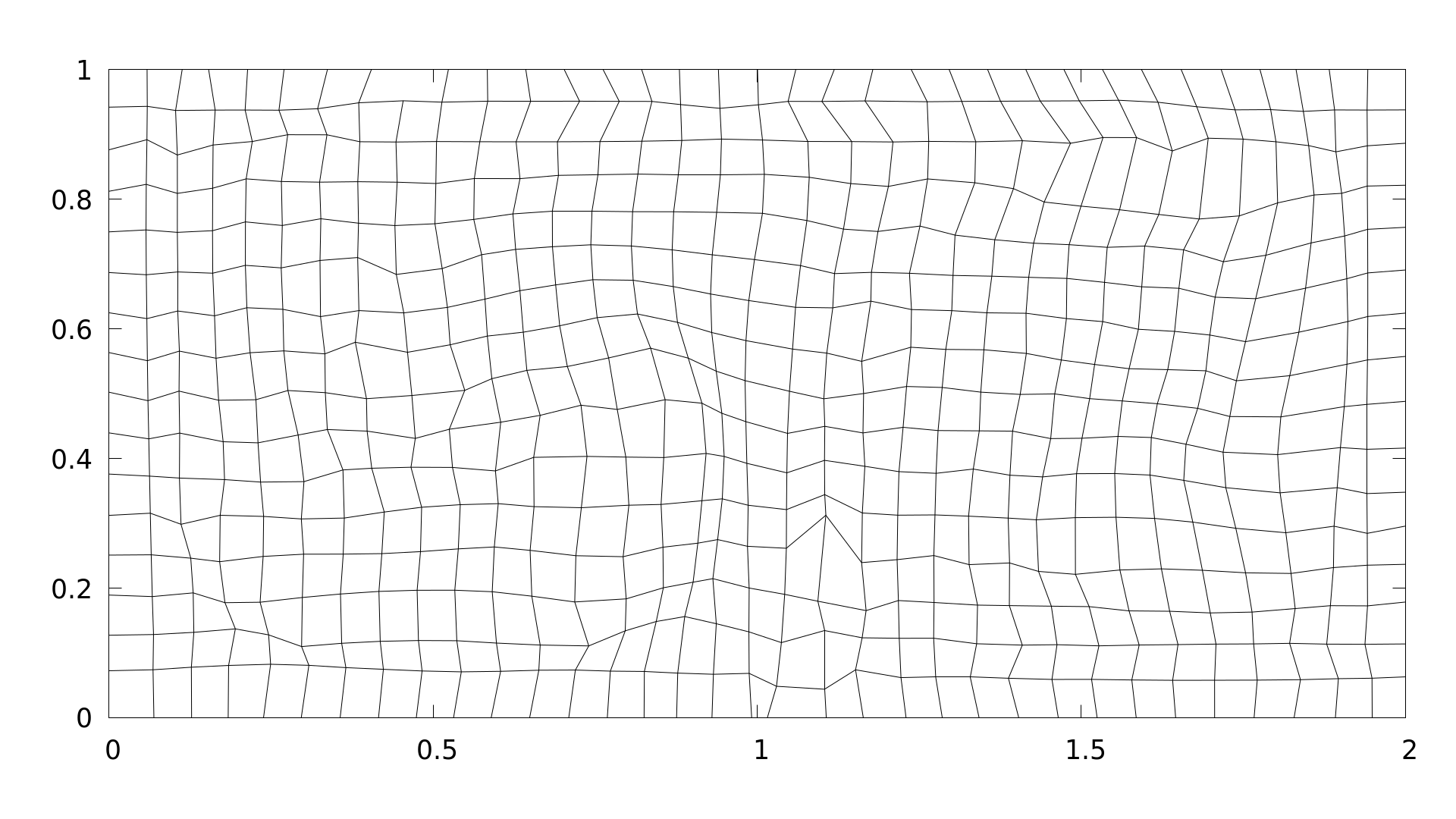}
  \label{fig:dg_h_grid_2_1}
  \end{subfigure}
 \hspace{-0.38cm}
 \vspace{-0.8cm}
  \begin{subfigure}[][][t]{0.26\textwidth}
    \includegraphics[width= \textwidth]{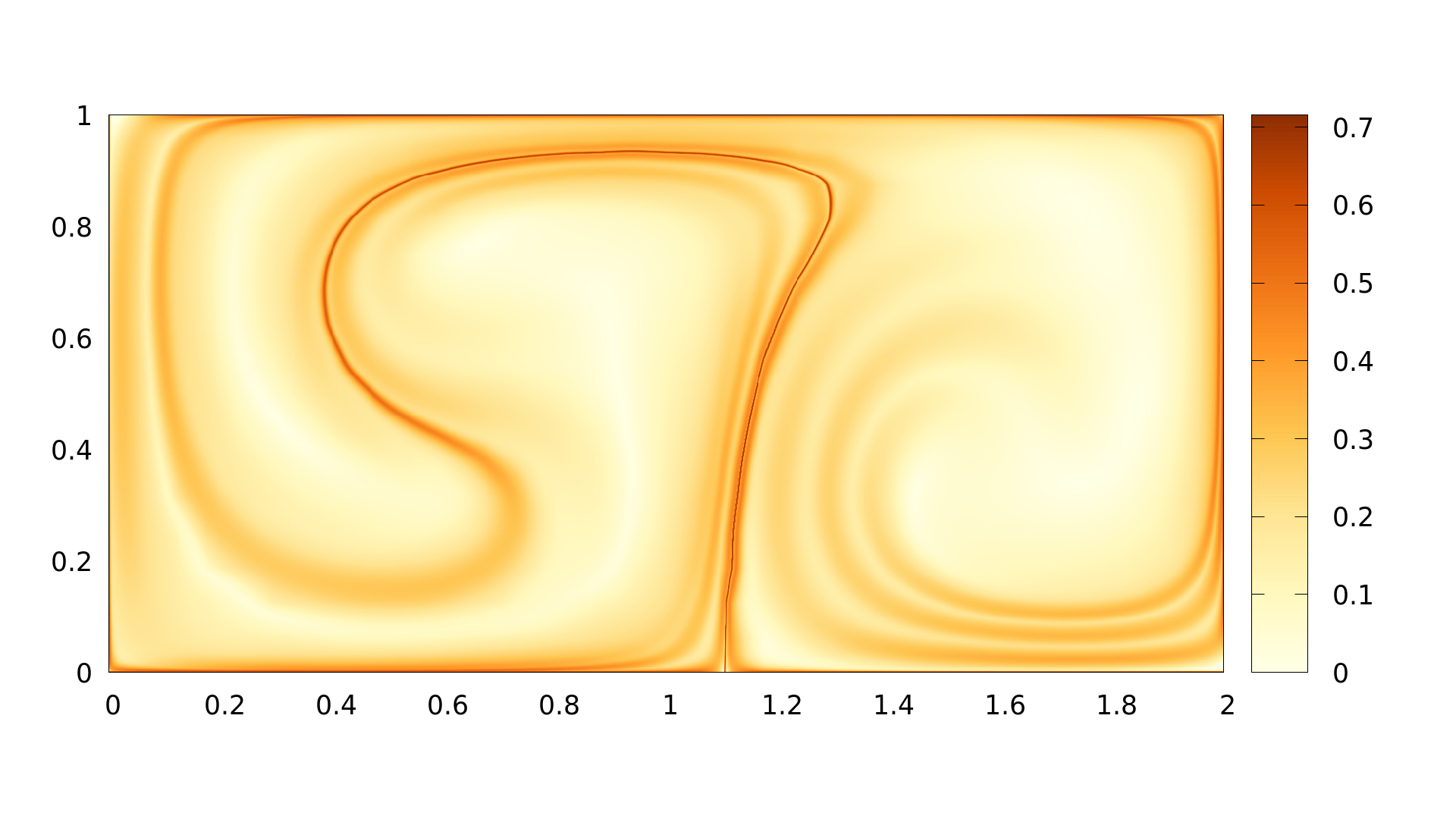}
  \label{fig:dg_h_ftle_2_1}
  \end{subfigure}
    \begin{subfigure}[][][t]{0.22\textwidth}
    \includegraphics[width= \textwidth]{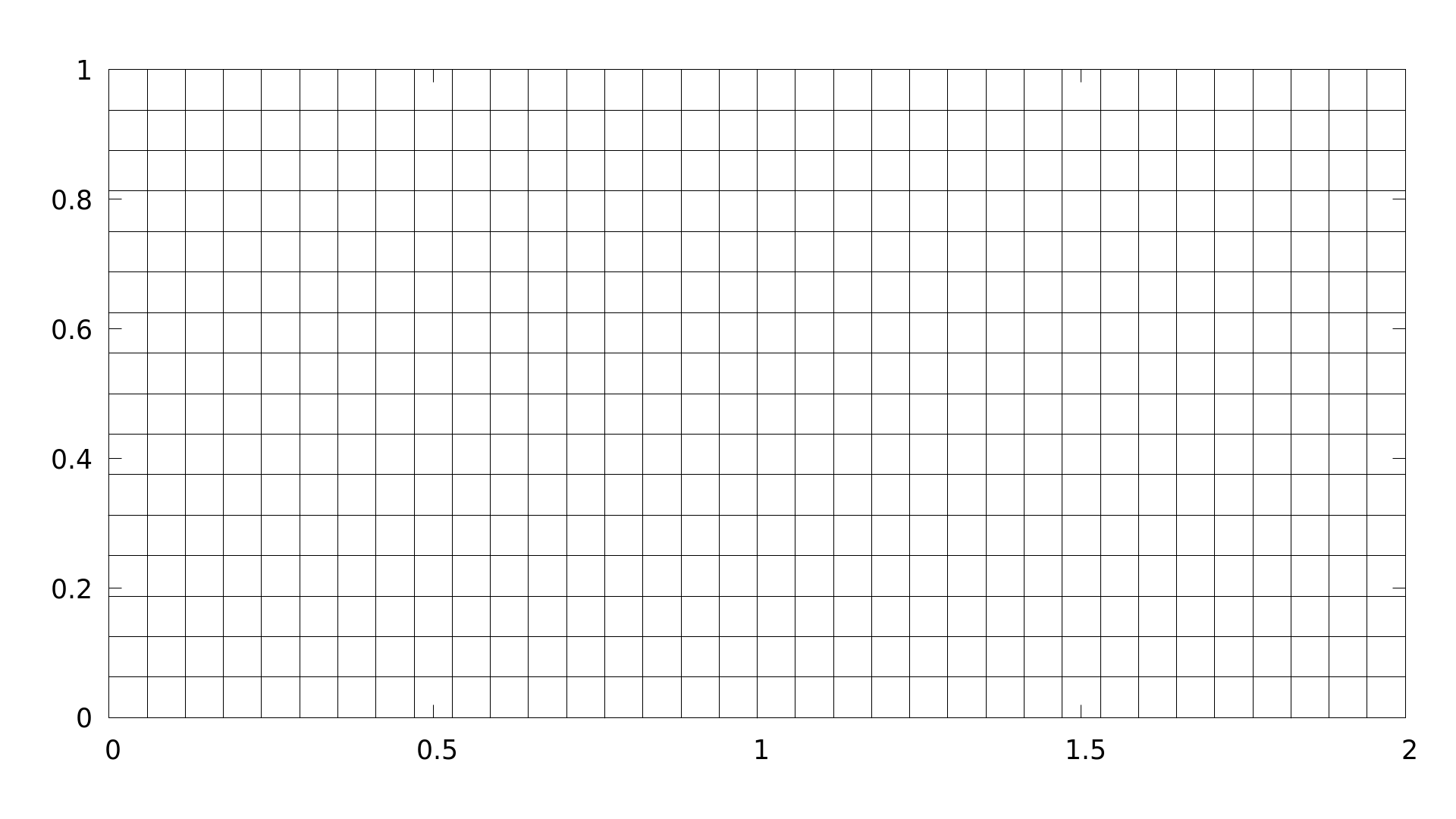}
  \label{fig:dg_h_grid_2_2}
  \end{subfigure}
 \hspace{-0.38cm}
 \vspace{-0.8cm}
  \begin{subfigure}[][][t]{0.26\textwidth}
    \includegraphics[width= \textwidth]{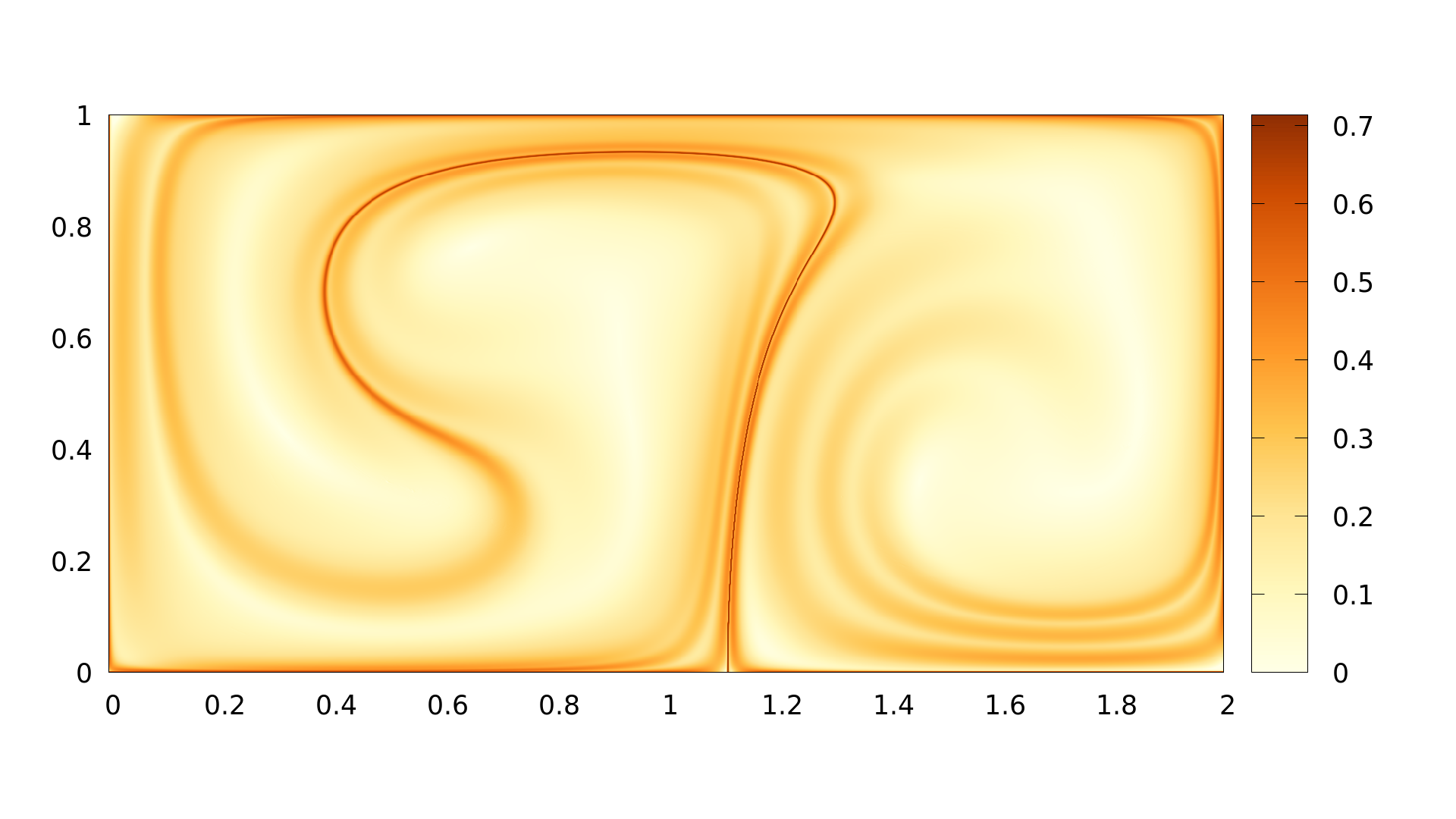}
  \label{fig:dg_h_ftle_2_2}
  \end{subfigure}
      \begin{subfigure}[][][t]{0.22\textwidth}
    \includegraphics[width= \textwidth]{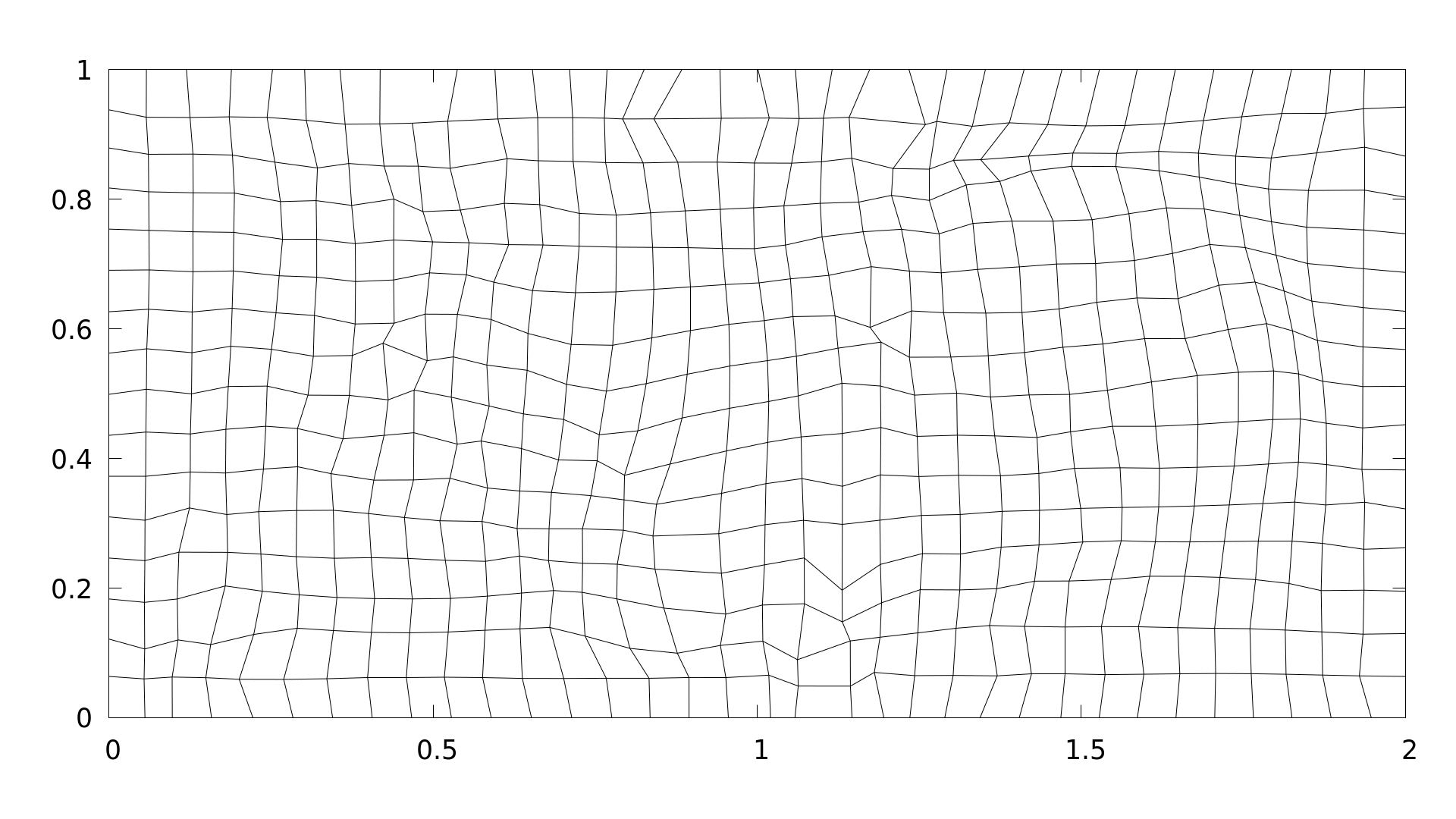}
  \label{fig:dg_h_grid_2_3}
  \end{subfigure}
 \hspace{-0.38cm}
 \vspace{-0.8cm}
  \begin{subfigure}[][][t]{0.26\textwidth}
    \includegraphics[width= \textwidth]{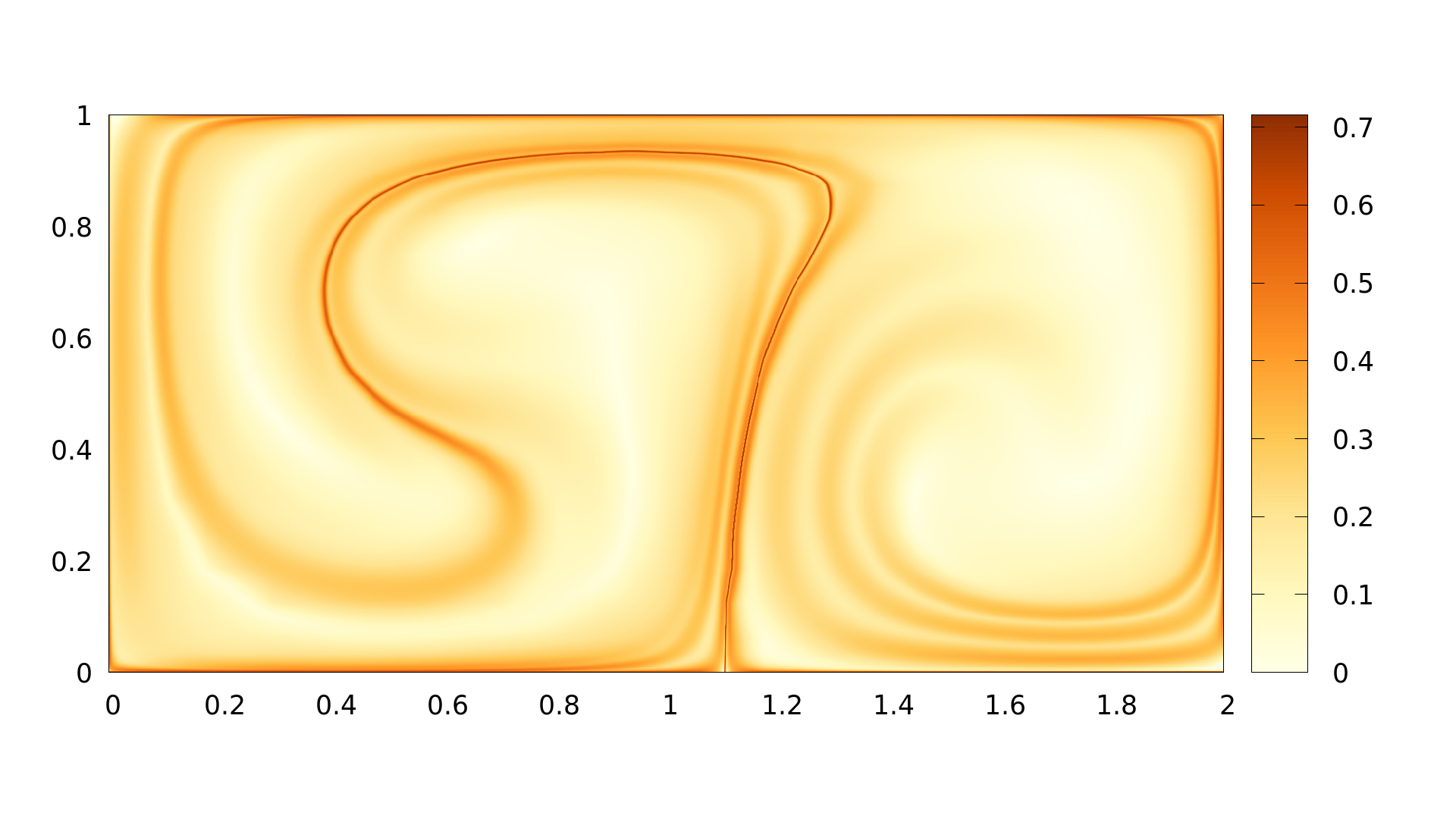}
  \label{fig:dg_h_ftle_2_3}
  \end{subfigure}
      \begin{subfigure}[][][t]{0.22\textwidth}
    \includegraphics[width= \textwidth]{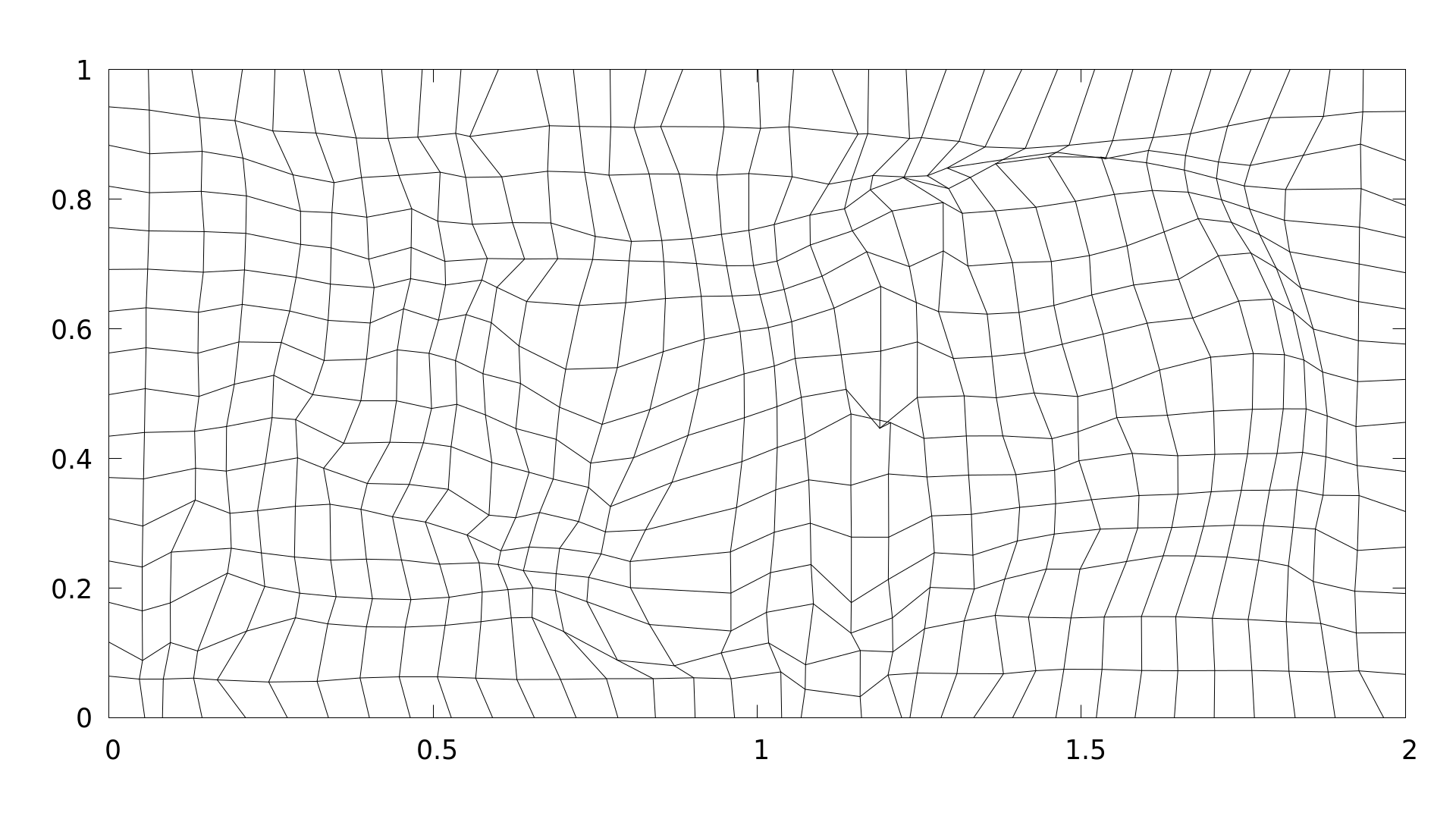}
  \label{fig:dg_h_grid_2_4}
  \end{subfigure}
 \hspace{-0.38cm}
  \vspace{-0.5cm}
  \begin{subfigure}[][][t]{0.26\textwidth}
    \includegraphics[width= \textwidth]{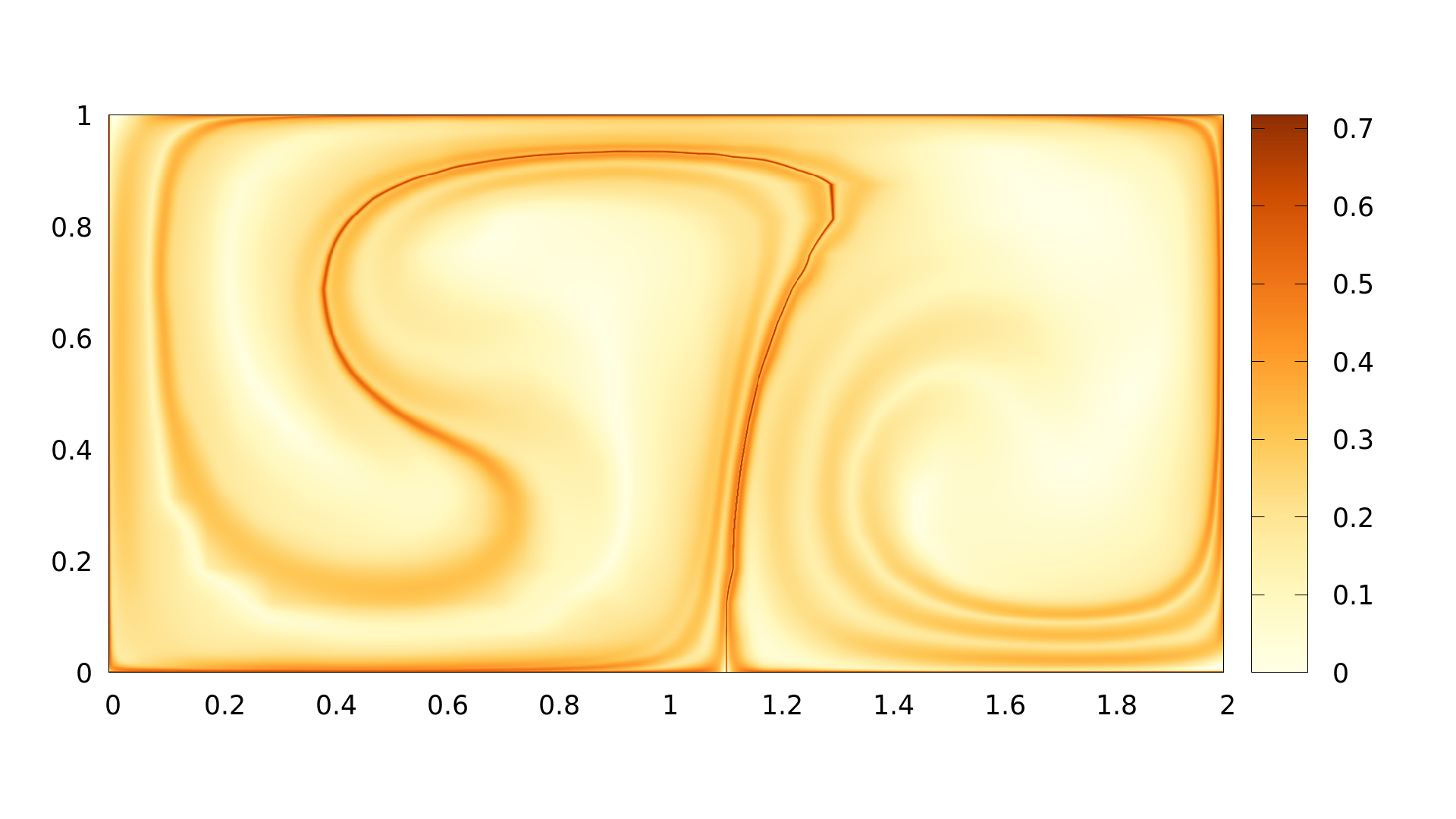}
  \label{fig:dg_h_ftle_2_4}
  \end{subfigure}
    \caption{
    Double Gyre ensemble with $\tau = 11$.
    Left column from top to bottom:
    Optimal domain displacement onto median flow $\phi_m$ with $m=25$ visualized as grid deformation:
    $\vp_{m,1}$,  $\vp_{m,13}$, $\vp_{m,m}$, $\vp_{m,38}$, $\vp_{m,50}$ \\
    Right column from top to bottom:
    Optimal displaced FTLE ensemble member onto median flow 25:
    $ \overline{\mbox{FTLE}}_1$, $ \overline{\mbox{FTLE}}_{13}$, $ \overline{\mbox{FTLE}}_{25}$, $ \overline{\mbox{FTLE}}_{38}$, $ \overline{\mbox{FTLE}}_{50}$
  \label{fig:dg_h_2}
}

 \end{figure}

 To give further insight of the optimal domain displacement,
 we want to give an example of the Double Gyre optimal domain displacements.
 Figure \ref{fig:dg_h_2} shows the difference of the domain alignments along the ensembles. 
 Note that ensembles closer to the median tend to have smaller domain displacements too.
 The resulting domain displaced FTLE fields are similar to the FTLE field of the selected median flow.

\pagebreak

 \subsection{Red Sea}
\begin{figure}[h]
  \centering
  \begin{subfigure}{0.24\textwidth}
   \begin{tikzpicture}[spy using outlines={rectangle,black,magnification=2,size=1.3cm, connect spies}]
      \node[anchor=south west,inner sep=0] (image) at (0,0) {\includegraphics[width= \textwidth,trim={11cm 0 11cm 0},clip]{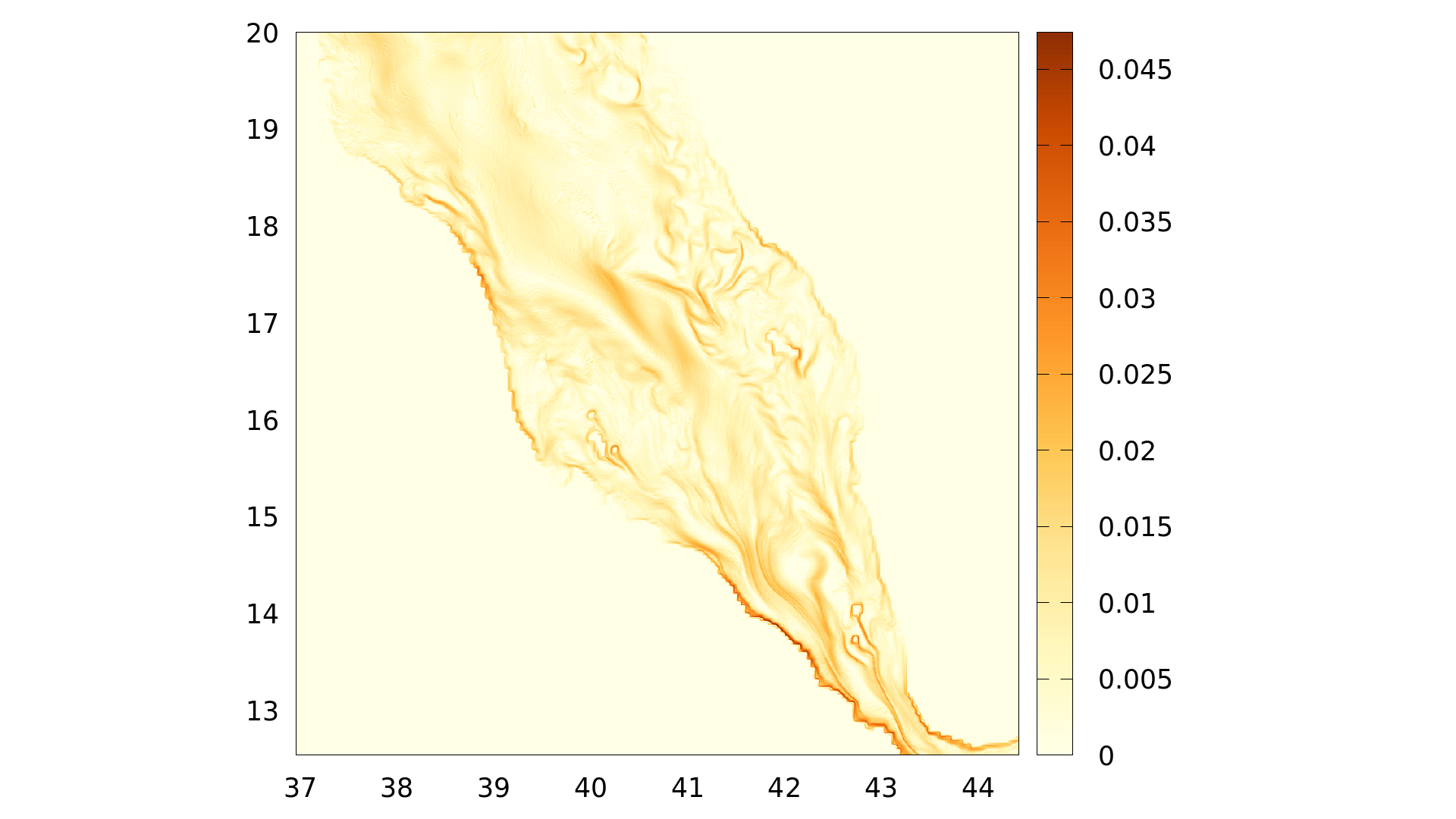}};%
        \begin{scope}[node distance=-1.8mm and -1.2mm, x={(image.south east)},y={(image.north west)}]%
        \coordinate (A) at (0.25,0.85);
        \coordinate (magA) at (0.21, 0.26);
        \spy on (A) in node at (magA);
        
        \coordinate (B) at (0.63,0.2);
        \coordinate (magB) at (0.65, 0.78);
        \spy on (B) in node at (magB);
      \end{scope}%
    \end{tikzpicture}
  \label{fig:goa_ftle_0}
  \end{subfigure}
      \begin{subfigure}{0.24\textwidth}
   \begin{tikzpicture}[spy using outlines={rectangle,black,magnification=2,size=1.3cm, connect spies}]
      \node[anchor=south west,inner sep=0] (image) at (0,0) {\includegraphics[width= \textwidth,trim={11cm 0 11cm 0},clip]{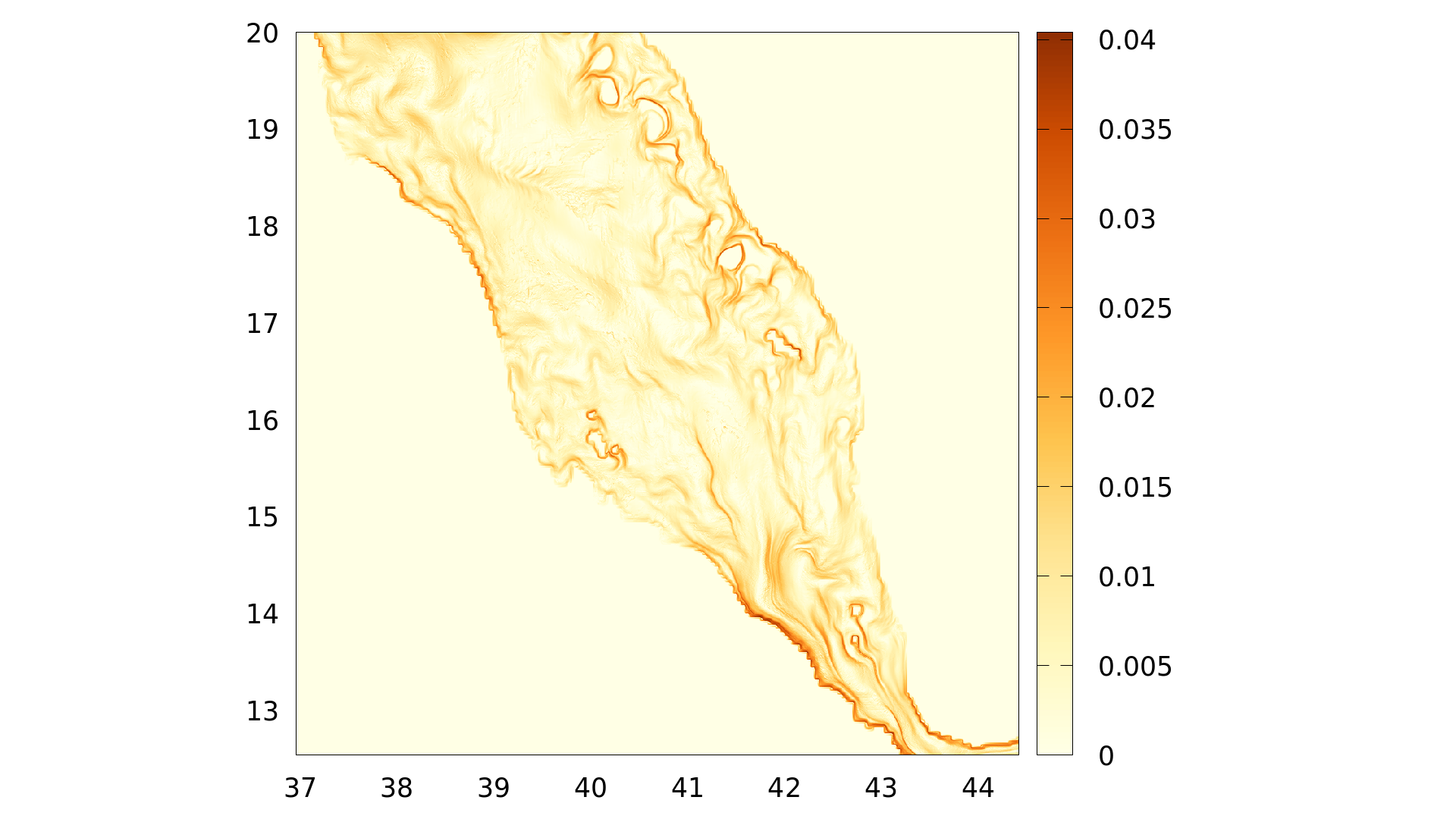}};%
        \begin{scope}[node distance=-1.8mm and -1.2mm, x={(image.south east)},y={(image.north west)}]%
        \coordinate (A) at (0.25,0.85);
        \coordinate (magA) at (0.21, 0.26);
        \spy on (A) in node at (magA);
        
        \coordinate (B) at (0.63,0.2);
        \coordinate (magB) at (0.65, 0.78);
        \spy on (B) in node at (magB);
      \end{scope}%
    \end{tikzpicture}
  \label{fig:goa_ftle_1}
  \end{subfigure}

    \begin{subfigure}{0.24\textwidth}
   \begin{tikzpicture}[spy using outlines={rectangle,black,magnification=2,size=1.3cm, connect spies}]
      \node[anchor=south west,inner sep=0] (image) at (0,0) {\includegraphics[width= \textwidth,trim={11cm 0 11cm 0},clip]{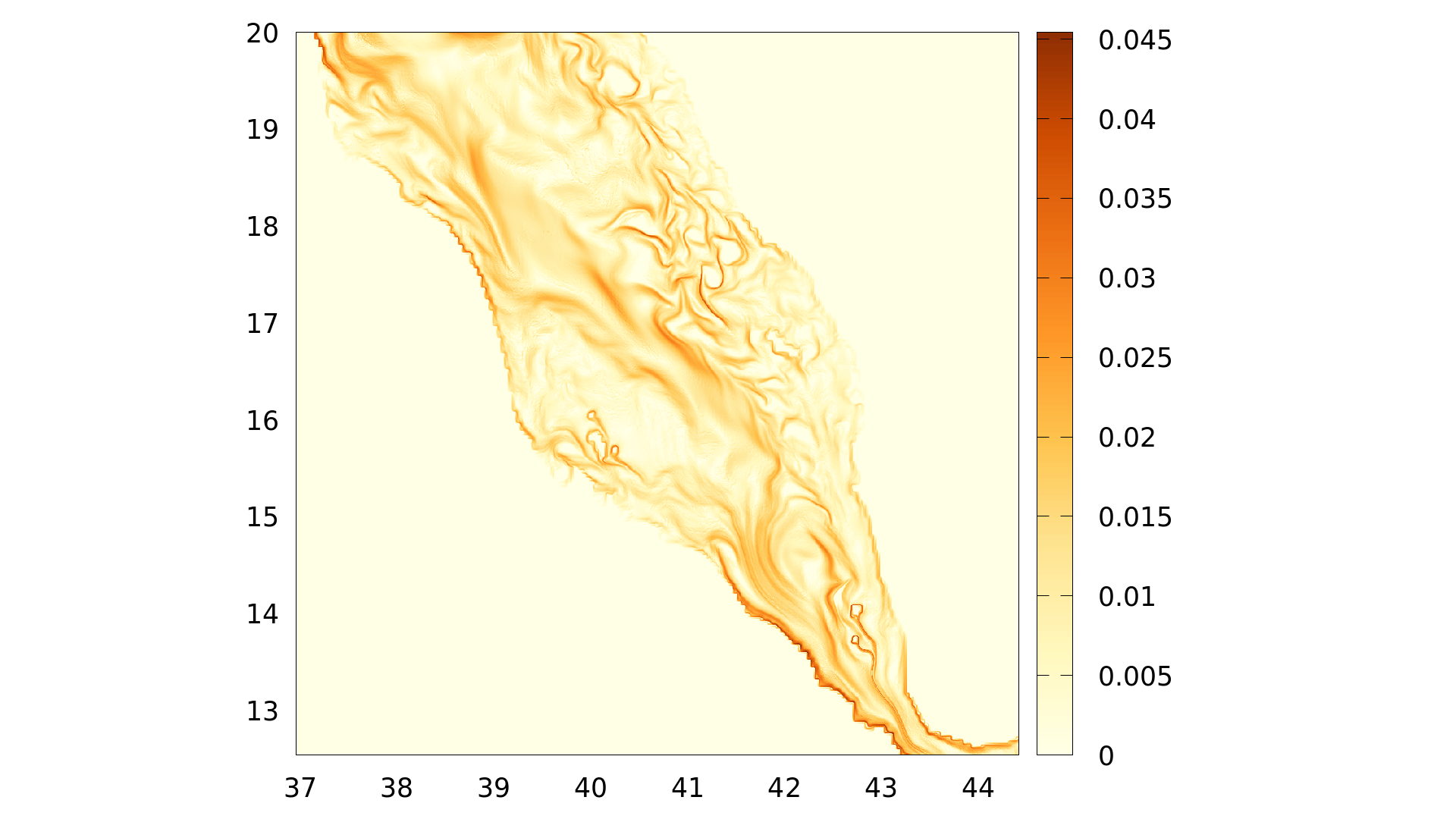}};%
        \begin{scope}[node distance=-1.8mm and -1.2mm, x={(image.south east)},y={(image.north west)}]%
        \coordinate (A) at (0.25,0.85);
        \coordinate (magA) at (0.21, 0.26);
        \spy on (A) in node at (magA);
        
        \coordinate (B) at (0.63,0.2);
        \coordinate (magB) at (0.65, 0.78);
        \spy on (B) in node at (magB);
      \end{scope}%
    \end{tikzpicture}
  \label{fig:goa_ftle_3}
  \end{subfigure}
      \begin{subfigure}{0.24\textwidth}
   \begin{tikzpicture}[spy using outlines={rectangle,black,magnification=2,size=1.3cm, connect spies}]
      \node[anchor=south west,inner sep=0] (image) at (0,0) {\includegraphics[width= \textwidth,trim={11cm 0 11cm 0},clip]{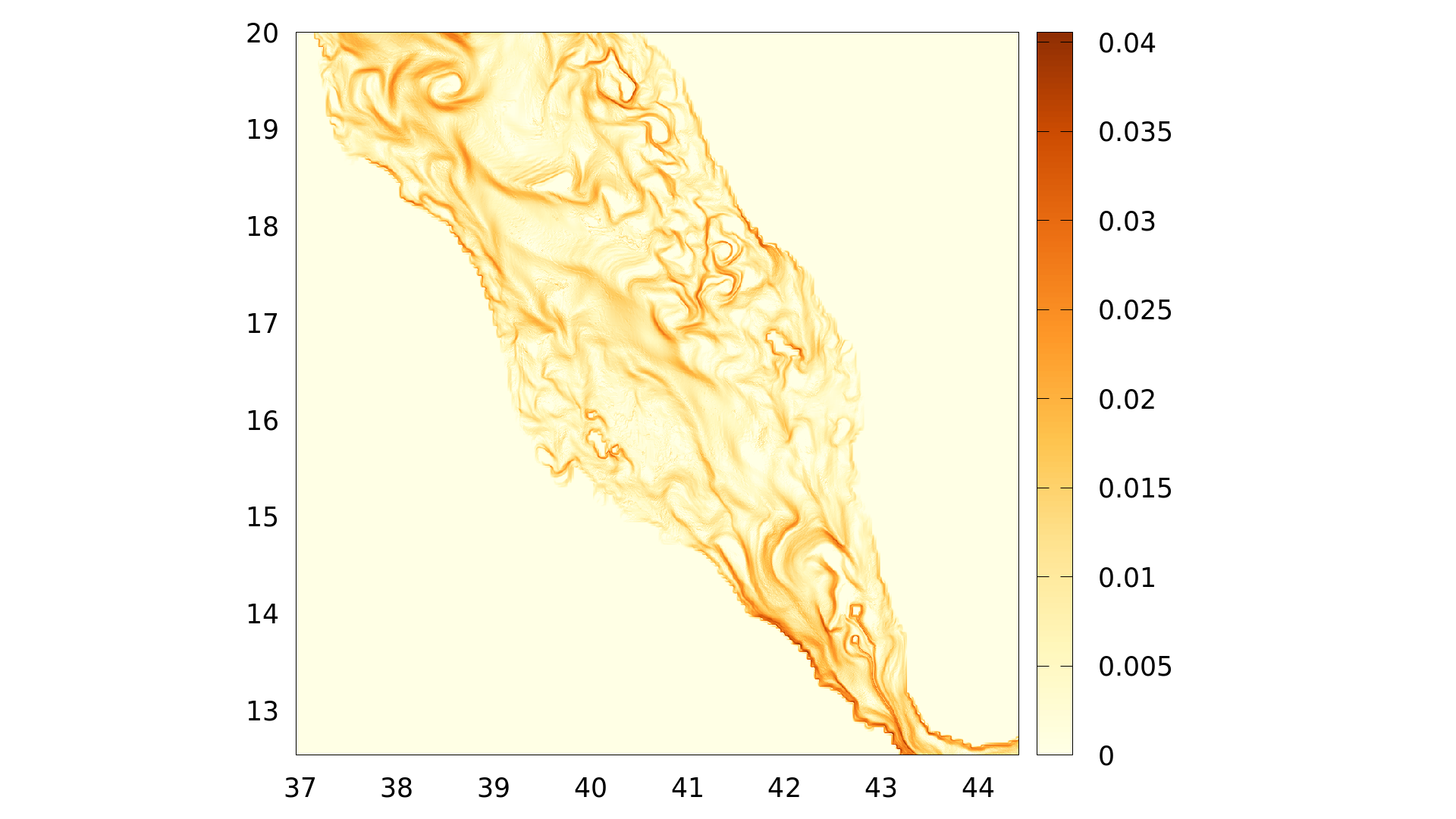}};%
        \begin{scope}[node distance=-1.8mm and -1.2mm, x={(image.south east)},y={(image.north west)}]%
        \coordinate (A) at (0.25,0.85);
        \coordinate (magA) at (0.21, 0.26);
        \spy on (A) in node at (magA);
        
        \coordinate (B) at (0.63,0.2);
        \coordinate (magB) at (0.65, 0.78);
        \spy on (B) in node at (magB);
      \end{scope}%
    \end{tikzpicture}
  \label{fig:goa_ftle_4}
  \end{subfigure}
    \caption{
    Red Sea ensemble with $\tau = 48$ which relates to an integration time of two days.
    Top row from left to right:
    FTLE of the Red Sea ensemble data set 1 and 2. \\
    Bottom row from left to right:
    FTLE of the Red Sea ensemble data set 13 and 16. \\
}
\label{fig:red_sea_ftle}
 \end{figure}

We apply our method to all 
50 members of the Red Sea Data set from the SCIVis-Contest 2020 \cite{SciVisContest2020}.
In particular, we take a closer look at the south of the Red Sea and the Bab-el-Mandeb strait.
Figure \ref{fig:red_sea_ftle} shows the FTLE fields of the 4 ensemble members, revealing that the ensembles are quite divers and different in some regions. 
The red sea has a relative constant inflow through the Bab-el-Mandeb strait, the differences are low in this region along the ensemble members. 
This can also be observed in the magnified area of the Bab-el-Mandeb strait.
 This changes drastically for the northern part where ensembles are quite different. 
Those differences can especially be observed in the magnified northern region of the Red Sea. 
 Some showing low to no seperation at all \ref{fig:red_sea_ftle}(top left and top right), while other show higher seperation (bottom left) or even eddies (bottom right).
 We consider this as an extreme case for our approach in trying to find alignments of the ensemble members. 

 \begin{figure}[h]
  \centering
  \begin{subfigure}{0.24\textwidth}
   \begin{tikzpicture}[spy using outlines={rectangle,black,magnification=2,size=1.3cm, connect spies}]
      \node[anchor=south west,inner sep=0] (image) at (0,0) {\includegraphics[width= \textwidth,trim={11cm 0 11cm 0},clip]{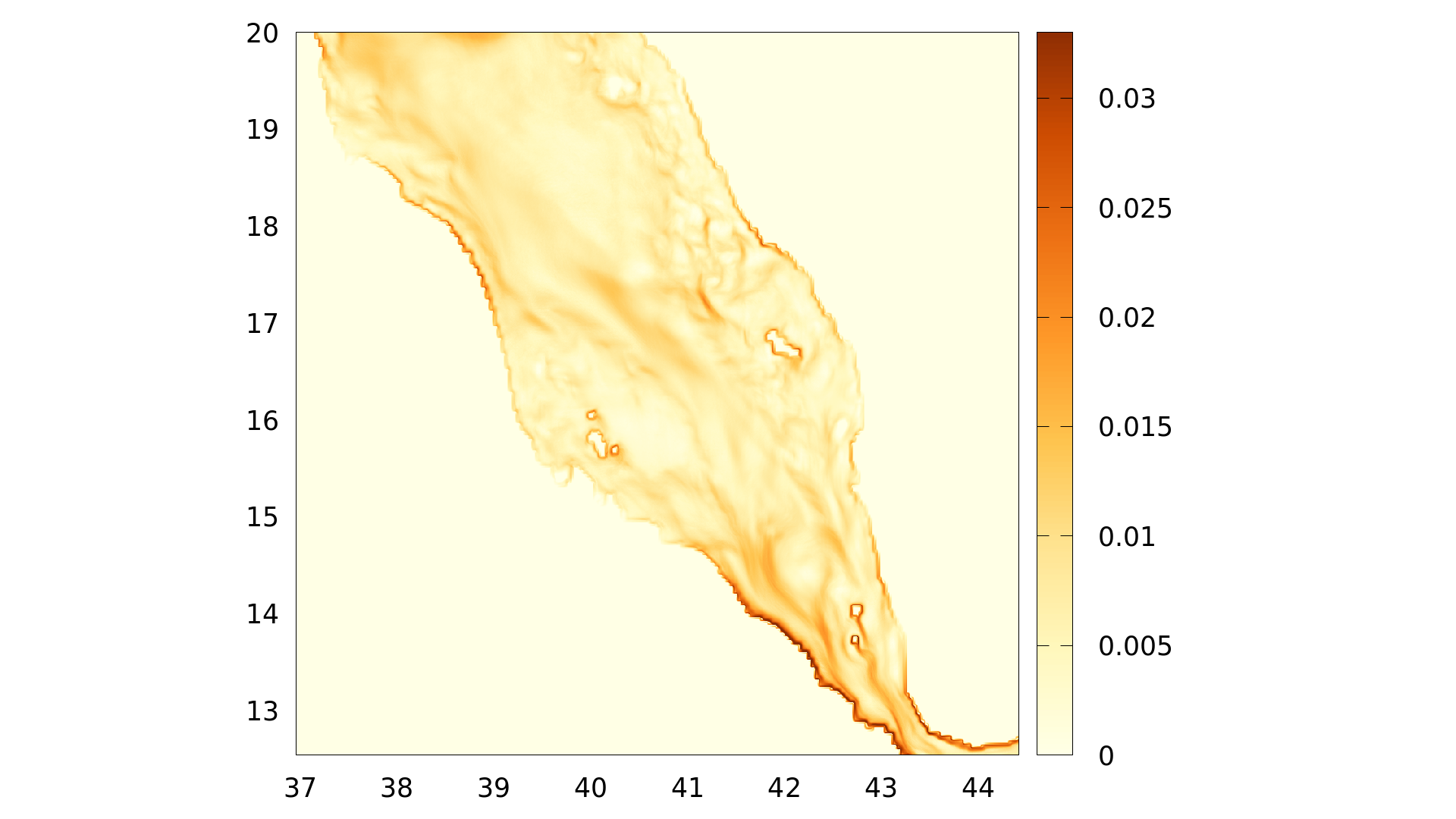}};%
        \begin{scope}[node distance=-1.8mm and -1.2mm, x={(image.south east)},y={(image.north west)}]%
        \coordinate (A) at (0.25,0.85);
        \coordinate (magA) at (0.21, 0.26);
        \spy on (A) in node at (magA);
        
        \coordinate (B) at (0.63,0.2);
        \coordinate (magB) at (0.65, 0.78);
        \spy on (B) in node at (magB);
      \end{scope}%
    \end{tikzpicture}
  \label{fig:goa_d-ftle}
  \end{subfigure}
      \begin{subfigure}{0.24\textwidth}
   \begin{tikzpicture}[spy using outlines={rectangle,black,magnification=2,size=1.3cm, connect spies}]
      \node[anchor=south west,inner sep=0] (image) at (0,0) {\includegraphics[width= \textwidth,trim={11cm 0 11cm 0},clip]{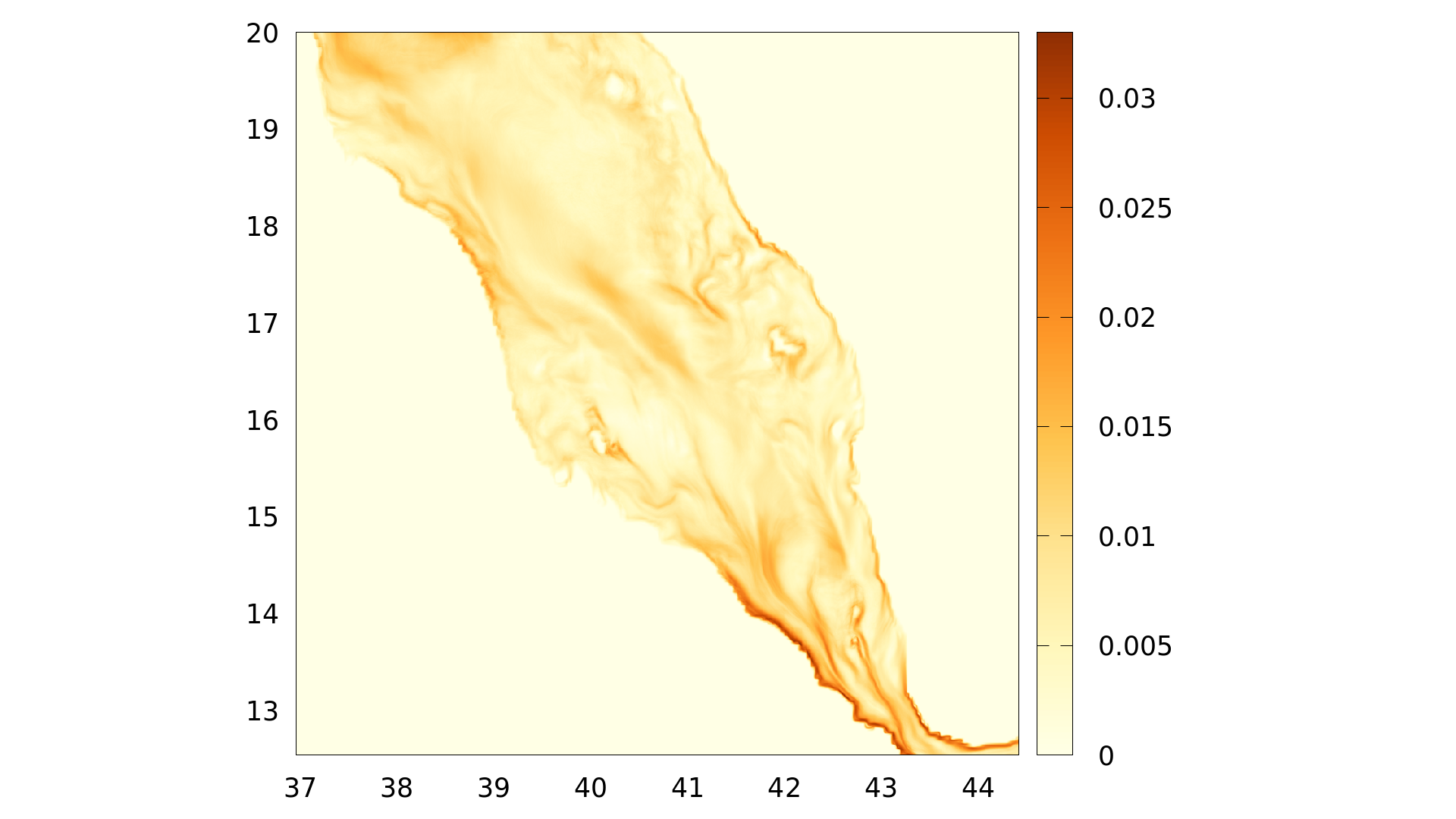}};%
        \begin{scope}[node distance=-1.8mm and -1.2mm, x={(image.south east)},y={(image.north west)}]%
        \coordinate (A) at (0.25,0.85);
        \coordinate (magA) at (0.21, 0.26);
        \spy on (A) in node at (magA);
        
        \coordinate (B) at (0.63,0.2);
        \coordinate (magB) at (0.65, 0.78);
        \spy on (B) in node at (magB);
      \end{scope}%
    \end{tikzpicture}
  \label{fig:goa_aligned-d-ftle}
  \end{subfigure}
      \begin{subfigure}{0.24\textwidth}
   \begin{tikzpicture}[spy using outlines={rectangle,black,magnification=2,size=1.3cm, connect spies}]
      \node[anchor=south west,inner sep=0] (image) at (0,0) {\includegraphics[width= \textwidth,trim={11cm 0 11cm 0},clip]{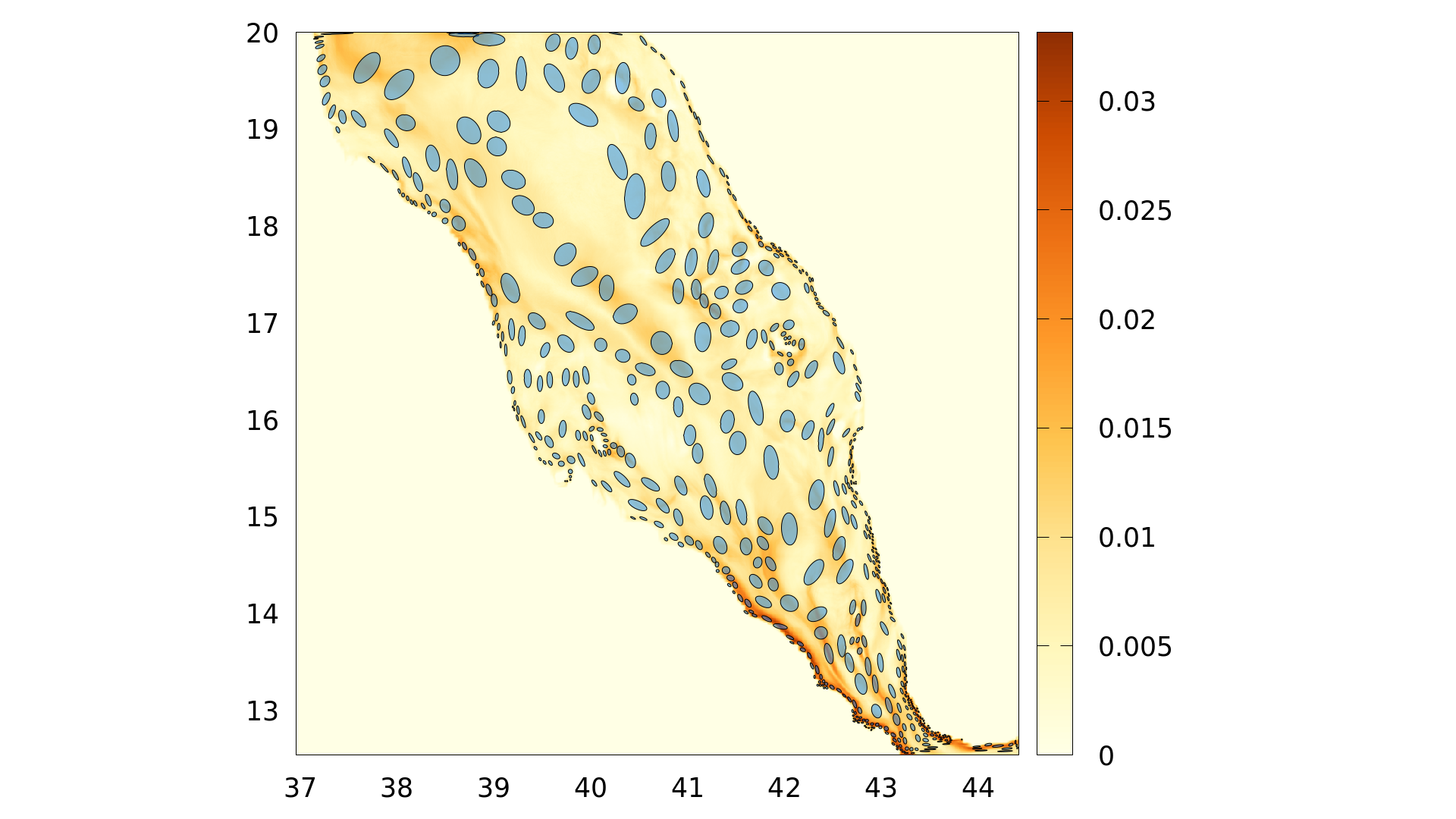}};%
        \begin{scope}[node distance=-1.8mm and -1.2mm, x={(image.south east)},y={(image.north west)}]%
        \coordinate (A) at (0.25,0.85);
        \coordinate (magA) at (0.21, 0.26);
        \spy on (A) in node at (magA);
        
        \coordinate (B) at (0.63,0.2);
        \coordinate (magB) at (0.65, 0.78);
        \spy on (B) in node at (magB);
      \end{scope}%
    \end{tikzpicture}
  \label{fig:goa_dd-d-ftle_35}
  \end{subfigure}
      \begin{subfigure}{0.24\textwidth}
   \begin{tikzpicture}[spy using outlines={rectangle,black,magnification=2,size=1.3cm, connect spies}]
      \node[anchor=south west,inner sep=0] (image) at (0,0) {\includegraphics[width= \textwidth,trim={11cm 0 11cm 0},clip]{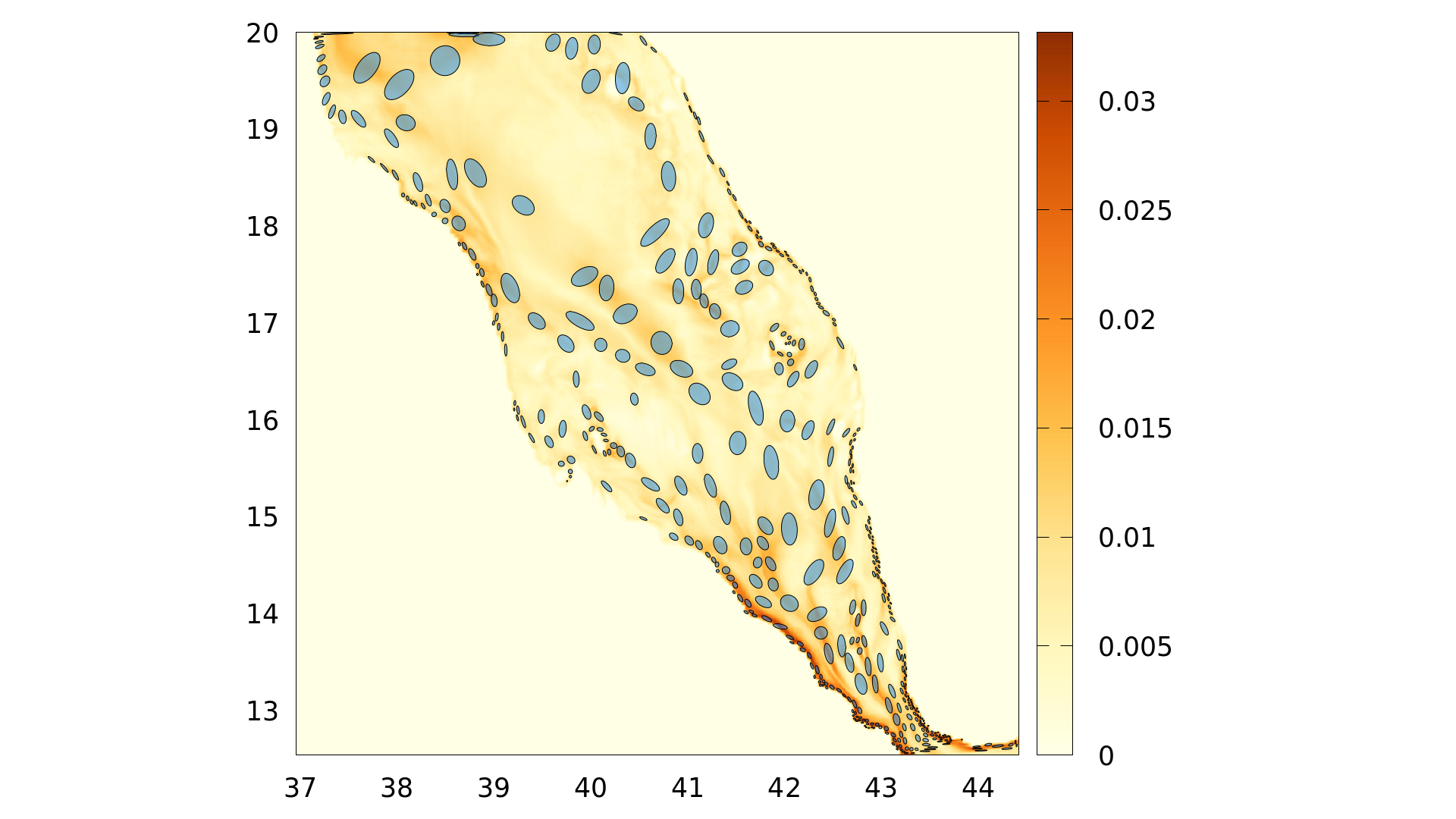}};%
        \begin{scope}[node distance=-1.8mm and -1.2mm, x={(image.south east)},y={(image.north west)}]%
        \coordinate (A) at (0.25,0.85);
        \coordinate (magA) at (0.21, 0.26);
        \spy on (A) in node at (magA);
        
        \coordinate (B) at (0.63,0.2);
        \coordinate (magB) at (0.65, 0.78);
        \spy on (B) in node at (magB);
      \end{scope}%
    \end{tikzpicture}
  \label{fig:goa_dd-d-ftle_25}
  \end{subfigure}
  \vspace{-0.5cm}
     \caption{
	Red Sea data set with 50 ensemble members; (top left) D-FTLE revealing different kind of ridges;
	(top right) D-FTLE of the domain displaced ensemble members gives slightly sharper ridges than (top left) especially in the Bab-el-Mandeb strait;
	(bottom left) D-FTLE of the domain displaced ensemble members with ellipses of the $\mC$ field shown for FTLE values obove 18\% and 25\% (bottom right) of the value range.
      \label{fig:red_sea_dd}
      }
  \end{figure}
  
  Figure \ref{fig:red_sea_dd} (top left) shows D-FTLE for the ensemble members, revealing a blurry image for most of the domain. 
  This is predictable due to the large differences among the ensembles, especially in the northern part of the Red Sea.
  Nevertheless there are slightly better results for the region of the Bab-el-Mandeb strait.
  \ref{fig:red_sea_dd} (top right) shows the optimal domain displaced D-FTLE. 
  It can be observed that there is no improvement for the northern part of the Red Sea. 
  As close to zero alignment is possible in this region, our method gives equal results to the standard method.
  Nevertheless it can be observed that in the region of the Bab-el-Mandeb strait, alignment is possible resulting in slightly sharper FTLE ridges.
  The bottom Images of figure \ref{fig:red_sea_dd} shows the uncertainty of the underlying aligned D-FTLE fields. 
  Those ellipses underline the interpretation of higher uncertainty in the northern part while 
  ellipses aligning with the ridges in the Bab-el-Mandeb strait describing more certainty of location than in seperation strength.
  The magnified region is also showing a FTLE ridge covered by ellipses aligning along it while those ellipses get brighter northwards, 
  showing a change of uncertainty in this direction.

 \subsection{Wind data over the Indian Sea}
\begin{figure}[H]
  \centering
    \vspace{-1.4cm}
  \begin{subfigure}{0.5\textwidth}
    \includegraphics[width= \textwidth]{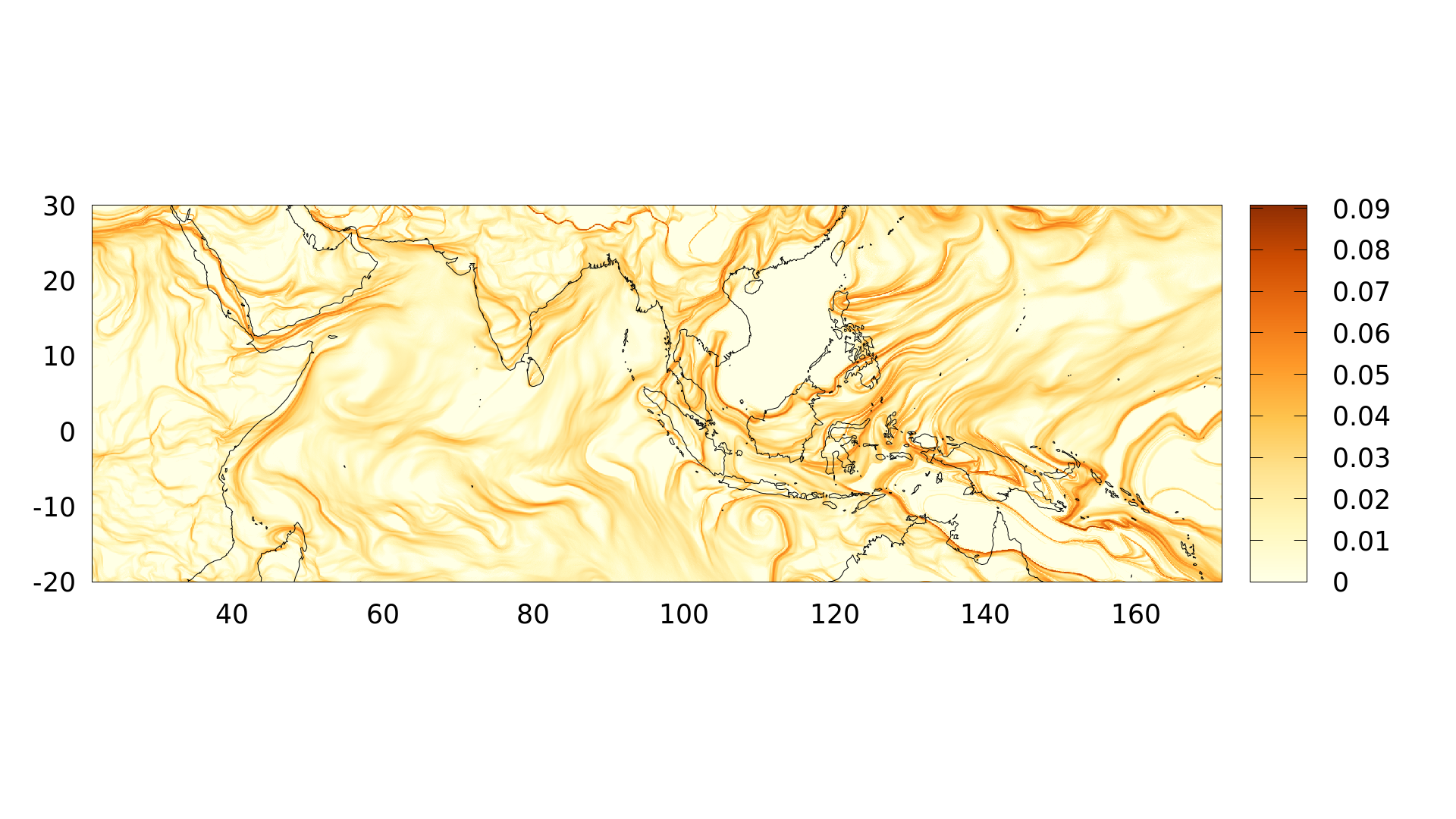}
  \label{fig:dg_2_dd_d-ftle}
  \end{subfigure}
  \hfill
  \vspace{-3.0cm}
      \begin{subfigure}{0.5\textwidth}
    \includegraphics[width= \textwidth]{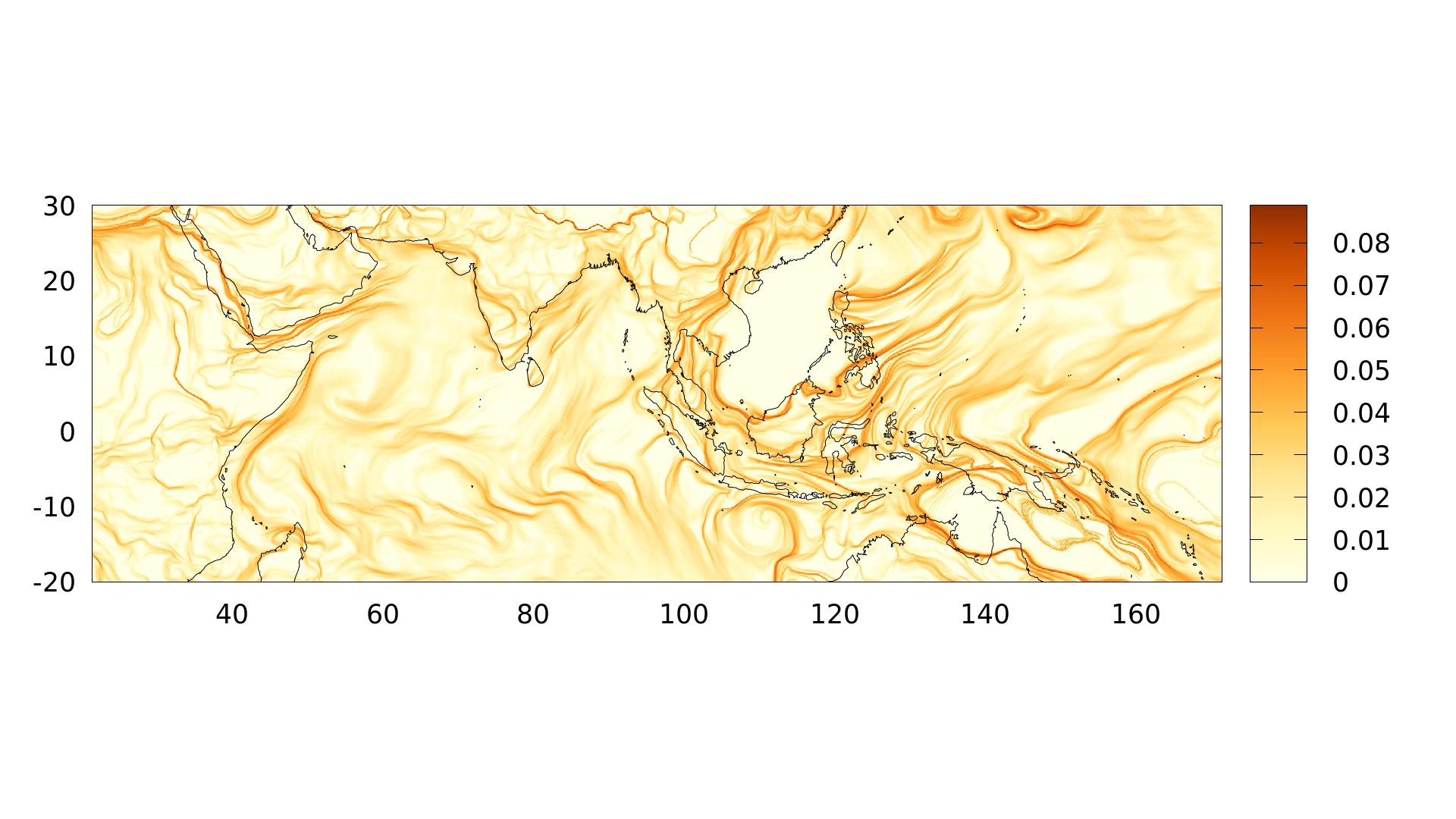}
  \label{fig:dg_2_variance_dd-dftle}
  \end{subfigure}
  \hfill
  \vspace{-2.cm}
     \caption{
	FTLE of ensemble member 1 and 2 of the GEFS data set.
      \label{fig:is_ftle}
      }
  \end{figure}

\begin{figure}[H]
  \centering
 \begin{tikzpicture}[spy using outlines={rectangle,black,magnification=4,size=3.5cm, connect spies}]
      \node[anchor=south west,inner sep=0] (image1) at (0,0) {\includegraphics[width=1.\linewidth, interpolate]{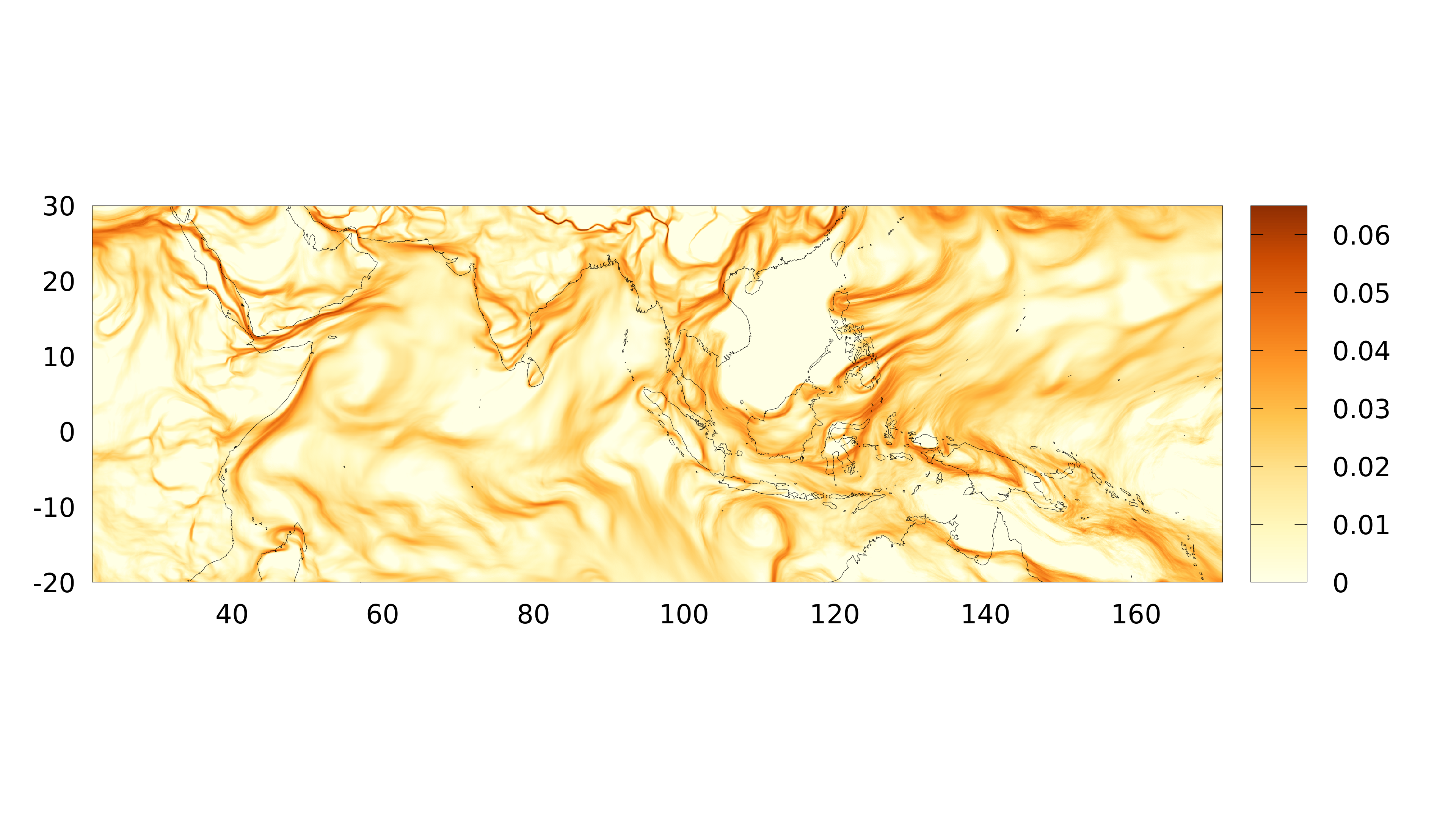}};%
        \begin{scope}[node distance=-1.8mm and -1.2mm, x={(image1.south east)},y={(image1.north west)}]%
        
        \coordinate (A) at (0.62,0.55);
        \coordinate (magA) at (0.26, -0.15);
        \spy on (A) in node at (magA);

      \end{scope}%
      \node[anchor=south west,inner sep=0] (image2) at (0,-6.5) {\includegraphics[width=1.\linewidth, interpolate]{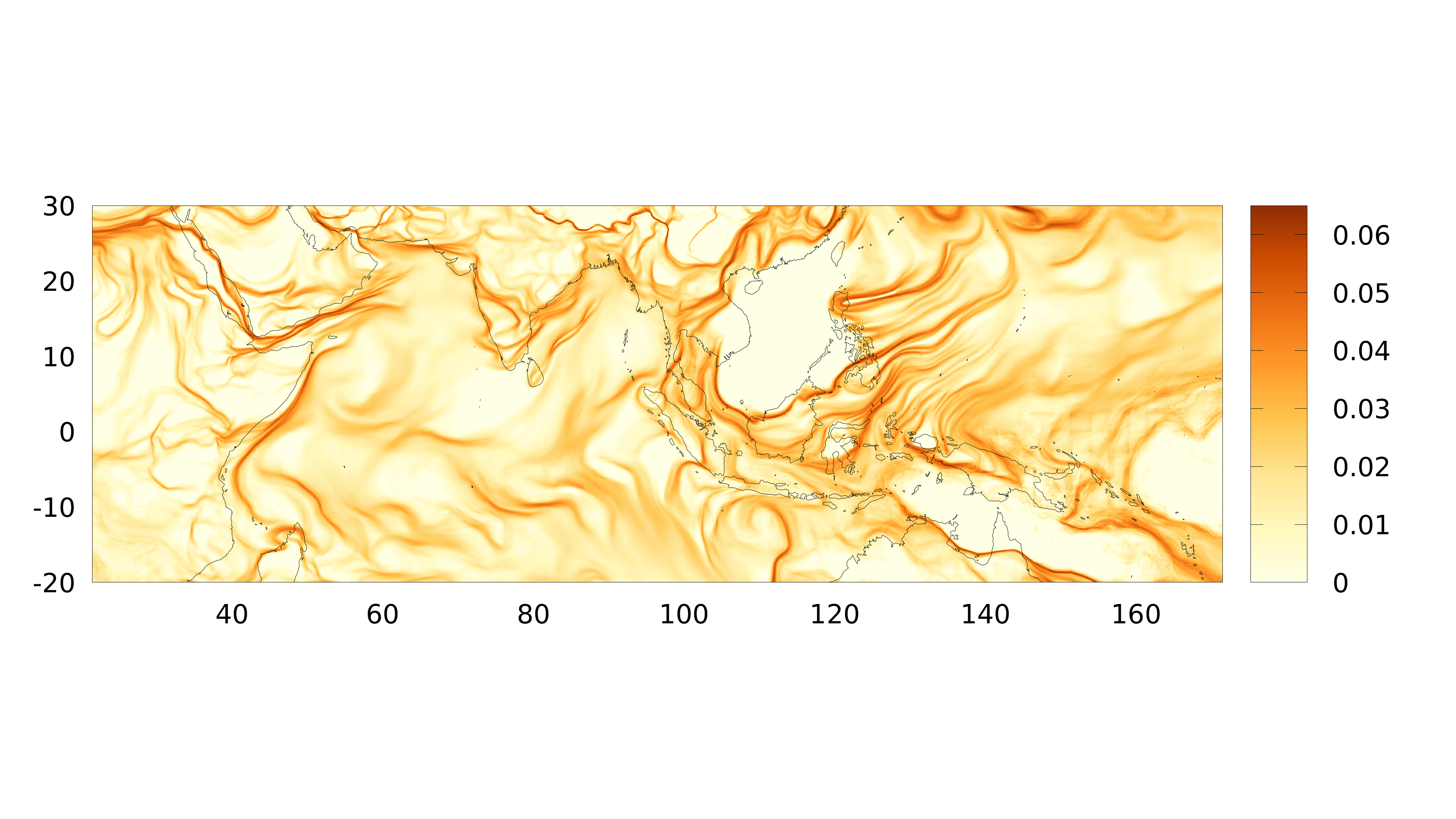}};%
        \begin{scope}[node distance=-1.8mm and -1.2mm, x={(image2.south east)},y={(image2.north west)}]%
        
        \coordinate (B) at (0.62,-0.18);
        \coordinate (magB) at (0.7, -2.5); 
        \spy on (B) in node at (magB);
      \end{scope}%
	\node[anchor=south west,inner sep=0] (image3) at (0,-9.2) {\includegraphics[width=1.\linewidth, interpolate]{Indian_Sea_domain-displaced_d-ftle_4k.png}};%
    \end{tikzpicture}%
  \vspace{-1.cm}
    \caption{
    GEFS ensemble with $\tau = 48$ which relates to an integration time of two days.
    From top to bottom:
    D-FTLE, 
    optimal domain displaced D-FTLE without ellipses, ellipses on FTLE values above 25\% of the value range.
}
 \label{fig:is}
 \end{figure}

We apply our approach to all 
21 members of the Global Ensemble Forecast System (GEFS) of wind flow. 
Although the data set includes the whole earth, we are having a closer look at the 
wind flow close to the Indian Sea.
Figure \ref{fig:is_ftle} shows two FTLE fields of the (GEFS) ensemble set which are distinct while showing similar structures.
There is also a difference in maximum separation strength between these two ensembles recognizable by the range of color bars. 
Figure \ref{fig:is} (top) shows D-FTLE for the whole 21 ensemble members. 
Due to the lower number of ensembles and less variation than the data set of the Red Sea, D-FTLE still produces reasonable images. 
Nevertheless most of the ridges are blurry and tend to fade out rather quickly.
We magnify the region of the Philippines to take a closer look and compare our results \ref{fig:is}(middle).
Sharper ridges can be observed with our method not only in the magnified area but the whole domain. 
The domain displaced D-FTLE fields with ellipses (bottom) gives also insight about the amount and source of uncertainty for each ridge. 
As most of the ellipses align with the ridge direction, the source of uncertainty is dominated by uncertainty in ridge separation strength. 

  \subsection{Performance}
  The computation times for the examples used in this paper are shown in table \ref{tab:timings}.
  $n$ denotes the number of ensembles while $\psi$ Res. and $\vp$ Res. cover the spatial resolution of the flow and displace maps.
  Note, that timings for the flow map covers the time to compute for all ensembles.
  Same holds for $\vp$ which shows the total compute time for all displace maps.
  All images shown here were computed on a 10 times higher resolution. 
  As the computation of the optimal domain displacement is a one time preprocess which can be saved for later use, 
  one can create higher resolution optimal domain displaced D-FTLE fields at a later time.

  \begin{center}
    \vspace{-0.2cm}
  \begin{table}[H]
\begin{tabular}{|m{0.20\linewidth} m{0.02\linewidth} m{0.15\linewidth}  m{0.1\linewidth}  m{0.13\linewidth}  m{0.15\linewidth}  |}
 \hline
  Flow & $n$ & $\psi$ Res. & $\vp$ Res. & $\psi$ Time & $\vp$ Time\\
 \hline\hline
 Steady DG & 50 & $(75, 150)$ & $(17, 35)$ &   0.25 sec. & 2521 sec.\\
 \hline
  Unsteady DG & 50 & $(75, 150)$ & $(17, 35)$ &   0.25 sec. & 2524 sec.\\
 \hline
   Red Sea & 50 & $(112, 112)$ & $(41, 41)$ &   3.02 sec. & 15113 sec.\\
 \hline
  Indian Sea & 21 & $(100, 300)$ & $(21, 61)$ &   4.13 sec. & 7564 sec.\\
 \hline
\end{tabular}
 \caption{
	Runtime measurements for the different data sets that were used. 
	$\psi$ Time is the time for a single iteration towards the optimal domain displacement for a single flow map. 
	$\vp$ Time is the summed time for all computations in order to find the optimal domain displacement for all ensemble members. 
	This depends on the maximum iterations for the optimization which we set to 100. 
	All measurements were taken on a Ryzen 9 3950x 16-core processor.
	\label{tab:timings}
}
\end{table}
\end{center}

\section*{Acknowledgments} 
\vspace{-0.1cm}
The work was partially supported by DGF grant TH 692/17-1 
\vspace{-0.2cm}

\section{Discussion}
\label{sec:discussion}
\paragraph*{Alternative approaches.}
We propose to align flow maps.
Instead, one may consider standard methods for image alignment applied
to the FTLE fields of the ensemble members.
Unfortunately, image-processing approaches cannot be easily applied
here:
they are typically based on the extraction and the alignment of point
features in the image.
FTLE images, however, show curved features.
Therefore, the proposed flow map alignment gave more robust results
than image-based FTLE alignment or alignment of the flow map gradients.

\begin{figure}[H]
  \centering
  \begin{subfigure}[][][t]{0.243\textwidth}
    \includegraphics[width= \textwidth]{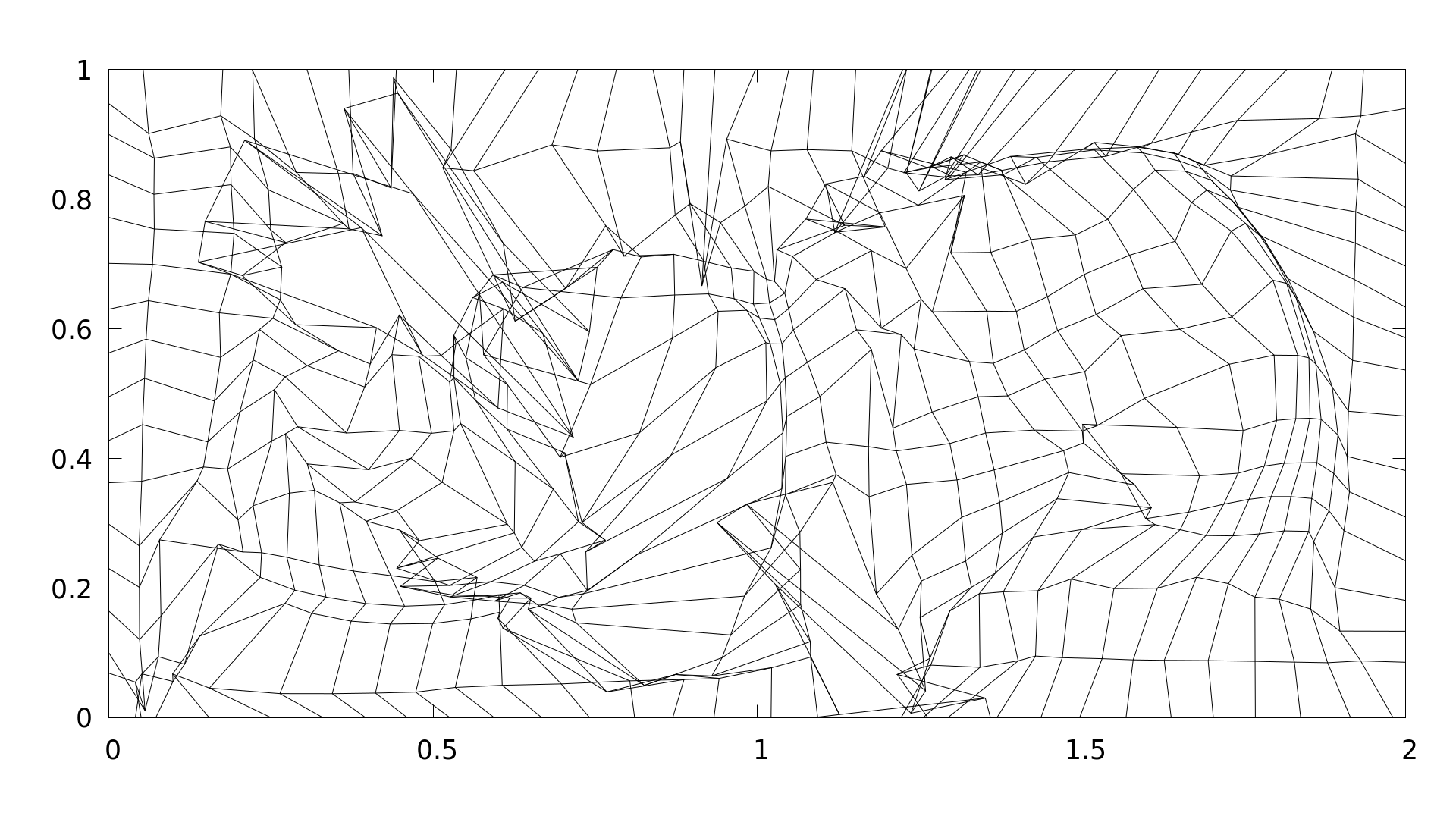}
  \label{fig:dg_rho_0_0}
  \end{subfigure}
 \hspace{-0.38cm}
 \vspace{-0.5cm}
  \begin{subfigure}[][][t]{0.243\textwidth}
    \includegraphics[width= \textwidth]{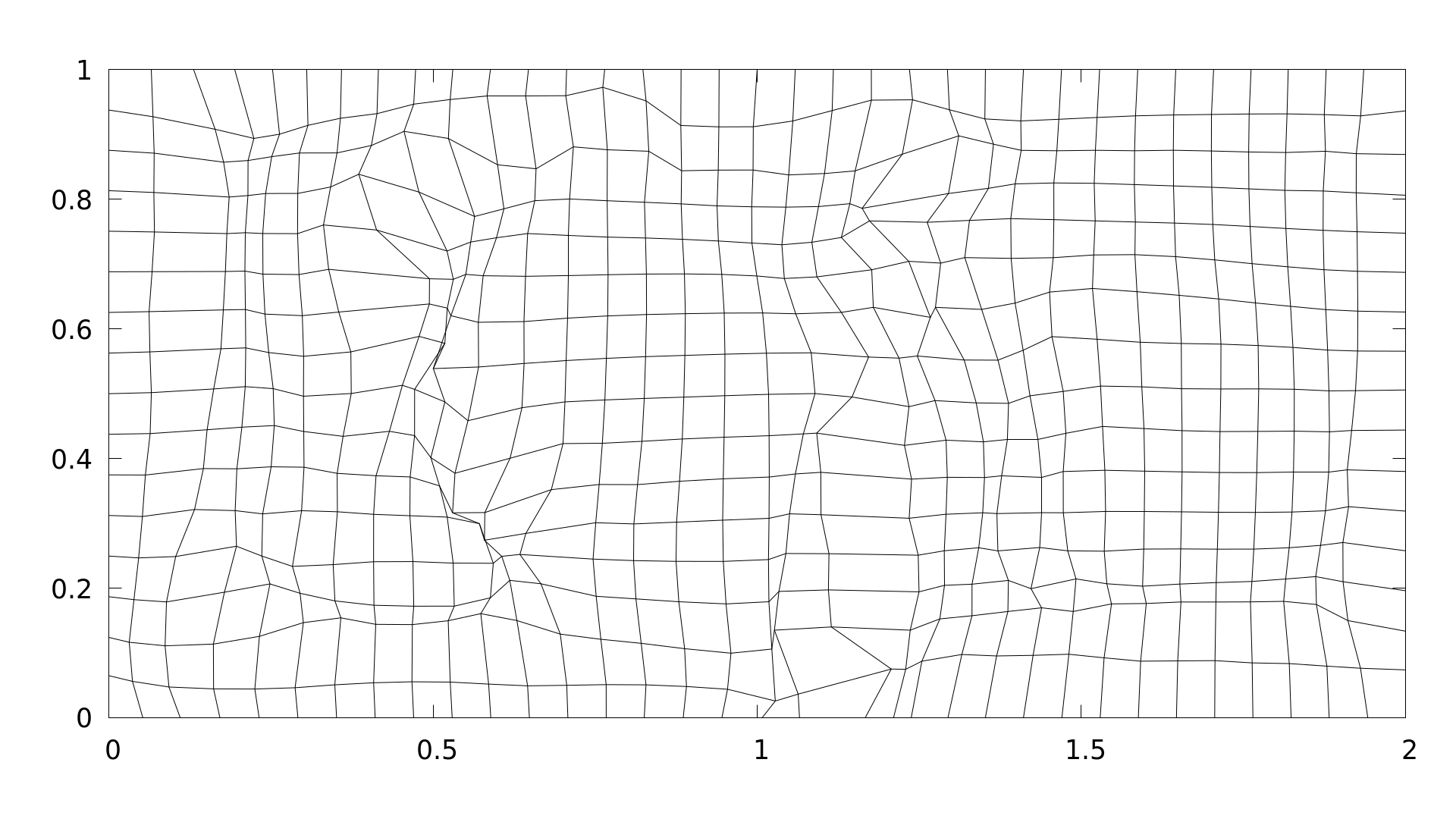}
  \label{fig:dg_rho_0_5}
  \end{subfigure}
  \begin{subfigure}[][][t]{0.243\textwidth}
    \includegraphics[width= \textwidth]{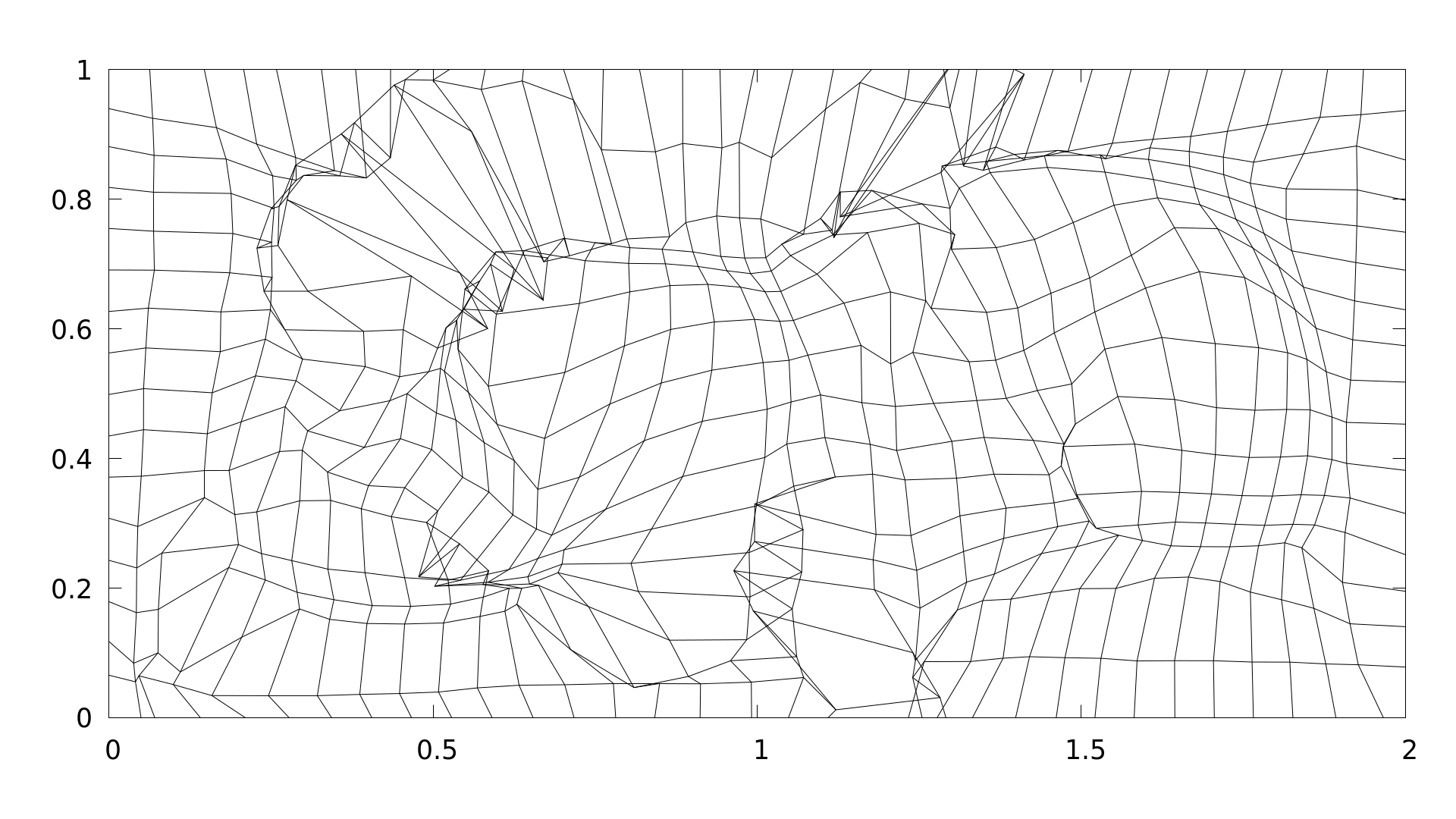}
  \label{fig:dg_rho_0_05}
  \end{subfigure}
 \hspace{-0.38cm}
 \vspace{-0.5cm}
  \begin{subfigure}[][][t]{0.243\textwidth}
    \includegraphics[width= \textwidth]{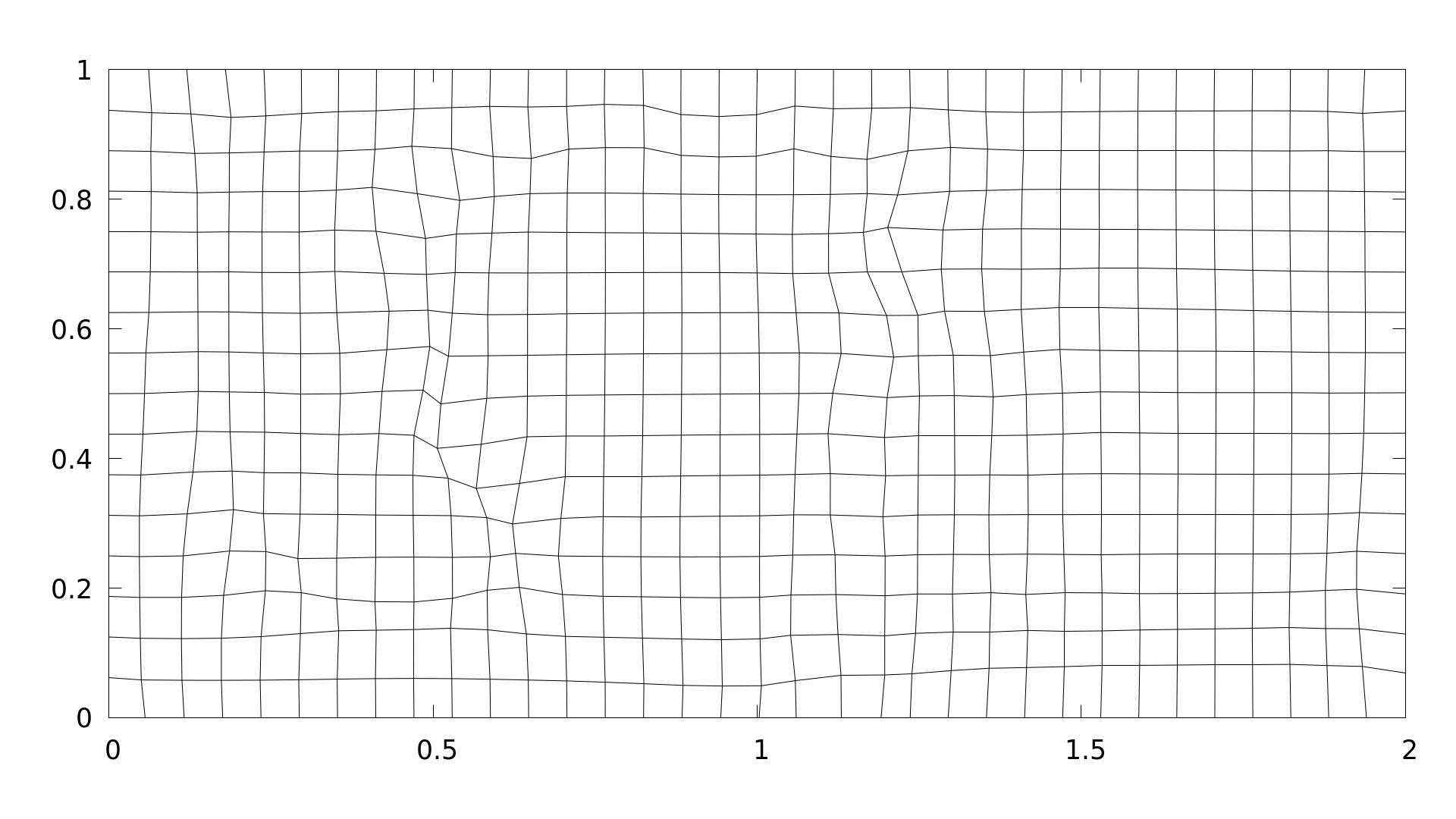}
  \label{fig:dg_rho_0_6}
  \end{subfigure}
    \begin{subfigure}[][][t]{0.243\textwidth}
    \includegraphics[width= \textwidth]{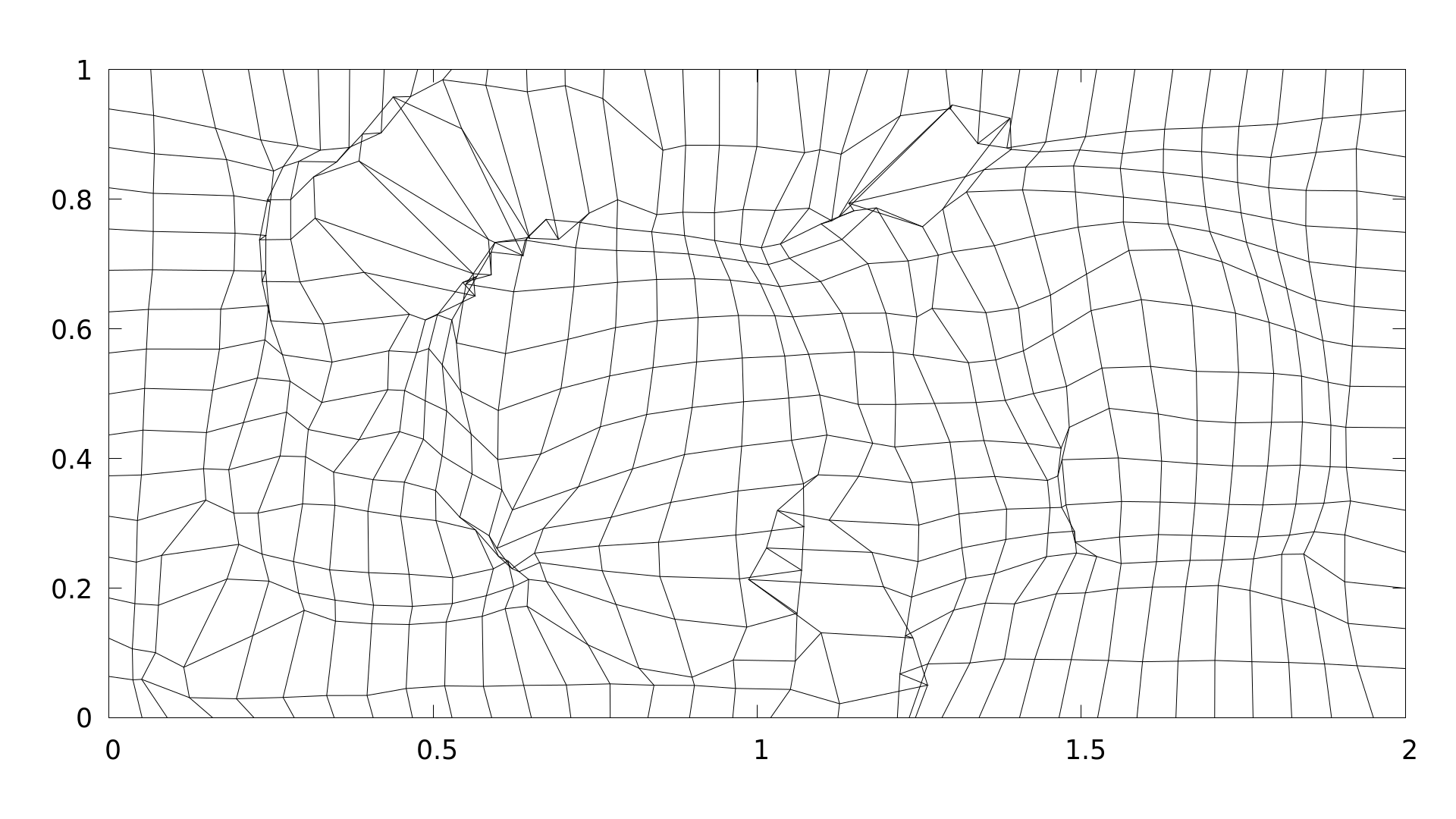}
  \label{fig:dg_rho_0_1}
  \end{subfigure}
 \hspace{-0.38cm}
 \vspace{-0.5cm}
  \begin{subfigure}[][][t]{0.243\textwidth}
    \includegraphics[width= \textwidth]{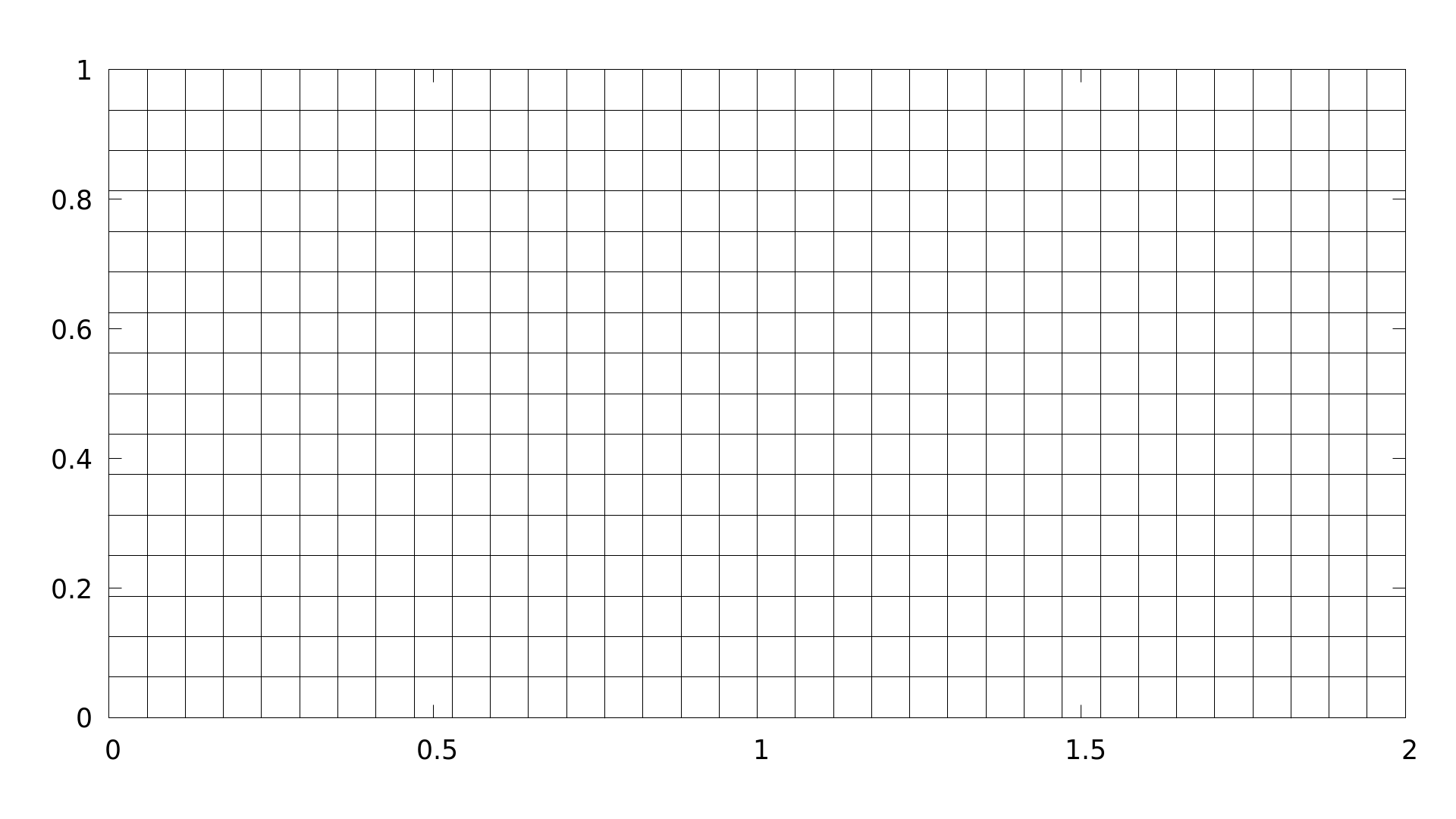}
  \label{fig:dg_rho_0_7}
  \end{subfigure}

    \caption{
      $\vp_{1,50}$ for Double Gyre ensemble at $\tau = 11$ with different weights $\rho$
      for the smoothness term. \\
      Left column from top to bottom:
      $\rho = 0, 0.2, 0.4$.\\
      Right column from top to bottom:
      $\rho = 0.6, 0.8, 1$.
    \label{fig:dg_rho}
  }
 \end{figure}

\paragraph*{Parameters.}
Our method requires some parameters:
the weight $\rho$ determines the balance between alignment and
rigidness (and also smoothness) for computing the domain
displacements.
Figure \ref{fig:dg_rho} shows a parameter study.
In summary, we require a significant influence of the rigidity term
$B$.
We chose a value of $\rho=0.2$, which worked for all examples.
Small variations of $\rho$ had no significant effect on the results.

Furthermore, we need to prescribe the grid defining the distortion.
The number of nodes defines the degrees of freedom of the displacement
function.
As the displacements are expected to be smooth, this grid size can be
chosen independently from the data resolution.
The choice is also a balance between accurate alignment and
computational cost, which increases with the degrees of freedom.
We chose 30--60 nodes along the larger extent of the rectangular
domains.

Note that we do not enforce bijective displacements by additional
penalty terms.
Figure \ref{fig:dg_rho} shows ``flipped'' cells in the
deformed grid for small $\rho$, which results in a displacement
mapping that cannot be inverted.
This may be a rare case for our choice of $\rho$, but it is a
perfectly possible case.
However, our approach does not strictly require a bijective
deformation of the domain.
We only evaluate the ``forward'' deformation, but never the inverse.

\paragraph*{3D case.}
We describe our approach for 2D flows.
The extension to 3D flows poses additional challenges to the general
setting, the numerical optimization, performance and finally also to
the visual representation.
However, we do not see any fundamental issue that prevents an extension
of the approach to 3D.
We leave the extension to 3D for future research.

\pagebreak
\bibliographystyle{eg-alpha-doi.bst}
\bibliography{literature.bib}


\end{document}